\documentclass[preprint2]{aastex63}

\usepackage{amsmath}
\usepackage{CJK}

\shorttitle{The Stick Filament in Orion A}
\shortauthors{Kong et al.}

\begin{document}
\begin{CJK*}{UTF8}{gbsn}

\title{The CARMA-NRO Orion Survey: Filament Formation via Collision-Induced Magnetic Reconnection - The Stick in Orion A}

\author{Shuo Kong (孔朔)}
\affiliation{Steward Observatory, University of Arizona, Tucson, AZ 85719, USA}
\affiliation{Department of Astronomy, Yale University, New Haven, CT 06511, USA}

\author{Volker Ossenkopf-Okada}
\affiliation{I.~Physikalisches Institut, Universit\"at zu K\"oln,
              Z\"ulpicher Str. 77, D-50937 K\"oln, Germany}

\author{H\'ector G. Arce}
\affiliation{Department of Astronomy, Yale University, New Haven, CT 06511, USA}

\author{John Bally}
\affiliation{Department of Astrophysical and Planetary Sciences, University of Colorado, Boulder, Colorado, USA}

\author{\'Alvaro S\'anchez-Monge}
\affiliation{I.~Physikalisches Institut, Universit\"at zu K\"oln,
              Z\"ulpicher Str. 77, D-50937 K\"oln, Germany}

\author{Peregrine McGehee}
\affiliation{Department of Earth and Space Sciences, College of the Canyons, Santa Clarita, CA 91355}

\author{S\"umeyye Suri}
\affiliation{Max Planck Institute for Astronomy,  Koenigstuhl 17, 69117 Heidelberg, Germany}

\author{Ralf S. Klessen}
\affiliation{Universit\"{a}t Heidelberg, Zentrum f\"{u}r Astronomie, Institut f\"{u}r Theoretische Astrophysik, Albert-Ueberle-Str. 2, 69120 Heidelberg, Germany}
\affiliation{Universit\"{a}t Heidelberg, Interdisziplin\"{a}res Zentrum f\"{u}r Wissenschaftliches Rechnen, Im Neuenheimer Feld 205, 69120 Heidelberg, Germany}

\author{John M. Carpenter}
\affiliation{Joint ALMA Observatory, Alonso de C\'ordova 3107 Vitacura, Santiago, Chile}

\author{Dariusz C. Lis}
\affiliation{Jet Propulsion Laboratory, California Institute of Technology, 4800 Oak Grove Drive, Pasadena, CA 91109, USA}

\author{Fumitaka Nakamura}
\affiliation{National Astronomical Observatory of Japan, 2-21-1 Osawa, Mitaka, Tokyo 181-8588, Japan}

\author{Peter Schilke}
\affiliation{I.~Physikalisches Institut, Universit\"at zu K\"oln,
              Z\"ulpicher Str. 77, D-50937 K\"oln, Germany}

\author{Rowan J. Smith}
\affiliation{Jodrell Bank Centre for Astrophysics, School of Physics and Astronomy, University of Manchester, Oxford Road, Manchester M13 9PL, UK}

\author{Steve Mairs}
\affiliation{East Asian Observatory, 660 N. A'ohoku Place, Hilo, Hawaii, 96720, USA}

\author{Alyssa Goodman}
\affiliation{Harvard-Smithsonian Center for Astrophysics, 60 Garden Street, MS 42, Cambridge, MA 02138, USA}

\author{Mar\'ia Jos\'e Maureira}
\affil{Max-Planck-Institute for Extraterrestrial Physics (MPE), Giessenbachstr. 1, D-85748 Garching, Germany}

\begin{abstract}
A unique filament is identified in the {\it Herschel} maps of the Orion A giant molecular cloud. The filament, which, we name the Stick, is ruler-straight and at an early evolutionary stage. 
Transverse position-velocity diagrams show two velocity components closing in on the Stick. The filament shows consecutive rings/forks in C$^{18}$O(1-0) channel maps, which is reminiscent of structures generated by magnetic reconnection. We propose that the Stick formed via collision-induced magnetic reconnection (CMR). We use the magnetohydrodynamics (MHD) code Athena++ to simulate the collision between two diffuse molecular clumps, each carrying an anti-parallel magnetic field. The clump collision produces a narrow, straight, dense filament with a factor of $>$200 increase in density. The production of the dense gas is seven times faster than free-fall collapse. The dense filament shows ring/fork-like structures in radiative transfer maps. Cores in the filament are confined by surface magnetic pressure. CMR can be an important dense-gas-producing mechanism in the Galaxy and beyond.
\end{abstract}


\section{Introduction}
\end{CJK*}

Filaments in molecular clouds play a key role in star formation. However, how exactly filaments form and evolve remains an open question. Recent theoretical work shows that filaments can form as a result of supersonic turbulence \citep[e.g.,][]{2016MNRAS.457..375F}, or through shear \citep[e.g.,][]{2013A&A...556A.153H,2020MNRAS.492.1594S}, or gravitational instability \citep[e.g.,][]{1998ApJ...506..306N}, or accretion of low density gas along magnetic field lines \citep[e.g.,][]{2016A&A...586A.136P}. Observationally, systematic surveys of filamentary structures are providing valuable information about filaments in molecular clouds \citep{2010A&A...518L.104M,2011A&A...529L...6A,2014prpl.conf...27A,2019A&A...623A.142S}. 

Large scale mapping observations of the Orion A molecular cloud through {\it Herschel}, JCMT, and the CARMA-NRO Orion survey \citep[][hereafter SK15, L16, K18, respectively]{2015A&A...577L...6S,Lane2016,2018ApJS..236...25K}
revealed a very prominent filament (hereafter the ``Stick'') with a straight morphology and rings and forks visible in individual channel maps. The ring/fork structures resemble those formed in magnetic reconnection (hereafter MR) simulations \citep[see, e.g., Figure 1 of][]{2011ApJ...735..102K}. In a region with anti-parallel magnetic fields, MR happens and magnetic islands (plasmoids) form. Material piles up in the islands when the field loses energy. The field topology changes and more reconnection follows, ending up in the ring/fork structure. Motivated by this, we explore the possibility that the Stick formed via MR.

The role of MR has been extensively discussed in \citet{1996MNRAS.279.1251L,1999ApJ...517..700L,1999ApJ...511..193V,2012ApJ...757..154L,2014SSRv..181....1L}. In this paper, we focus on modeling the formation of the Stick filament in the context of MR. We utilize magnetohydrodynamics (MHD) simulations with the Athena++ code \citep{2020ApJS..249....4S} to model the process. By reproducing key features in the Stick, we show that MR can be a viable mechanism for the formation of this peculiar filament. 

\section{Observational Features of the Stick}\label{sec:obs}

\begin{figure*}[htb!]
\centering
\epsscale{1.15}
\plotone{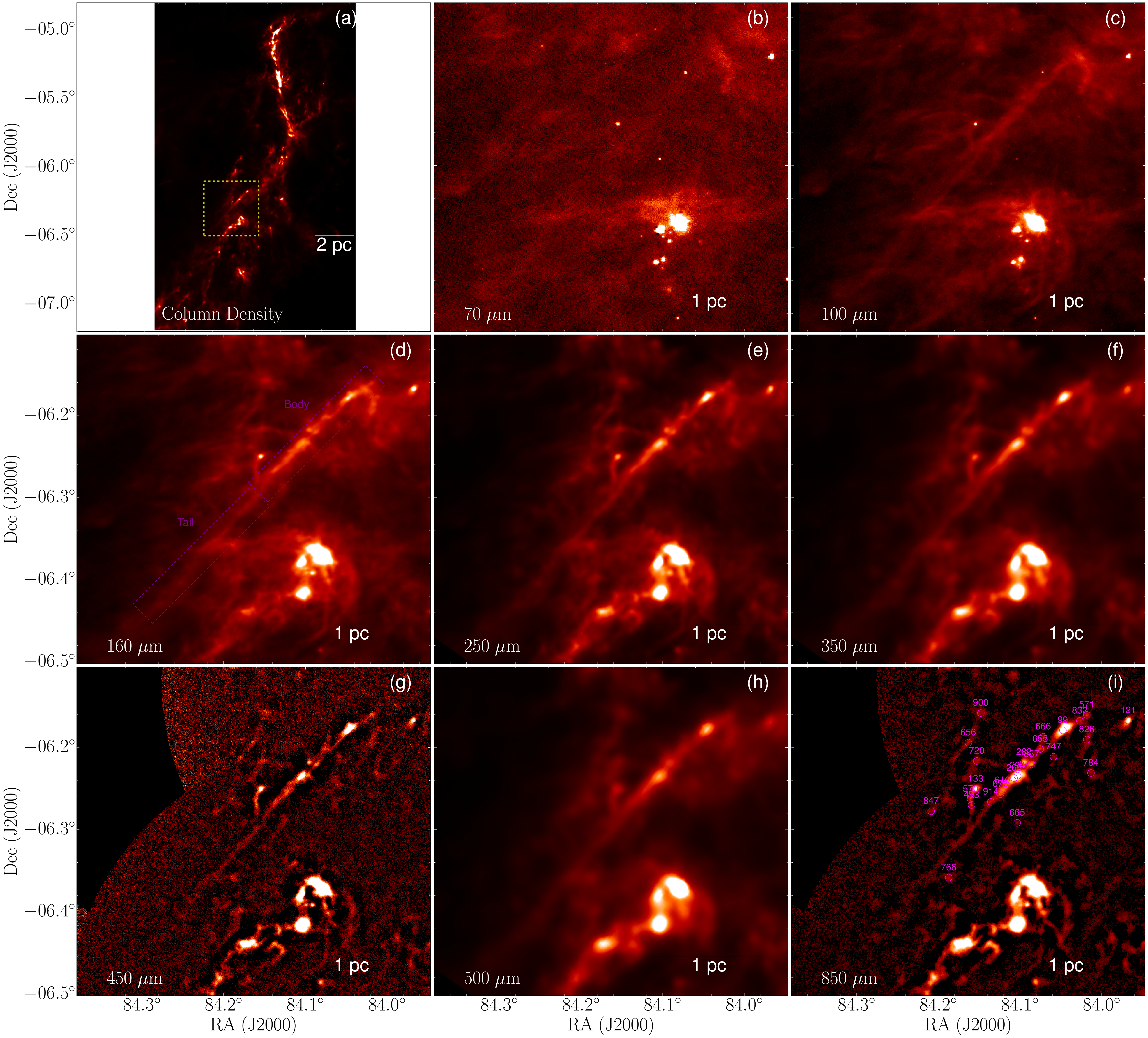}
\caption{
{\bf (a):} Column density map based on 
{\it Herschel} maps (SK15). The color stretch 
ranges from 0 to 10$^{23}$ cm$^{-2}$.
The yellow dashed box shows the 
field-of-view of other panels.
{\bf (b), (c), (d), (e), (f), (h):} 
Far infrared images from {\it Herschel} data.
The purple dashed rectangles in panel (d)
show the Stick body and tail.
{\bf (g), (i):} JCMT continuum images
from L16. The magenta circles are cores around the Stick (L16, also see Figure \ref{fig:c18omom0}). 
\label{fig:omc6}}
\end{figure*}

Using dust continuum data from SK15, we noticed a very particular filament  at the southern end of the famous integral-shaped filament \citep[ISF,][]{1987ApJ...312L..45B} in the Orion A molecular cloud. Figure \ref{fig:omc6} shows a summary view of the Stick at different wavelengths. It is the diagonal filament (from northwest to southeast) in each panel. Panel (a) shows an overall view of the Orion A cloud that includes the ISF and the L1641-N region. Panels (b)-(i) show the zoom-in view of the Stick at multiple wavelengths from 70 $\mu$m to 850 $\mu$m (SK15 and L16).

As shown in Figure \ref{fig:omc6}, the Stick is narrow, long, and straight. We divide the Stick into two parts (marked with two purple rectangles in Figure \ref{fig:omc6}d): 1) the bright body in the northwest; and 2)  the faint tail in the southeast. Each rectangle in the figure has a length of 1.4 pc and a width of 0.23 pc. We adopt a 390 pc distance to the Orion A cloud \citep[see][]{2018AJ....156...84K,2018A&A...619A.106G}. The aspect ratio for the body is, therefore, greater than 6 (the actual filament is narrower than the rectangle), and greater than 12 for the entire Stick. The \textit{Herschel} color temperature map derived by SK15 shows that the Stick is very cold, with dust temperatures of only $\sim$ 15 K. Table \ref{tab:stick} lists some basic physical properties of the Stick.

\begin{deluxetable*}{ccccccccc}[htb!]
\tablecolumns{9}
\tablewidth{0pt}
\tablecaption{Stick parameters \label{tab:stick}}
\tablehead{
\colhead{Parts} &
\colhead{$\overline{N}_H$} &
\colhead{$N_{\rm H,peak}$} &
\colhead{$M$} &
\colhead{$L$} &
\colhead{$m_L$} & 
\colhead{$\overline{T}_d$}\\
\colhead{} & 
\colhead{} & 
\colhead{} & 
\colhead{M$_\odot$} & 
\colhead{pc} & 
\colhead{M$_\odot$ pc$^{-1}$} & 
\colhead{K}}
\startdata
Body & 3.1$\times10^{22}$ cm$^{-2}$ & 1.4$\times10^{23}$ cm$^{-2}$ & 110 & 1.4 & 81 & 16 \\
~ & 0.073 g cm$^{-2}$ & 0.33 g cm$^{-2}$ & ~ & ~ & ~ & ~ \\
~ & 350 M$_\odot$ pc$^{-2}$ & 1600 M$_\odot$ pc$^{-2}$ & ~ & ~ & ~ & ~ \\
~ & $A_V\sim15$ mag & $A_V\sim70$ mag & ~ & ~ & ~ & ~ \\
Tail & 1.3$\times10^{22}$ cm$^{-2}$ & 2.9$\times10^{22}$ cm$^{-2}$ & 47 & 1.4 & 34 & 17 \\
~ & 0.030 g cm$^{-2}$ & 0.068 g cm$^{-2}$ & ~ & ~ & ~ & ~ \\
~ & 140 M$_\odot$ pc$^{-2}$ & 320 M$_\odot$ pc$^{-2}$ & ~ & ~ & ~ & ~ \\
~ & $A_V\sim6.5$ mag & $A_V\sim14$ mag & ~ & ~ & ~ & ~ \\
Overall & 2.2$\times10^{22}$ cm$^{-2}$ & 1.4$\times10^{23}$ cm$^{-2}$ & 160 & 2.8 & 58 & 16 \\
~ & 0.051 g cm$^{-2}$ & 0.33 g cm$^{-2}$ & ~ & ~ & ~ & ~ \\
~ & 250 M$_\odot$ pc$^{-2}$ & 1600 M$_\odot$ pc$^{-2}$ & ~ & ~ & ~ & ~ \\
~ & $A_V\sim11$ mag & $A_V\sim70$ mag & ~ & ~ & ~ & ~ 
\enddata
\tablecomments{The calculation is within the purple dashed boxes in Figure \ref{fig:omc6}(d).}
\end{deluxetable*}

\begin{figure*}[htb!]
\centering
\epsscale{1.}
\plotone{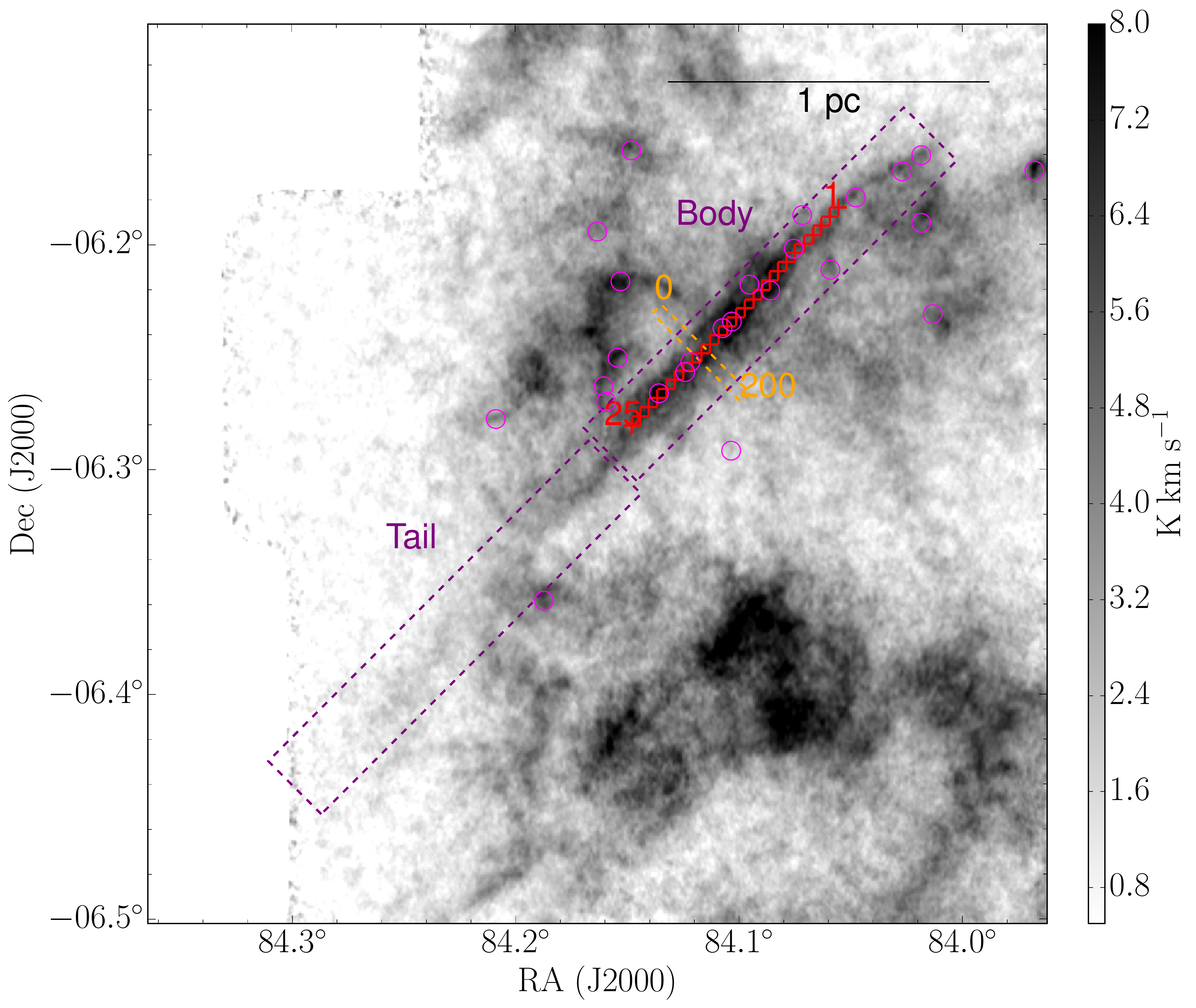}
\caption{
Integrated intensity map for C$^{18}$O(1-0).
The two black dashed rectangles mark the Stick
body and tail (same as Figure \ref{fig:omc6}d).
The red crosses show the central points of 
the PV-cuts. The orange dashed rectangle shows an
example of the PV-cut.
The orange numbers ``0'' and ``200'' at the two ends
of the PV-cut are offsets in arcsec.
The red numbers ``1'' and ``25'' mark the 25
PV-diagrams shown later in Figure \ref{fig:stickpv}.
The magenta circles are cores around the Stick (same as Figure \ref{fig:omc6}i).
\label{fig:c18omom0}}
\end{figure*}

L16  defined ten cores on the Stick filament using the JCMT 850 $\mu$m image (see Figures \ref{fig:omc6}(i) and \ref{fig:c18omom0}). All sub-mm cores on the main part of the Stick are starless based on L16. Core 666 is protostellar (L16) but at the edge of the Stick. A recent high-resolution study does not show any evidence of outflows from these cores \citep{2020ApJ...896...11F}. The lack of protostellar activity in the Stick cores suggests that they formed fairly recently. \citet[][hereafter K17]{2017ApJ...846..144K} used the GBT GAS survey \citep{2017ApJ...843...63F} to find that the majority of the cores are not gravitationally bounded but pressure confined. The Stick is probably at an early evolutionary stage. For reference, the northernmost core 99 has a free-fall time of $\sim 5.0\times10^4$ yr (a mass of 1.9 M$_\odot$ and an effective radius of 0.026 pc, see L16, K17). 

Recently, K18 published $^{12}$CO, $^{13}$CO and C$^{18}$O (1-0)  maps of the Orion A cloud as part of the CARMA-NRO Orion Survey. Subsequently, \citet{2019A&A...623A.142S} conducted a comprehensive study of the filaments using the  C$^{18}$O map from K18, where the Stick filament clearly stands out. In Figure \ref{fig:c18omom0}, we show the Stick in the C$^{18}$O integrated intensity map from K18. The C$^{18}$O data has a beam size of $10\arcsec\times8\arcsec$ and a spectral resolution of 0.22 km s$^{-1}$ (see K18 for more details). The C$^{18}$O emission nicely matches the filament at far infrared wavelengths. It is also clear that the Stick body has stronger C$^{18}$O emission while the Stick tail is  marginally  detected. 

\subsection{Temperature and CO abundances in the Stick}\label{subsec:stickta}

\begin{figure*}[htb!]
\centering
\epsscale{1.1}
\plotone{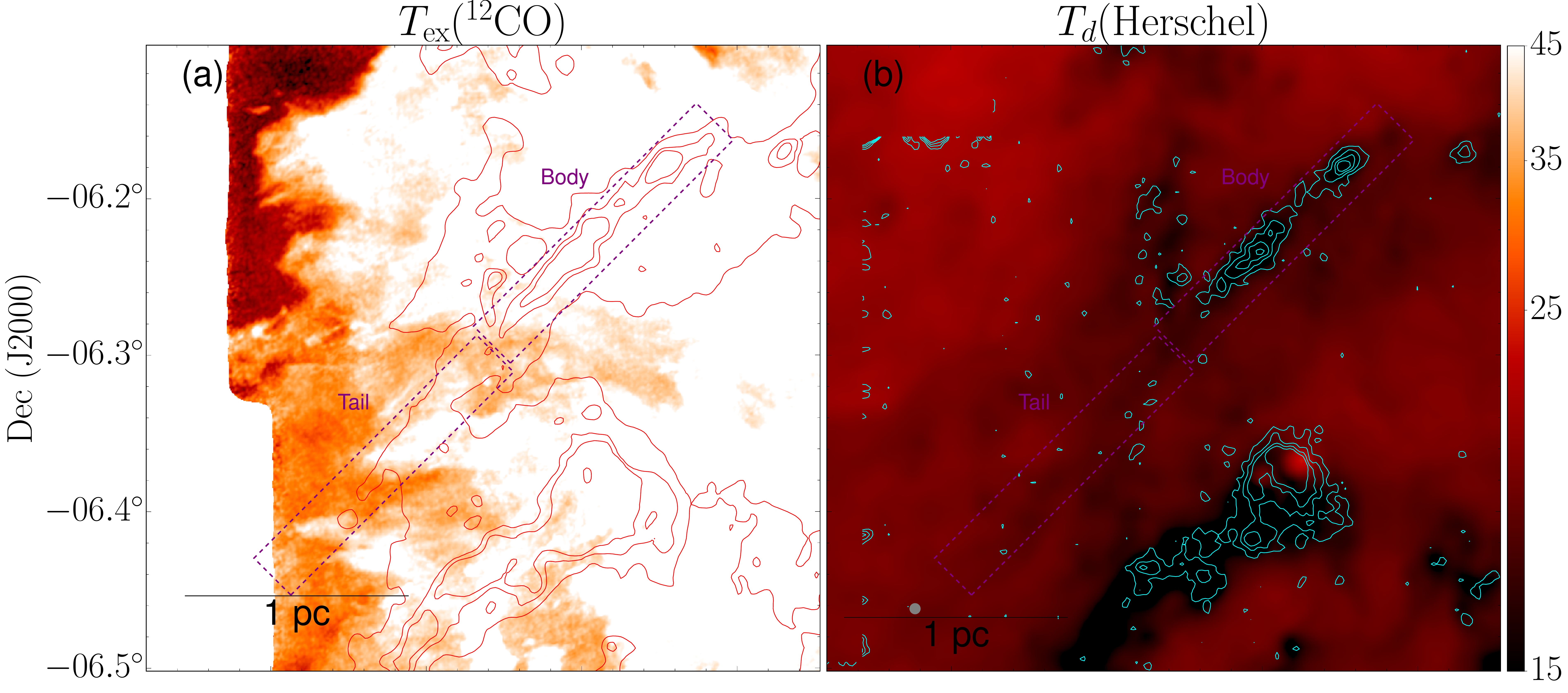}\\
\plotone{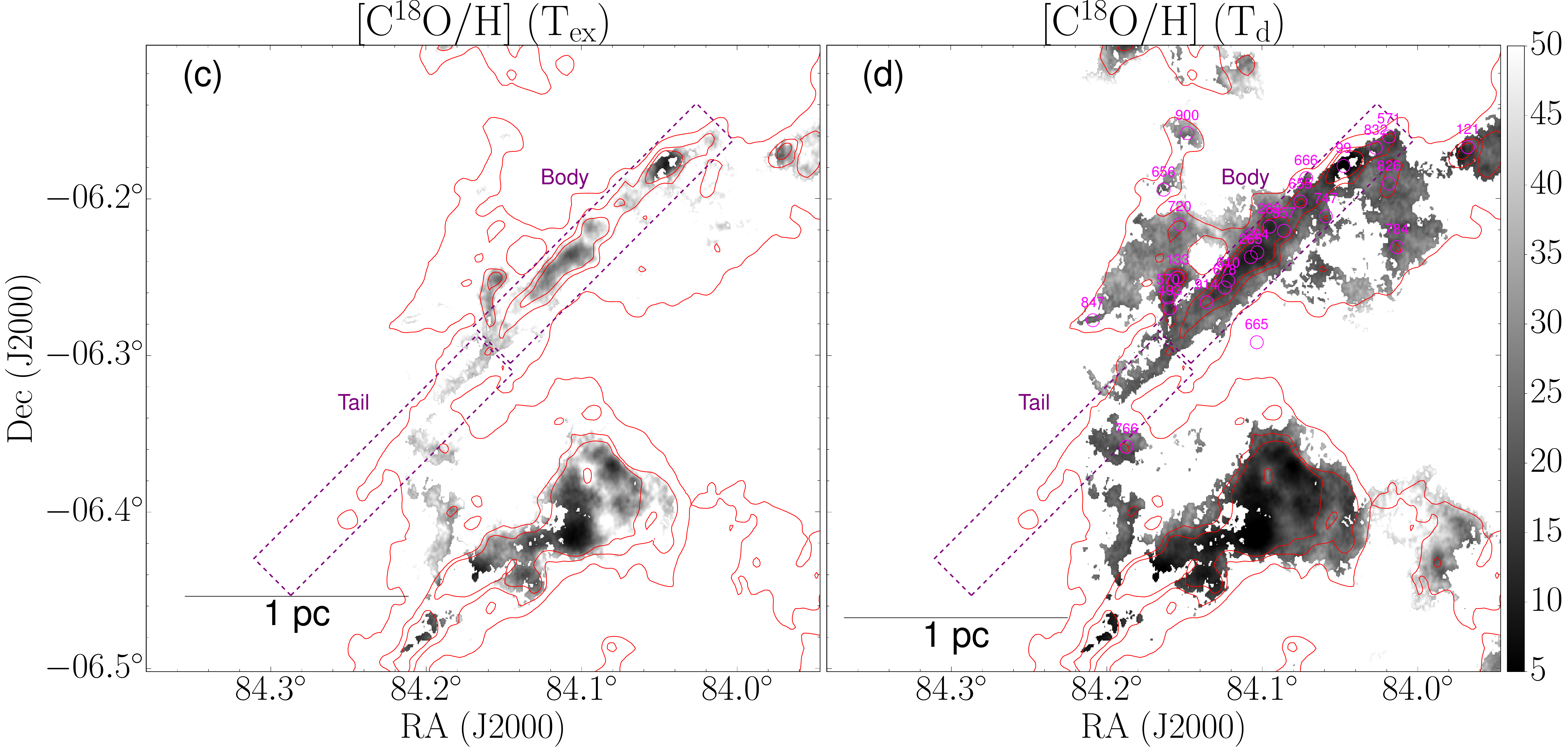}
\caption{
{\bf (a):}
$^{12}$CO excitation temperature map (K18) in square root scale. The color bar unit is K. The two purple dashed rectangles mark the Stick body and tail (same as Figure \ref{fig:omc6}d). The red contours show 1.4, 2.8, 4.2 $\times10^{22}$ cm$^{-2}$ column density levels (from the {\it Herschel} data in Figure \ref{fig:omc6}a).
{\bf (b):} 
{\it Herschel} dust temperature map (SK15) in square root scale. The cyan contours show the N$_2$H$^+$(1-0) integrated intensity at 0.7, 1.4, 2.1, 2.8 K km s$^{-1}$.
{\bf (c):}
C$^{18}$O abundance map [C$^{18}$O/H]. The color is scaled up by 10$^{8}$. The excitation temperature is assumed to be that of $^{12}$CO. The red contours are the same as panel (a).
{\bf (d):} 
The same as panel (c), but the excitation temperature is assumed to be the {\it Herschel} dust temperature (SK15). The magenta circles are cores around the Stick (L16). The red contours are the same as panel (a).
\label{fig:texc18oabun}}
\end{figure*}

In Figure \ref{fig:texc18oabun}(a), we show the excitation temperature map for the Stick (based on the $^{12}$CO emission, see K18). In Figure \ref{fig:texc18oabun}(b), we show the dust temperature based on {\it Herschel} results (\citealt{2014A&A...566A..45L}; SK15). The two panels have the same color scale. The $^{12}$CO emission, which is mostly optically thick, likely traces the warmer surface of the molecular cloud, so the excitation temperature is significantly higher than the {\it Herschel} dust temperature. In panels (c) and (d), we show the C$^{18}$O abundance [C$^{18}$O/H]. The C$^{18}$O column density is computed following K18, assuming that the C$^{18}$O(1-0) emission is optically thin. The computation is based on the temperatures in panels (a) and (b), respectively. Then, we divide the C$^{18}$O column density by the {\it Herschel} total column density (SK15) to obtain the abundances. In \S\ref{app:A} Figure \ref{fig:hist_t_abu}, we show that the column density based on the {\it Herschel} dust temperature is more robust than that based on the $^{12}$CO excitation temperature, which overestimates the column density by a factor of $\sim$2.

We can see in Figure \ref{fig:texc18oabun}(d) that the central region of the Stick body has lower [C$^{18}$O/H] than the outer region. This implies depletion in the dense cores, which is consistent with their low temperatures. Based on Figure \ref{fig:hist_t_abu}(b), the average C$^{18}$O abundance based on $T_{\rm dust}$ is about 2.0$\times10^{-7}$. If we assume a canonical C$^{18}$O abundance of 2.4$\times10^{-7}$ \citep{2008ApJ...680..371W}, we can see that overall the Stick body shows negligible C$^{18}$O depletion. As noted above, this is not the case for  the dense cores in the Stick.
For example, core 99 has the lowest abundance of $\sim$5.0$\times10^{-8}$
(see Figure \ref{fig:texc18oabun}(d)). Thus its CO depletion factor is $\sim$5. Core 294 has a C$^{18}$O abundance of $\sim$9.0$\times10^{-8}$, corresponding to a depletion factor of $\sim$3. Low CO depletion is expected in early evolutionary stages \citep{2005ApJ...619..379C}, and favors our earlier conjecture that the Stick filament is very young. 

\subsection{Kinematics of the Stick}\label{subsec:kin}

\begin{figure*}[htb!]
\centering
\epsscale{1.1}
\plotone{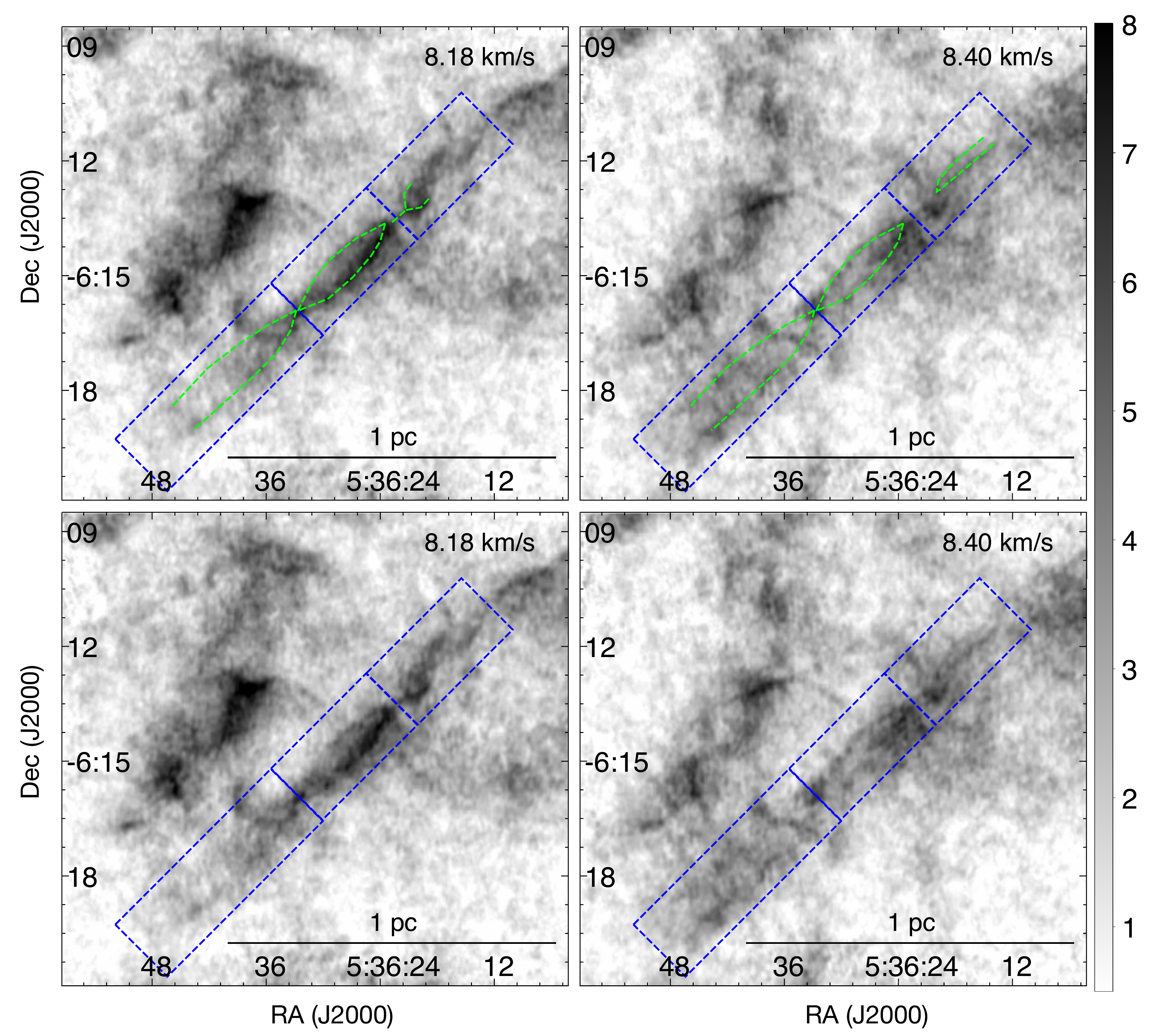}
\caption{
Channel maps for C$^{18}$O(1-0). The color bar has the unit of K. The number at the top right shows the velocity in km s$^{-1}$. The blue dashed boxes show the enclosing rectangles of the three rings/forks (see \S\ref{subsec:kin}). The first and second rows are the same, except that the green dashed curves in the first row highlight the rings/forks.
\label{fig:chan18}}
\end{figure*}

Figure \ref{fig:chan18} shows the C$^{18}$O channel maps that allow us to investigate the kinematics of the Stick. Here we only include the two channels in which the rings/forks are mostly visible (the complete channel maps for C$^{18}$O is in K18 Figure 28). The most striking features are the small arc-like filaments forming a chain of two-pronged structures, resembling tuning forks. We can see three of such structures enclosed in the three blue dashed rectangles. 
There are two tuning forks on two ends projecting toward northwest and southeast, while the one in the middle is more like a complete ring. Hereafter we simply call them ``ring/fork'' structures. The rings/forks are quite puzzling, but we believe they must be related to the formation mechanism of the Stick. 

\begin{figure*}[htb!]
\centering
\epsscale{1.1}
\plotone{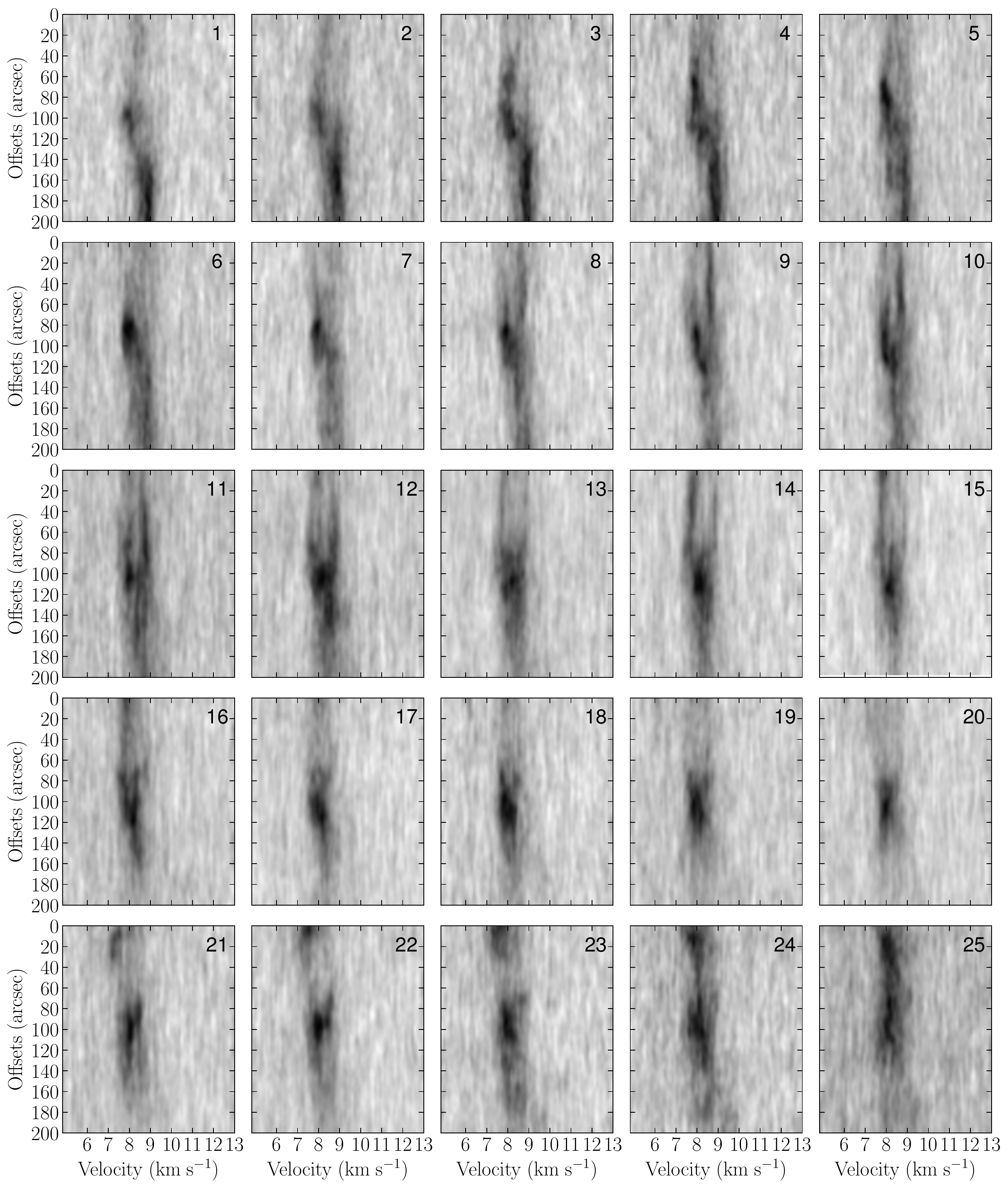}
\caption{
PV-diagrams for C$^{18}$O(1-0) transverse to the filament main axis.
The number at the top right is the PV-cut number in
Figure \ref{fig:c18omom0}. The zero offset is at the
northeast end of the red rectangle in Figure \ref{fig:c18omom0}.
\label{fig:stickpv}}
\end{figure*}

To probe the transverse kinematics of the filament, we make PV-diagrams along cuts perpendicular to the filament. 
Each path (PV-cut) is centered on the filament and extends symmetrically to both sides of the filament (see Figure \ref{fig:c18omom0}). Figure \ref{fig:stickpv} shows the twenty-five PV-diagrams we made along the filament body. 
In most of the diagrams, we see two velocity components, especially at offsets less than  100\arcsec\ (see Figure \ref{fig:c18omom0})
and most clearly in panels 8-16.
These velocity components have a separation of $\sim$ 1 km s$^{-1}$, one at $\sim$ 8 km s$^{-1}$ and the other at $\sim$ 9 km s$^{-1}$. Interestingly, a similar kinematic feature is seen in the NH$_3$ data (see \S\ref{app:A} and Figure \ref{fig:stickpvnh3}).

\section{Modeling the Stick Formation}

Based on the observational results above, we can summarize several main characteristics of the Stick. 
The most unique features are its straight morphology and the ring/fork structures in channel maps (Figure \ref{fig:chan18}). None of the other filaments in Orion A has such a ruler-straight shape, especially since the environment in this cloud is very chaotic. Moreover, the Stick seems to be a standalone object, while other filaments are interconnected. We believe that the chain of rings/forks and the unconventional straightness are not a coincidence and that they point to the formation mechanism of the Stick (see below).

Other features of the Stick include the double velocity components, its young evolutionary status, and its high density ($\sim10^5$ cm$^{-3}$). The high density makes the Stick stand out from its environment. The aforementioned free-fall time for core 99 corresponds to an average core density of $n_{\rm H}=8\times10^5$ cm$^{-3}$ (K17). The high density is broadly consistent with the fact that core 99 shows C$^{18}$O depletion and detection in dense gas tracers, including NH$_3$ and N$_2$H$^+$. Figure \ref{fig:texc18oabun}(b) shows the N$_2$H$^+$(1-0) integrated intensity map (only the central hyperfine component is included). The Stick has notable detection in N$_2$H$^+$, especially toward the dense cores. Toward core 99, the N$_2$H$^+$(1-0) central component reaches 1 K brightness temperature. Given the critical density of N$_2$H$^+$(1-0), we expect the densest part of the Stick reaches $n_{\rm H} \sim 10^5$ cm$^{-3}$.

Here we explore the possibility that the Stick formed via magnetic reconnection (hereafter MR). The main motivation for the hypothesis is that magnetic reconnection creates ring/fork-like structures, very similar to those we see in the C$^{18}$O channel maps. 

\subsection{Simulation Setup and 2D Tests}\label{subsec:2dtest}

\begin{figure*}[htb!]
\centering
\epsscale{0.55}
\plotone{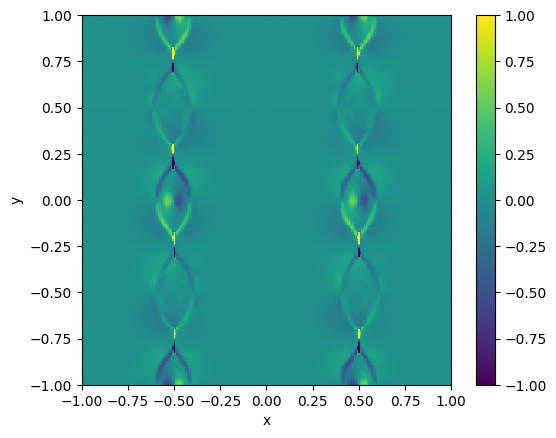}
\plotone{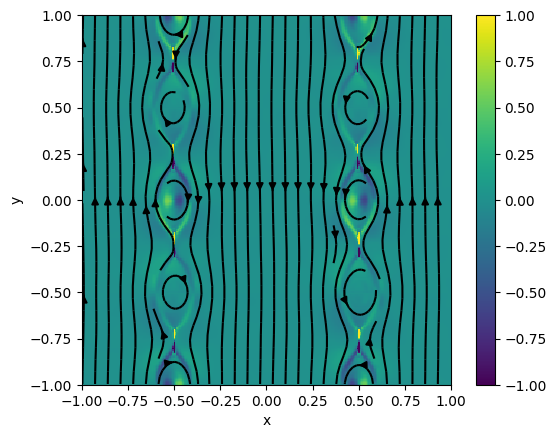}
\caption{
{\bf Left:} 
Velocity along y-axis in the simulation \mbox{MR\_2D}.
The color scale shows the velocity.
Note the ring/fork-like structures,
with knots in between.
{\bf Right:}
The color map is the same as left.
The arrows show the magnetic field
stream lines. All numbers are in code units.
\label{fig:mr2d}}
\end{figure*}

We use the latest public version of the Athena++ code \citep[version 19.0,][]{2020ApJS..249....4S}. In particular, we model the compressible, isothermal, inviscid MHD, with self-gravity and Ohmic resistivity. A second-order accurate van Leer (VL2) time integrator is used, with divergence-free constraints on magnetic fields at every time step ensured. A Courant-Friedrichs-Lewy (CFL) number of 0.3 is adopted in the simulations. A second-order piecewise linear method (PLM) is used for spatial reconstruction and the HLLD Riemann solver is adopted. To compute self-gravity, the Poisson equation is solved with the fast Fourier transforms (FFTs) module on the uniform Cartesian grid. The adaptive mesh refinement (AMR) capability is under development and it is not yet available (Tomida et al. in prep.). We use periodic boundary conditions for all dimensions. 

A mass density of $3.84\times10^{-21}$ g cm$^{-3}$ is set as the code unit. The density corresponds to $n_{\rm H_2}$=840 cm$^{-3}$ (assuming a mean molecular mass per H$_2$ of $\mu_{\rm H_2}=2.8 m_\textrm{H}$) and a free-fall timescale of 1.1 Myr. The code unit for time is 2.0 Myr. The length unit is 1.0 pc. The velocity unit is 0.51 km s$^{-1}$. With this setup, the gravitational constant is $G=1$. The magnetic field code unit is 3.1 $\mu$G. Hereafter, all numerical analyses will be in code units (accompanied by physical units in some cases). 

We start with exploratory runs in 2D. Figure \ref{fig:mr2d} shows the result from one simulation (hereafter \mbox{MR\_2D}). The simulation domain is a 2$\times$2 box with a 256$^2$ uniform grid. The density is set to 1 in the box. For $-0.5<x<0.5$, the magnetic field $B_y$ is set to -5 while $B_x$=0. The rest of the domain has $B_y$=5 and $B_x$=0. Hence, at $x=\pm$0.5, the field lines invert and MR happens. We apply a velocity perturbation along the x-axis to trigger the reconnection. The simulation setup basically follows the ``Current Sheet'' problem in \citet{2005JCoPh.205..509G}. The snapshot is taken at t=0.5. 

In Figure \ref{fig:mr2d}, we can see chains of rings forming at the interfaces of inverted B-fields (current sheets). The right panel of Figure \ref{fig:mr2d} shows how the ring-like structures form around the magnetic islands. MR takes place between the islands where magnetic energy dissipates into kinetic energy. The gas with the additional kinetic energy moves along the field line that has the ring-knot shape. Material from neighboring MR sites create the ring/fork shapes along the current sheet. Note that similar results have been reported in other numerical studies. For instance, see Figure 1 in \citet{2011ApJ...735..102K}.

We ran the simulation with different values of the Ohmic resistivity, $\eta_{\rm ohm}$, with value of 0.1, 0.01, $10^{-3}$, $10^{-4}$, $10^{-5}$.  We then compare these runs with a simulation without $\eta_{\rm ohm}$, i.e., only including the numerical resistivity. By visually inspecting the structures, we find the code numerical resistivity is $\la$ 10$^{-4}$ (1.5$\times10^{19}$ cm$^2$ s$^{-1}$). The result shown in Figure \ref{fig:mr2d} uses $\eta_{\rm ohm}=10^{-5}$. 

\subsection{Problems Going from 2D to 3D}

\begin{figure}[htb!]
\centering
\epsscale{1.1}
\plotone{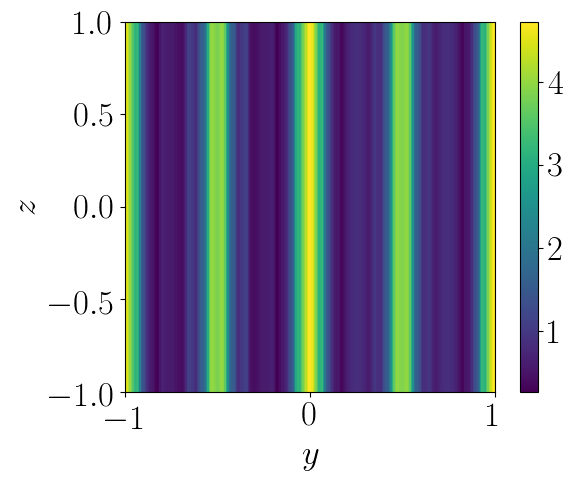}
\caption{
Density slice in y-z plane at x=-0.5 in the simulation \mbox{MR\_3D}.
\label{fig:mr3d}}
\end{figure}

While the 2D MR results show some promising features that resemble the Stick, there are several details that need to be investigated and discussed  if we attempt to explain the Stick formation with MR. First, it is hard to picture an initial condition of anti-parallel B-fields in a single molecular cloud (like Figure \ref{fig:mr2d}). The fields should stay side by side waiting for MR to happen. Probably the only plausible solution is to have two clouds coming from far away carrying anti-parallel B-fields. They meet and collide so MR could happen. Second, even if we assume cloud collision and simply extend the 2D simulation in \mbox{MR\_2D} to 3D by adding the z-axis, we will not form Stick-like filaments along the y-axis like \mbox{MR\_2D}. Each magnetic island will extend along the z-axis and become a cylinder along the z-axis because every x-y plane is essentially the same. Figure \ref{fig:mr3d} illustrates the cylinders by showing the y-z plane in a new simulation \mbox{MR\_3D}, which simply adds the third dimension to \mbox{MR\_2D}.

Third, also the most critical issue is that under the current configuration, the dense filament forms parallel to the magnetic fields. No matter how we try to ``manipulate'' the initial condition, we can hardly find a way to have an orthogonal component of the B-field relative to the filament. However, as the recent Planck polarization result showed, the plane-of-sky B-field orientation is almost perpendicular to the Stick filament \citep{2019A&A...629A..96S}. Moreover, we have known that the line-of-sight B-field orientation flips on both sides of the Orion A cloud \citep{1997ApJS..111..245H,2019A&A...632A..68T}. That is also incompatible with our initial conditions in \mbox{MR\_2D} or \mbox{MR\_3D}. 

Taking all these issues into consideration, perhaps the only compatible initial condition is that two molecular clumps collide at the location of the Stick. One clump carries B-fields pointing toward us while the other has B-fields pointing away from us. Thus we have flipped B-fields along the line of sight. Meanwhile, the B-fields of the two clumps are tilted along the x-axis (in our case the RA direction) so there is a plane-of-sky component  B-field that is orthogonal to the Stick. In fact, \citet{2012ApJ...746...25N} have seen the possibility of cloud-cloud collision in the L1641-N region (the bright region below the Stick).

\begin{figure}[htb!]
\centering
\epsscale{1.1}
\plotone{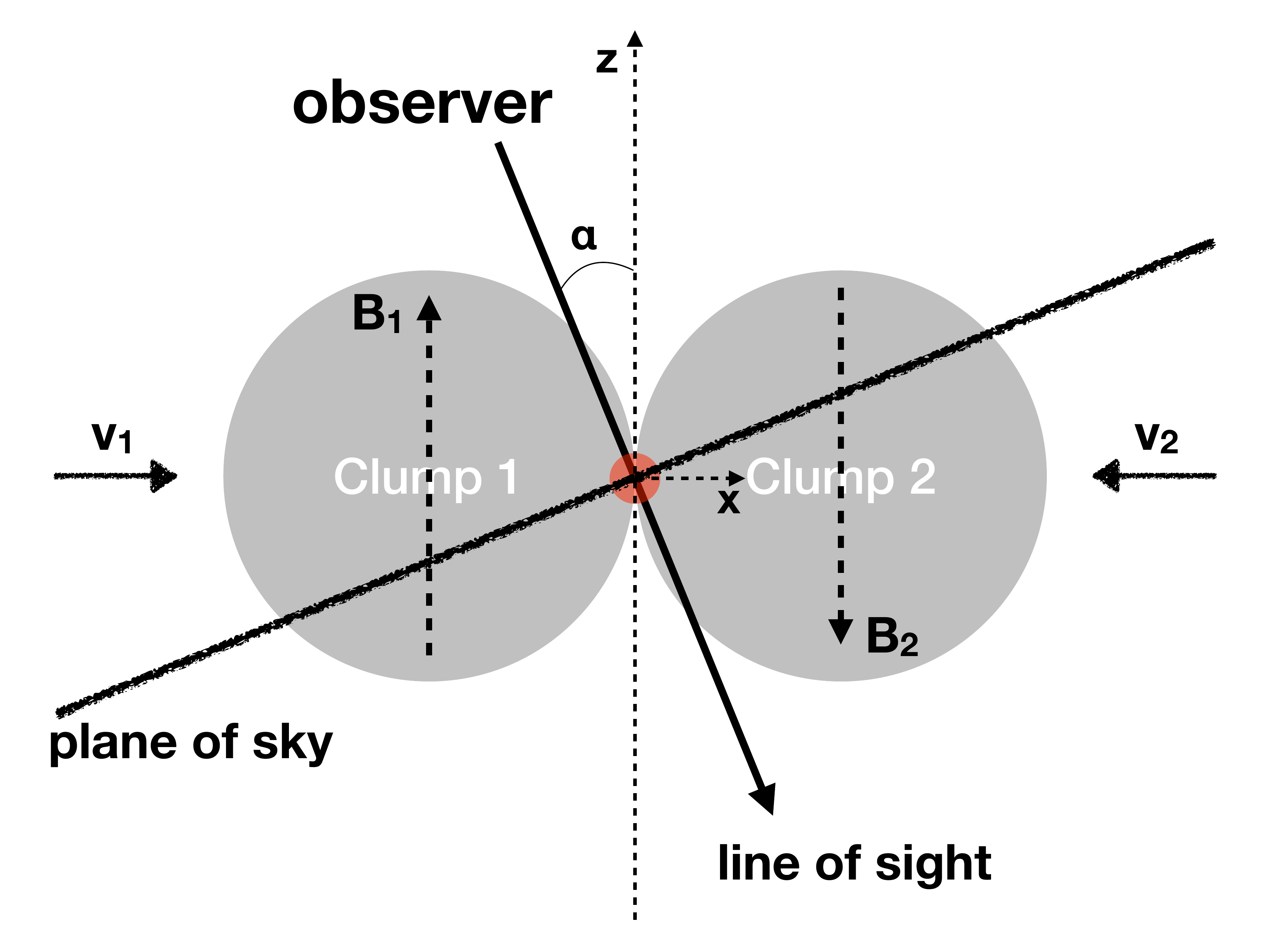}
\caption{
A schematic look of a possible initial condition of Stick formation. The Cartesian coordinate axes (x-axis and z-axis) are the dashed arrows with an origin at the Stick. The y-axis points into the screen and the filament is on the y-axis. The red circle in the middle shows the cross section of the Stick that stretches in and out of the screen. The two incoming clumps (with velocities $\bf{v_1}$ and $\bf{v_2}$) carry anti-parallel B-fields. We (as the observer) observe the collision at a tilted angle $\alpha$ (\S\ref{subsec:rt}).
\label{fig:ic}}
\end{figure}

Figure \ref{fig:ic} shows the schematic of the initial conditions of the fiducial model. Two clumps come together with collision velocities ${\bf v_1}$ and ${\bf v_2}$. Clump 1 carries a B-field pointing toward us (tilted) while clump 2 has a B-field in the opposite direction. The Stick is at the collision contact point and we (the observers) witness the collision at an angle through the clumps. In this setup, the line-of-sight B-field orientations are consistent with the Zeeman measurements from \citet{1997ApJS..111..245H} and the plane-of-sky B-field orientation is also consistent with the Planck result \citep{2019A&A...629A..96S}. Note that the polarization measurement of the plane-of-sky B-field orientation does not tell us the vector direction of the B-field. In our setup, we observe the plane-of-sky B-field orientation that is orthogonal to the Stick even though it is caused by anti-parallel B-fields. Now the question is, can we form a Stick-like filament as shown in Figure \ref{fig:ic}. It also has to match the observational characteristics, especially the ring/fork structures, the straight morphology, the elevated density, and the PV-diagrams.

\subsection{3D Modeling of the Stick}\label{subsec:icmrcol}

Here we follow the idea from the previous section to set up the simulation. First, we introduce the fiducial model \mbox{MRCOL} (short name for ``MR collision'', Figure \ref{fig:ic}). We adopt a uniform gas density of $\rho_1$=$\rho_2$=0.5 ($n_{\rm H_2}$=420 cm$^{-3}$) for the two clumps. The choice of density has two considerations. First, the Stick stands out from its surroundings, as shown earlier. So the colliding clumps (which are part of the immediate environment around the Stick) cannot be too dense. Second, we still want to limit the physical processes to the ``molecular regime'', meaning that MR creates the Stick from molecular gas (so the clump gas density cannot be too low). For CO-bright molecular gas a density of a few 100 cm$^{-3}$ is required \citep[e.g.][]{GloverClark2012}. The ambient gas density is set to $\rho_{\rm amb}$=0.05, i.e., a factor of 10 smaller than the clump density. 

We adopt an isothermal equation of state, where the gas temperature is set to 15 K. The choice of the temperature follows the maps of the dust temperature from {\it Herschel} and the NH$_3$ kinetic temperature from GBT (see \S\ref{subsec:stickta} and Figure \ref{fig:hist_t_abu}), which show the temperature in the region around the Stick uniformly has a value of about 15 K. The isothermal sound speed $c_s$ is 0.29 km s$^{-1}$, corresponding to $c_s$=0.58 in the code unit. The gas pressure is computed as $P=\rho c_s^2$. The clumps are colliding with a relative velocity of 2.0 km s$^{-1}$ ($v_{\rm 1,x}$=2.0 and $v_{\rm 2,x}$=-2.0 in code units).

The amplitude of the B-field is set to 10 $\mu$G (3.2 in code units). Of the eight HI velocity components shown in \citet{1997ApJS..111..245H}, component 7 best matches the $v_{\rm lsr}$ of the Stick at $\sim$8 km s$^{-1}$. Moreover, component 7 is the only one with absorption. Absorption features in HI spectra have been found to be associated with CO, $^{13}$CO, and C$^{18}$O gas in other molecular clouds \citep{2003ApJ...585..823L}. Interestingly, of the many observation points in \citet{1997ApJS..111..245H}, only four of them had B-field measurements for component 7 (see their Table 1). All of them are near the Stick location ($l$, $b$) $\approx$ (210\arcdeg, -19.5\arcdeg). Their line-of-sight B-fields are $\sim$ 10 $\mu$G. \citet{2012ARA&A..50...29C} summarized the relationship between B-field strength and gas density. At $n_H<10^3$ cm$^{-3}$, the B-field strength is $\sim$10 $\mu$G. Taking these into consideration, we set $|\bf{B_1}|=|\bf{B_2}|$ = 10 $\mu$G for the two clumps ($B_{\rm 1,z}=3.2$ and $B_{\rm 2,z}=-3.2$ in code unit).

\begin{figure}[htb!]
\centering
\epsscale{1.}
\plotone{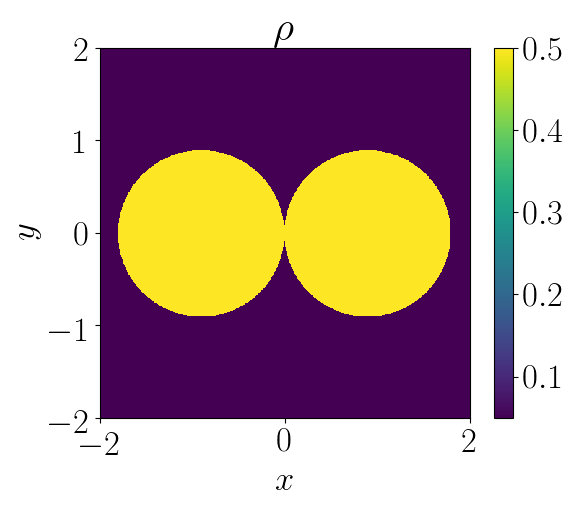}\\
\plotone{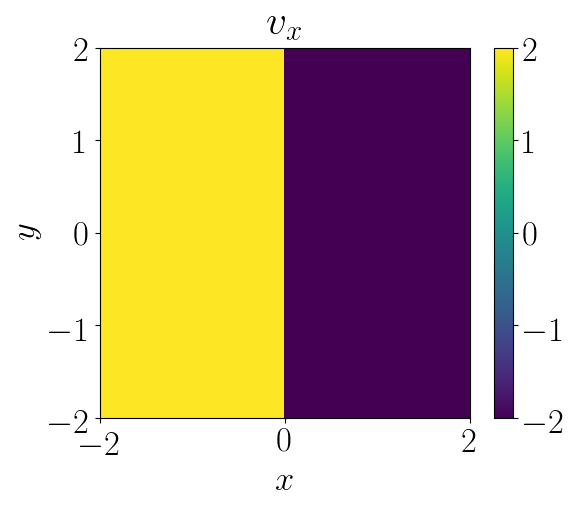}\\
\plotone{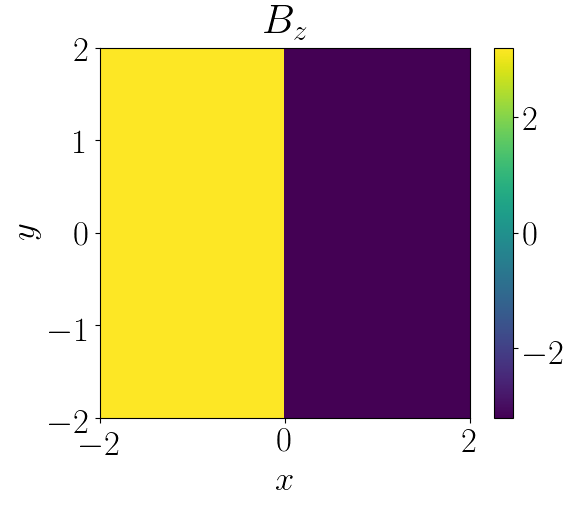}\\
\caption{
Initial conditions for \mbox{MRCOL} in code units (fiducial, see \S\ref{subsec:icmrcol} for the model description and \S\ref{subsec:2dtest} for unit conversion).
\label{fig:icmrcol}}
\end{figure}

Figure \ref{fig:icmrcol} shows the setup of the fiducial model \mbox{MRCOL}. The top panel shows a slice plot of density in the z=0 plane. Two spherical clumps are about to collide, each having a radius of 0.9 ($R_1=R_2=0.9$, corresponding to 0.9 pc). Their centers are set to $\bf{x_1}$ = (-0.9, 0, 0) (hereafter Clump1) and $\bf{x_2}$ = (0.9, 0, 0) (hereafter Clump2). The middle panel shows the colliding velocity. Clump1 and its ambient gas (x$<$0) are moving with a velocity $\bf{v_1}$ = (2.0, 0, 0.5) while Clump2 and its environment has $\bf{v_2}$ = (-2.0, 0, -0.5).
The velocity in the z-axis adds in a shear motion to the collision ($v_{\rm 1,z}=0.5$ and $v_{\rm 2,z}=-0.5$). It determines whether the collision is exactly head-on. This degree of freedom is important for cloud collisions because exact head-on collisions should be rare. One could also think of it as having a non-zero impact parameter. The choice of the shear velocity is rather arbitrary. We simply put in a relatively small value. The bottom panel shows the initial B-field. Clump1 and its environment (x$<$0) has $\bf{B_1}$ = (0, 0, 3.2) while Clump2 and its environment has $\bf{B_2}$ = (0, 0, -3.2). 

The Ohmic resistivity is set to $\eta_{\rm ohm}=0.001$ (1.5$\times10^{20}$ cm$^2$ s$^{-1}$) which is about an order of magnitude higher than the numerical resistivity ($\la10^{-4}$ in code unit or $\la1.5\times10^{19}$ cm$^2$ s$^{-1}$). 
This value of  $\eta_{\rm ohm}$ was chosen by increasing $\eta_{\rm ohm}$ from the numerical resistivity until the resulting filament shows the best match in the rings/forks structures seen in the molecular line maps of the Stick. 

\begin{deluxetable*}{@{\extracolsep{4pt}}cccccccccccc}[htb!]
\tablecaption{Models \label{tab:ic}}
\tablehead{
\colhead{Model} &
\colhead{Section} &
\colhead{Grid} &
\colhead{$\rho_{\rm amb}$} &
\multicolumn{4}{c}{Clump1} &
\multicolumn{4}{c}{Clump2} \\
\cline{5-8}
\cline{9-12}
\colhead{} &
\colhead{} &
\colhead{} &
\colhead{} & 
\colhead{$\rho_1$} & 
\colhead{$B_{\rm 1,y}$} &
\colhead{$B_{\rm 1,z}$} & 
\colhead{$v_{\rm 1,z}$} & 
\colhead{$\rho_2$} & 
\colhead{$B_{\rm 2,y}$} & 
\colhead{$B_{\rm 2,z}$} & 
\colhead{$v_{\rm 2,z}$}
}
\startdata
\mbox{MRCOL} (Fiducial) & \ref{subsec:icmrcol} & 512$^3$ & 0.05 & 0.5 & 0 & 3.2 & 0.5 & 0.5 & 0 & -3.2 & -0.5 \\
\mbox{MRCOL$_{\rho_0=0.7}$} & \ref{subsec:rt} & 512$^3$ & 0.05 & 0.7 & 0 & 3.2 & 0.5 & 0.7 & 0 & -3.2 & -0.5 \\
\mbox{COL\_noB} & \ref{subsec:mrcolre} & 512$^3$ & 0.05 & 0.5 & 0 & 0 & 0.5 & 0.5 & 0 & 0 & -0.5 \\
\mbox{COL\_sameB} & \ref{subsec:mrcolre} & 512$^3$ & 0.05 & 0.5 & 0 & 3.2 & 0.5 & 0.5 & 0 & 3.2 & -0.5 \\
\mbox{MRCOL\_uniform} & \ref{subsec:ph} & 512$^3$ & 0.5 & 0.5 & 0 & 3.2 & 0.5 & 0.5 & 0 & -3.2 & -0.5 \\
\mbox{MRCOL2D} & \ref{subsec:ph} & 512$^2$ & 0.05 & 0.5 & 3.2 & - & - & 0.5 & -3.2 & - & - \\
\mbox{MRCOL2D\_1024} & \ref{subsec:ph} & 1024$^2$ & 0.05 & 0.5 & 3.2 & - & - & 0.5 & -3.2 & - & - \\
\mbox{MRCOL2D\_2048} & \ref{subsec:ph} & 2048$^2$ & 0.05 & 0.5 & 3.2 & - & - & 0.5 & -3.2 & - & - \\
\mbox{MRCOL\_tiltB20} & \ref{subsec:kirkcores} & 512$^3$ & 0.05 & 0.5 & 0 & 3.2 & 0 & 0.5 & 1.0 & -3.0 & 0 \\
\mbox{MRCOL\_tiltB90} & \ref{subsec:kirkcores} & 512$^3$ & 0.05 & 0.5 & 0 & 3.2 & 0 & 0.5 & 3.2 & 0 & 0 \\
\mbox{MRCOL\_turb} & \ref{subsec:kirkcores} & 512$^3$ & 0.05 & 0.5 & 0 & 3.2 & 0 & 0.5 & 0 & -3.2 & 0
\enddata
\tablecomments{All values are in code units (\S\ref{subsec:2dtest}). Specifically, the density unit is $3.84\times10^{-21}$ g cm$^{-3}$ ($n_{\rm H_2}$=840 cm$^{-3}$). The time is 2.0 Myr. The length unit is 1.0 pc. The velocity unit is 0.51 km s$^{-1}$. The magnetic field unit is 3.1 $\mu$G. All dimensions have a length scale of 4 pc.}
\end{deluxetable*}

Table \ref{tab:ic} lists all models described in \S\ref{subsec:icmrcol} and beyond in this study.  The computation domain has a scale of 4 pc for each dimension. The coordinate system is defined in Figure \ref{fig:ic}. In all models, we adopt the Ohmic resistivity $\eta_{\rm ohm}=0.001$, the temperature of 15 K, and clumps radii of 0.9. Clump1 has a x-direction velocity $v_{\rm 1,x}=2.0$ and Clump2 has $v_{\rm 2,x}=-2.0$.

\subsection{Model Results}\label{subsec:mrcolre}

\begin{figure}[htb!]
\centering
\epsscale{1.}
\plotone{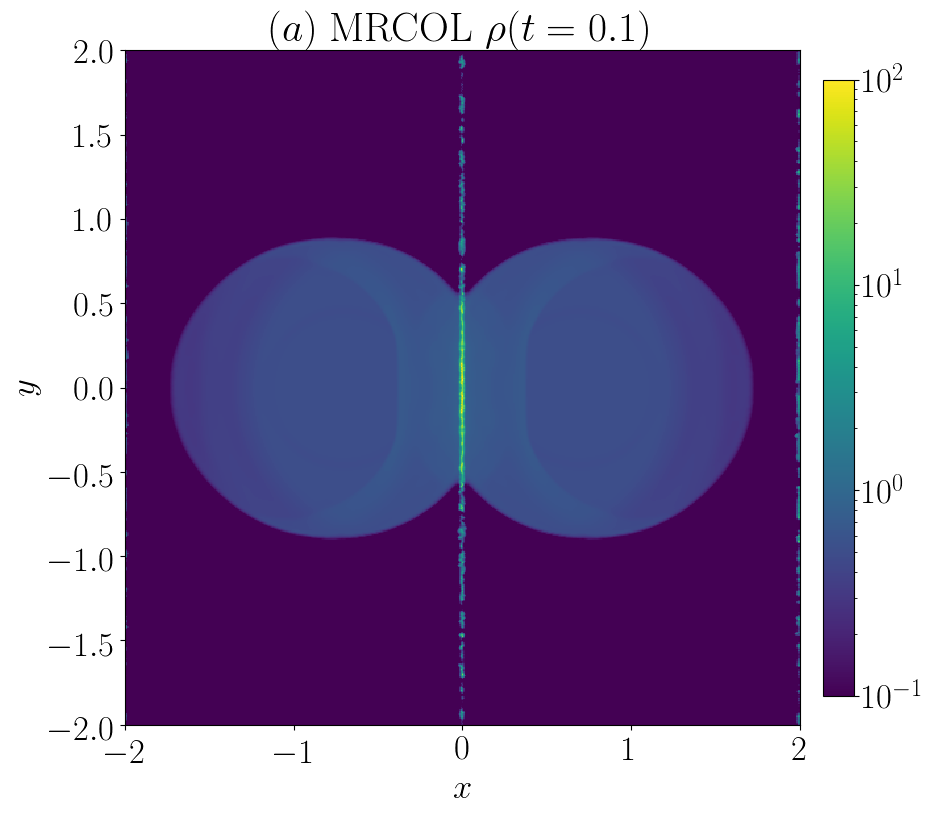}
\plotone{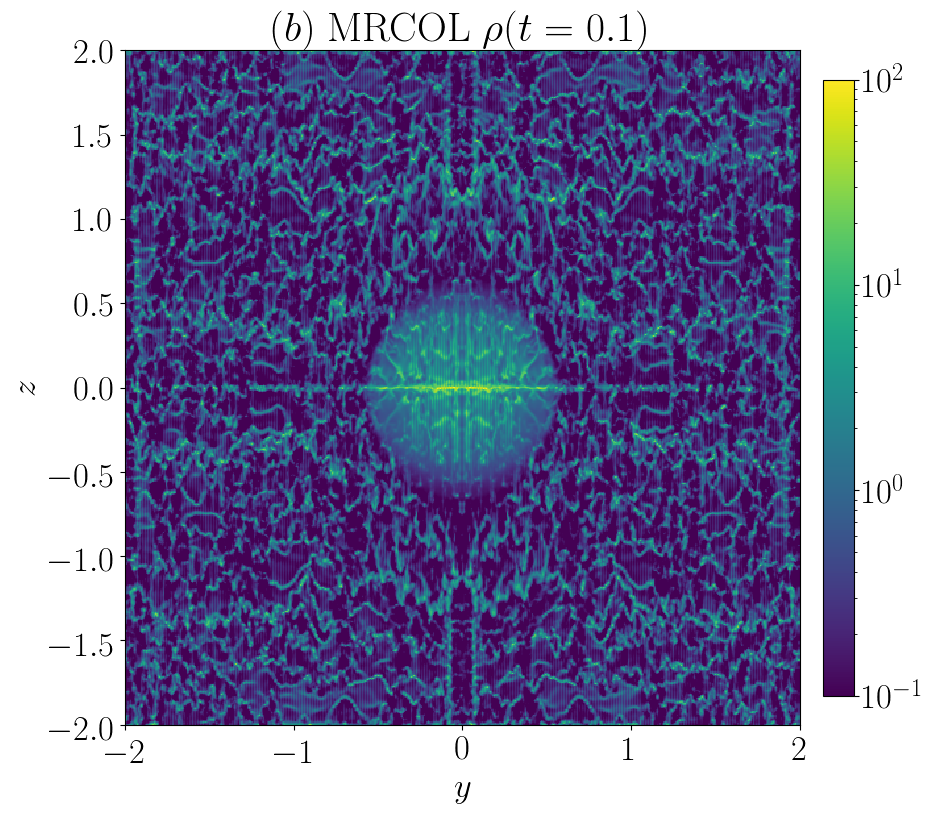}
\plotone{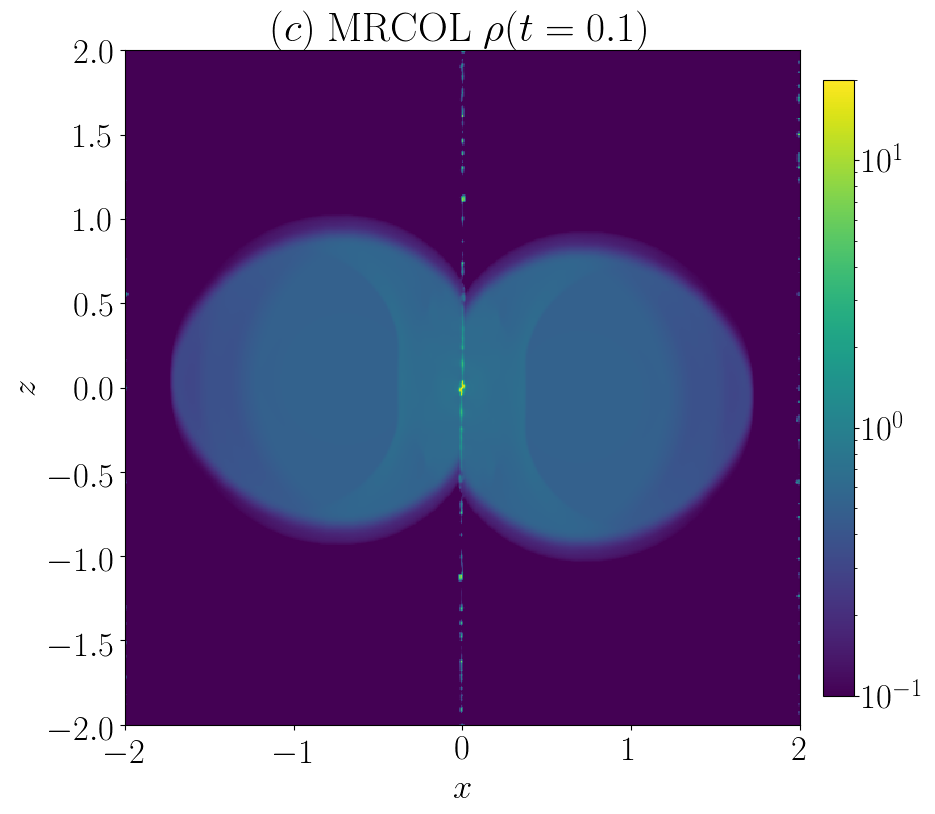}
\caption{
A snapshot of \mbox{MRCOL} density at t=0.1 (0.2 Myr). Panels (a), (b), (c) show the z=0, x=0, y=0 planes, respectively. The color scale is logarithmic. Panels (a)(b) have the same color range 0.1-100. Panel (c) color range is 0.1-10. See \S\ref{subsec:2dtest} for unit conversions.
\label{fig:mrcol}}
\end{figure}

Figure \ref{fig:mrcol} shows a snapshot of the  density output at t=0.1 (0.2 Myr) of the \mbox{MRCOL} model. In panel (a), Clump1 (left) and Clump2 (right) collide and create a density enhancement in the x=0 plane. The maximum density reaches $\rho_{\rm max}~\sim$ 138 ($n_{\rm H_2}=1.2\times10^5$ cm$^{-3}$). Recall that the clump density was initially 0.5 ($n_{\rm H_2}=420$ cm$^{-3}$) initially. The collision creates a density enhancement of more than two orders of magnitude. Interestingly, the high-density gas forms a filament. In panel (b), we show the x=0 plane where dense gas is generated. Indeed, we see the dense gas at the collision front concentrating in a filament, surrounded by a diffuse  ``pancake'' which reaches a  density of about 1. In the remaining area of the x=0 plane, there are many dense wiggles. Panel (c) shows the x-z plane that crosses the origin. Here we look at the cross-section of the filament. For this slice, the highest density (in the filament) is $\sim$20. Again we see the density enhancement in the x=0 plane. 

\begin{figure}[htb!]
\centering
\epsscale{1.}
\plotone{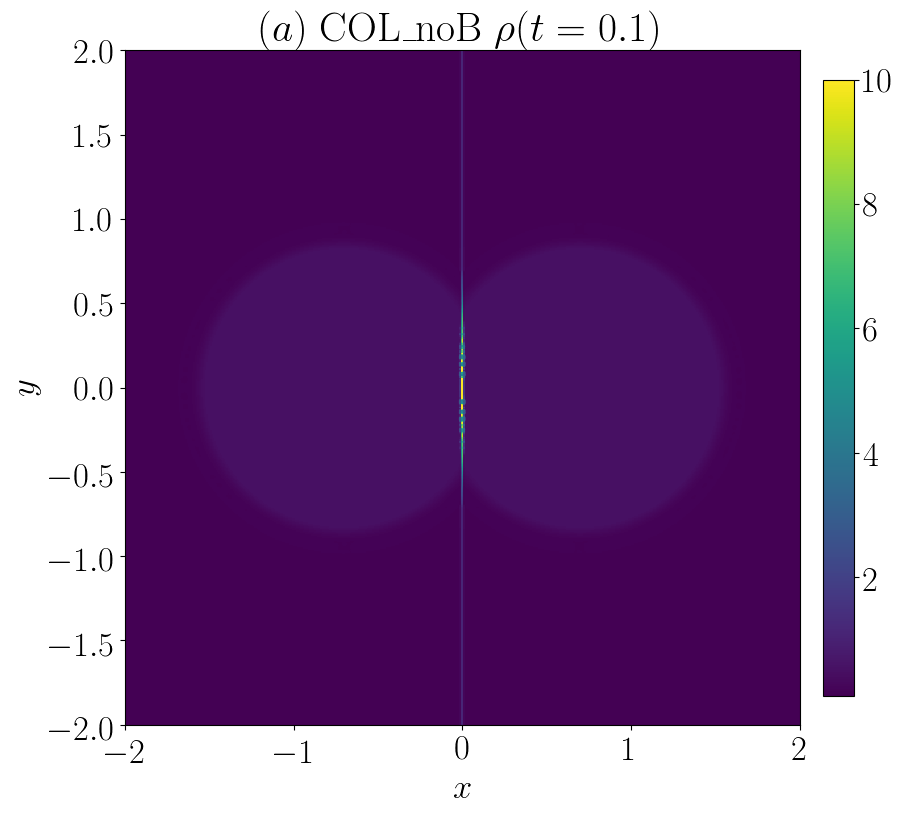}
\plotone{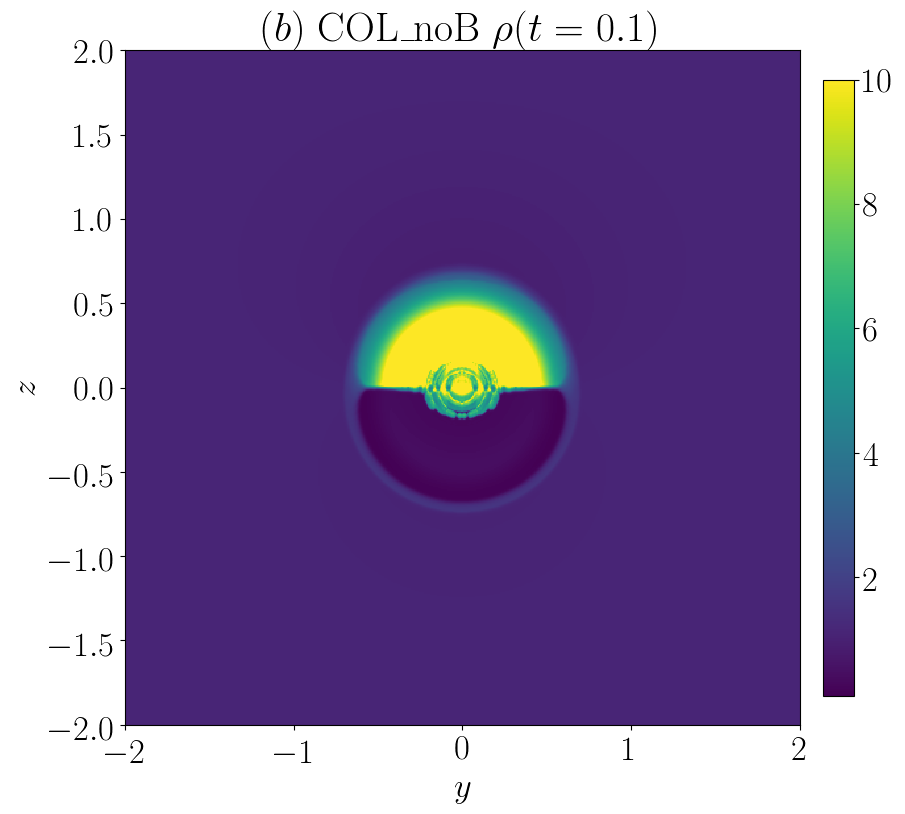}
\plotone{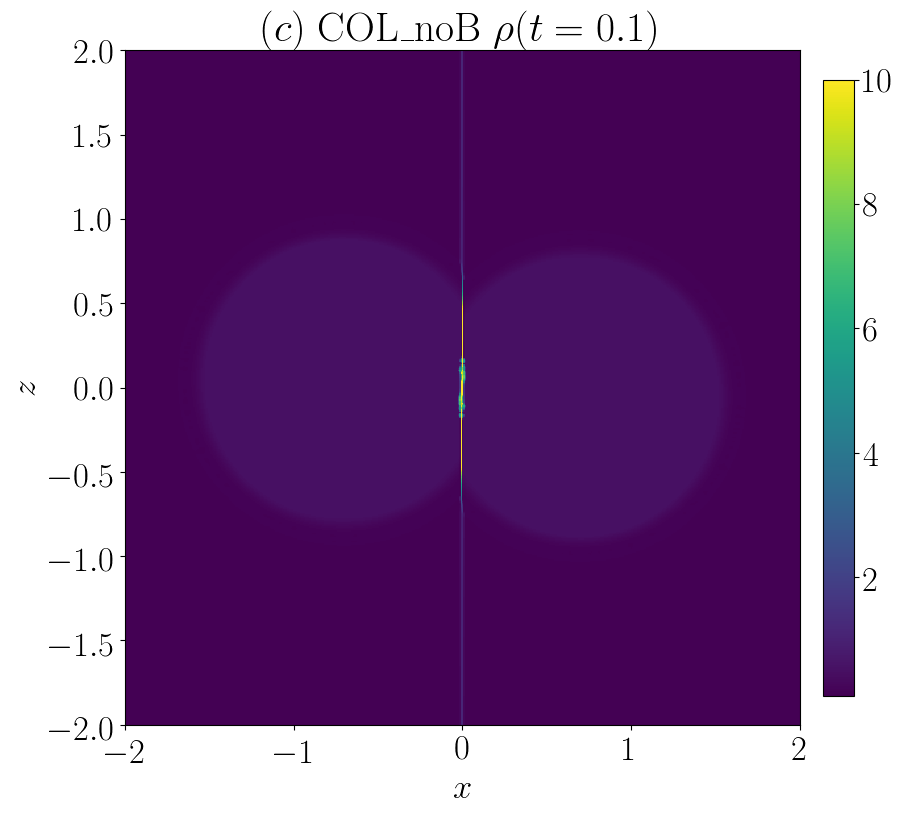}
\caption{
A snapshot of \mbox{COL\_noB} density at t=0.1 (0.2 Myr). Panels (a), (b), (c) show the z=0, x=0, y=0 planes, respectively. The color scale is linear, ranging from 0.1-10.  See \S\ref{subsec:2dtest} for unit conversions.
\label{fig:mrcol_noB}}
\end{figure}

For comparison, we show results from the same simulation but without B-fields (hereafter \mbox{COL\_noB}). Figure \ref{fig:mrcol_noB} shows the results. In panel (a), we see again a density enhancement in the x=0 plane. The strongest enhancement happens at the collision midplane between the two clumps. The highest density reaches $\rho_{\rm max}~\sim$ 27, i.e., a factor of 5 smaller than in \mbox{MRCOL}. The difference with the fiducial model is, as shown in panel (b), the collision in \mbox{COL\_noB} produces a pancake in the y-z plane, as the colliding clumps push material to the collision midplane. This is the compression phase in the clump-clump collision problem \citep{1970ApJ...159..277S,1970ApJ...159..293S,1997ApJ...491..216M,1998ApJ...497..777K}. In our simulation (\mbox{COL\_noB}), the pancake is tilted due to our inclusion of a shear velocity. This tilt is better shown in panel (c). 
In any case, there is no filament formed in \mbox{COL\_noB}. 

\begin{figure}[htb!]
\centering
\epsscale{1.}
\plotone{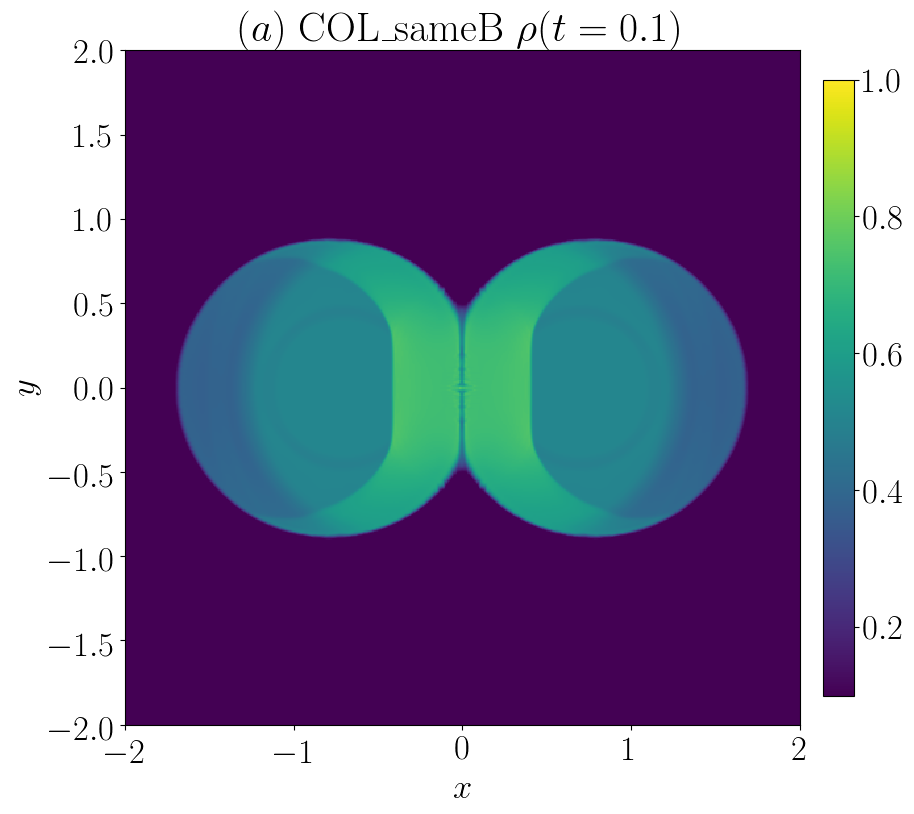}
\plotone{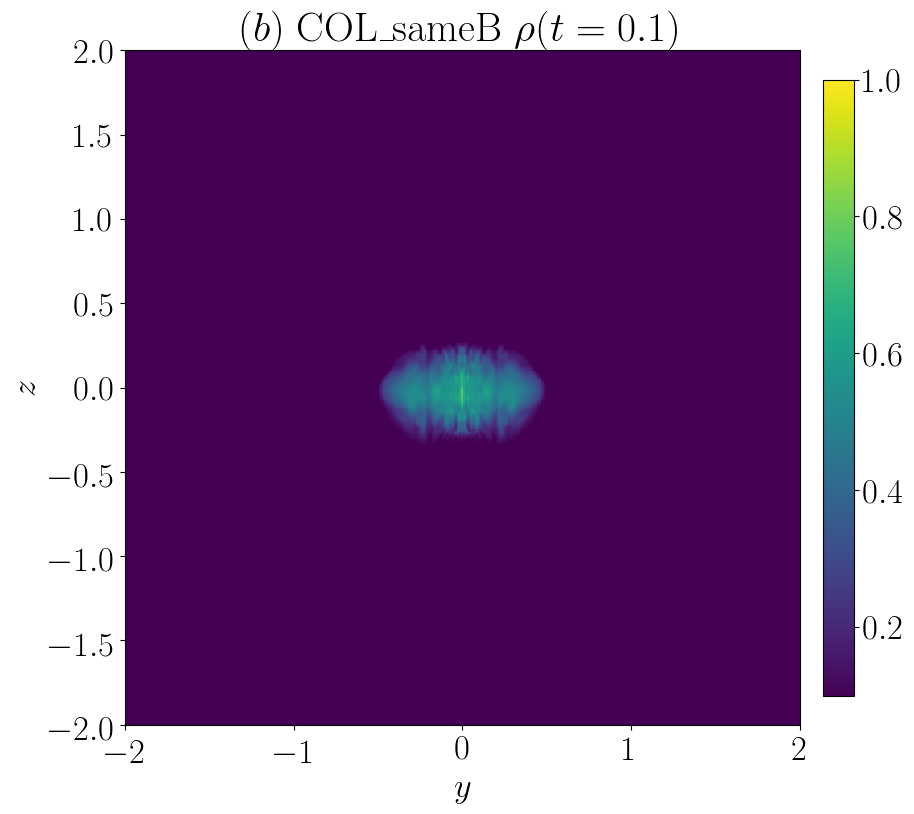}
\plotone{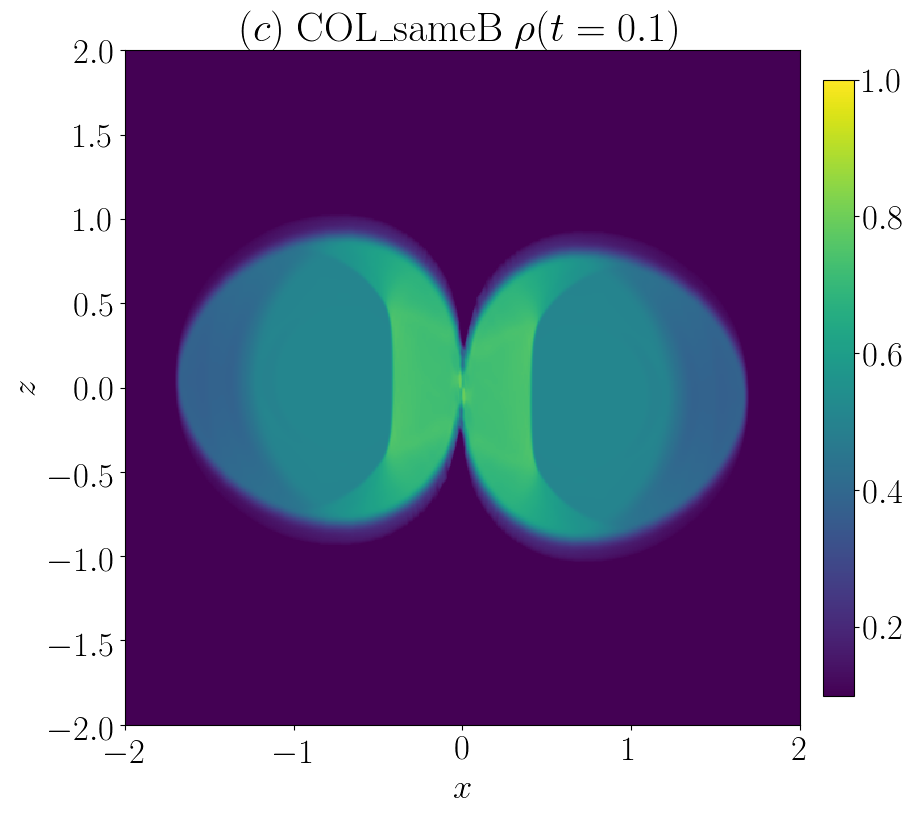}
\caption{
A snapshot of \mbox{COL\_sameB} density at t=0.1 (0.2 Myr). Panels (a), (b), (c) show the z=0, x=0, y=0 planes, respectively. The color scale is linear, ranging from 0.1-1.0. See \S\ref{subsec:2dtest} for unit conversions.
\label{fig:mrcol_sameB}}
\end{figure}

In Figure \ref{fig:mrcol_sameB}, we show results from the same simulation but with B-fields aligned in the z-direction (hereafter \mbox{COL\_sameB}). Magnetized cloud-cloud collisions have also been studied quite extensively \citep{1999ApJ...510..726M,2001A&A...379.1123M,2015MNRAS.453.2471B,2020ApJ...891..168W}. In our case, the clump collision in \mbox{COL\_sameB} is not able to enhance the gas density. The highest density in Figure \ref{fig:mrcol_sameB} reaches $\rho_{\rm max}~\sim$ 1. We can also see from panel (b) that no filament forms in the simulation. Therefore, comparing the three simulations (\mbox{MRCOL}, \mbox{COL\_noB}, \mbox{COL\_sameB}), we see that the simulation with the clumps with anti-parallel B-field is the only one that produces a high-density filament (instead of a pancake). In the following section (\S\ref{subsec:rt}), we first carry out a radiative transfer model of the filament and compare it with our observations. Then in \S\ref{subsec:ph}, we look into the physical process that gives rise to the formation of the filament in the \mbox{MRCOL} simulation.

\subsection{Radiative Transfer and Comparison to Observation}\label{subsec:rt}

To compare the simulation results with the observations, we simulate the line radiative transfer (RT) using the three-dimensional RT code SimLine3D \citep{Ossenkopf2002}. The code self consistently computes the excitation of the molecules including the effects of thermal excitation through collisions with H$_2$ and local and non-local radiative trapping, performs the ray tracing to calculate individual line profiles, and convolves the output with a simulated Gaussian telescope beam. Finally we add normally-distributed observational noise matching the rms of the observations. The code was extensively tested in the frame of the RT benchmark comparison \citep{vanZadelhoff2002}. Spectroscopic data for the molecules were taken from the Cologne Database for Molecular Spectroscopy \citep{Endres2016} and the collisional rate coefficients from \citet{Yang2010} assuming a thermal ortho/para ratio of molecular hydrogen \citep{LeBourlot1991}. We computed the position-position-velocity (PPV) cube for the C$^{18}$O(1-0) transition. 

For the dust RT we assumed optically thin emission of dust at 15~K using the emission coefficients of the OH5 dust model \citep{1994A&A...291..943O} applicable for dense clouds with ice depletion. The same dust properties were used by \citet{2015A&A...577L...6S} for the interpretation of the {\it Herschel} continuum data so that our model should be consistent with their analysis.

\begin{figure}[htb!]
\centering
\epsscale{1.}
\plotone{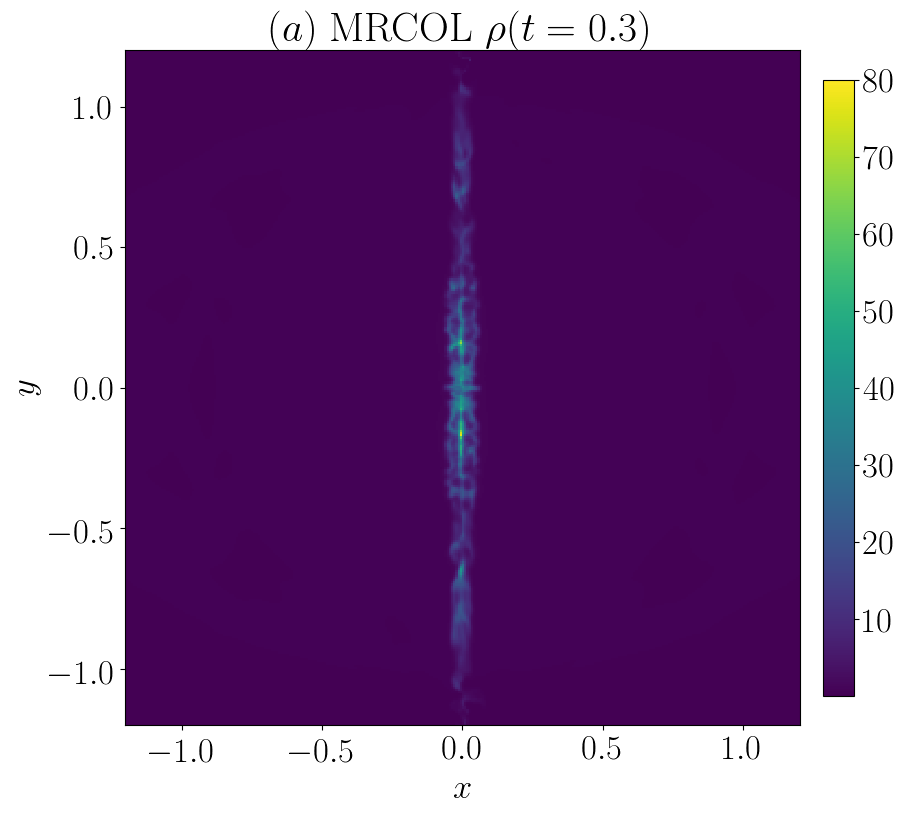}
\plotone{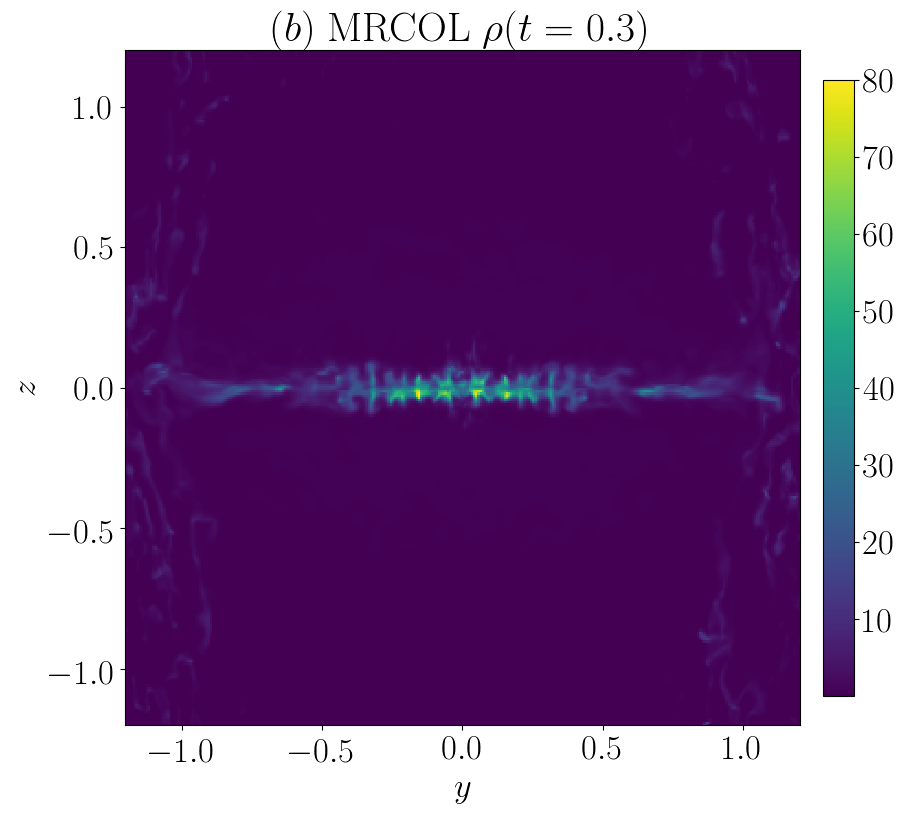}
\plotone{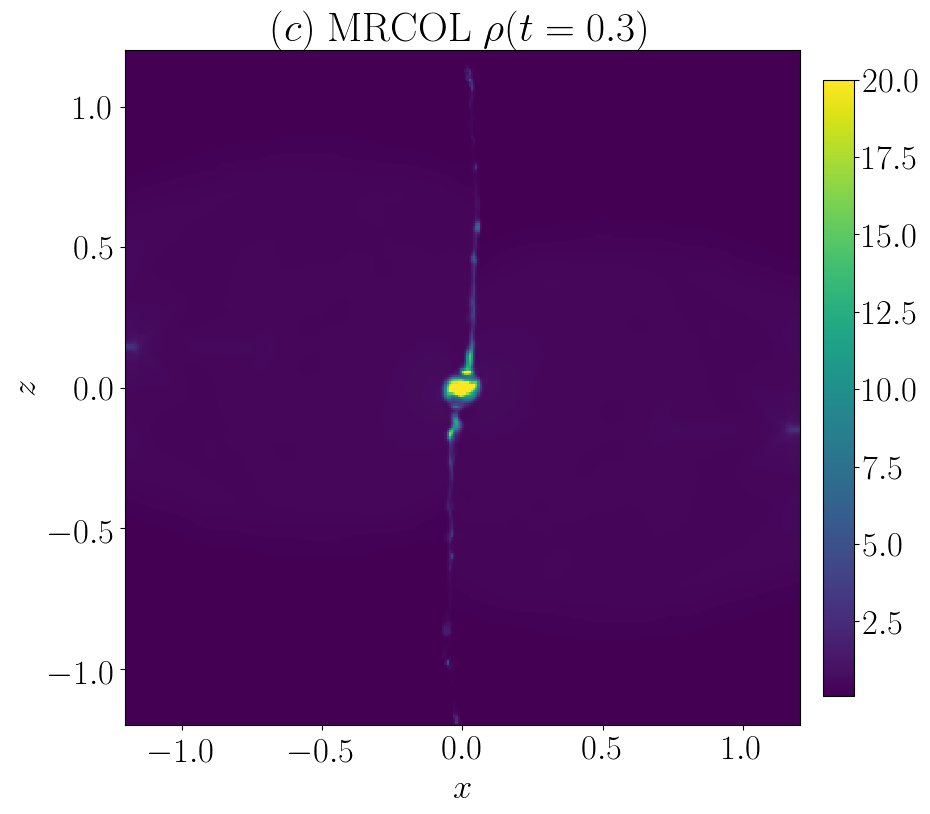}
\caption{
A snapshot of \mbox{MRCOL} density at t=0.3 (0.6 Myr). See \S\ref{subsec:2dtest} for unit conversions.
\label{fig:mrcol_t0p3}}
\end{figure}

To choose the simulation time step, we need to have some idea of the age of the  Stick. As discussed in \S\ref{sec:obs}, the Stick is very young. The highest CO depletion reaches a factor of $\sim$5. If we consider the astrochemical model by \citet{1993MNRAS.261...83H} which include the gas-phase and dust-surface chemistry and cosmic ray induced desorption (see their table 9), the gas-phase CO abundance [CO/H$_2$] reaches $6.1\times10^{-5}$ at $3\times10^5$ yr and  $1.4\times10^{-5}$ at $1\times10^6$ yr. Again, if we assume the canonical CO abundance [CO/H$_2$] of 2.4$\times10^{-4}$ \citep{2008ApJ...680..371W}, the chemical model gives a CO depletion factor of 3.9 at $3\times10^5$ yr and 17 at $1\times10^6$ yr. Note the physical condition in the \citet{1993MNRAS.261...83H} model adopted a temperature of 10 K and a density of $2\times10^4$ cm$^{-3}$. The Stick has a higher dust temperature of 16 K (Table \ref{tab:stick}), which favors CO desorption, but also a higher density, which supports CO depletion. If we assume these two effects cancel each other, then the Stick depletion factor of 5 indicates that the filament age is somewhere between 0.3-1 Myr. We thus adopt an age of 0.6 Myr (t=0.3) for comparison with the observations. As the depletion is mainly limited to the dense cores, we do not add this complexity  
to our simulations. We caution that we use the emission of the CO isotopologues to  trace the general gas distribution, and that we ignore the depletion in the RT computations (we only use the depletion to estimate an age for the Stick).  

\begin{figure}[htb!]
\centering
\epsscale{1.}
\plotone{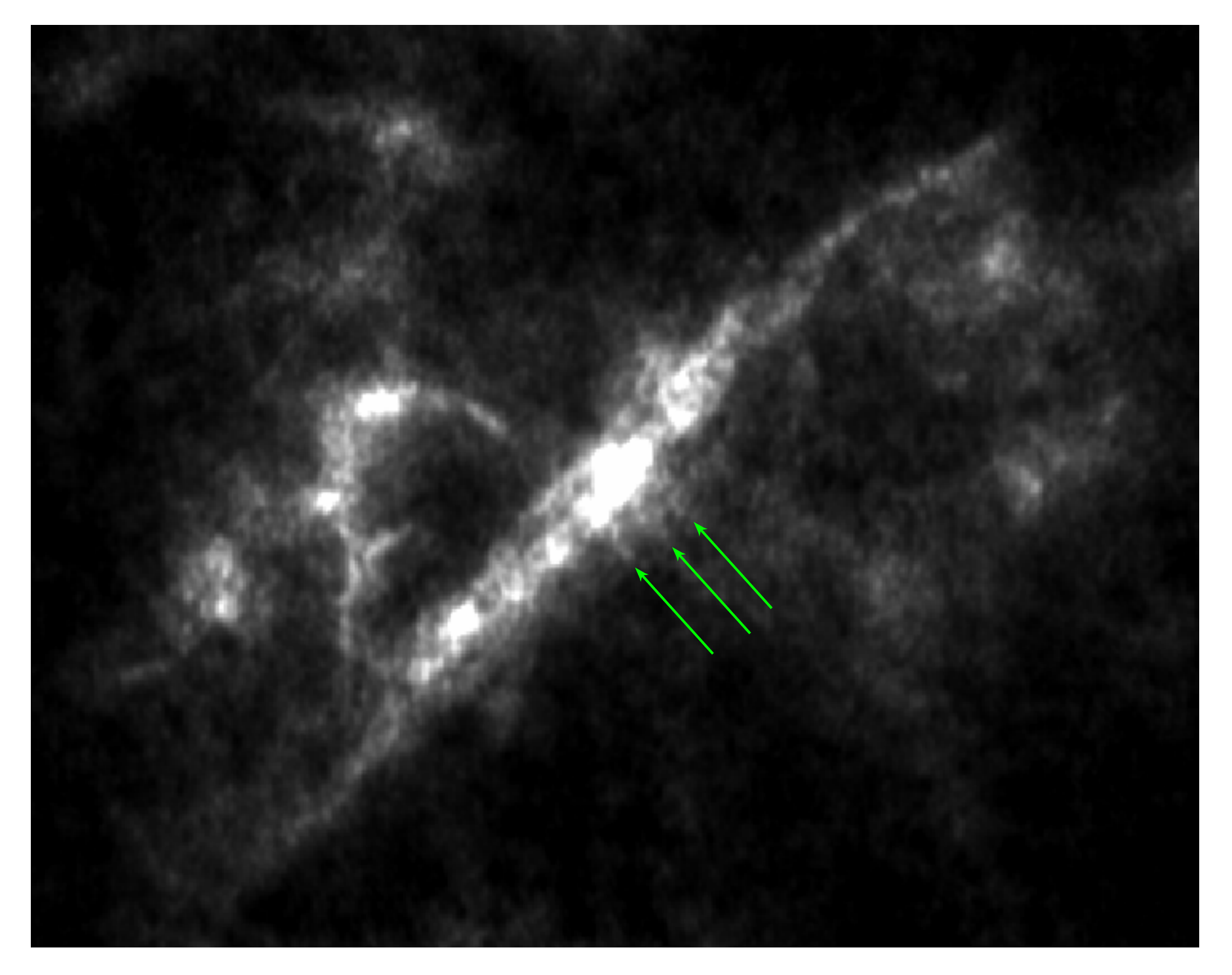}
\caption{
A zoom-in view of the C$^{18}$O integrated intensity map. The green arrow points to the three spikes.
\label{fig:stings}}
\end{figure}

Figure \ref{fig:mrcol_t0p3} shows a snapshot of \mbox{MRCOL} at t=0.3 (0.6 Myr). Again, we show slice maps for z=0 (panel a), x=0 (panel b), and y=0 (panel c) planes. In panel (a),  we can clearly see many rings/forks along the filament.
In panel (b), a filament (and not a pancake) forms in the collision midplane. Interestingly, the filament has many transverse spikes in the z-direction (the small structures stretching out of the filament), which look very similar to the observations. Figure \ref{fig:stings} shows a zoom-in view of the C$^{18}$O integrated intensity map from the observations. Note the spikes stretching out from the Stick (the arrows mark three of them). In Figure \ref{fig:mrcol_t0p3}(c), we see the cross-section of the filament. Again, a filament instead of a pancake forms at the midplane. 

\begin{figure*}[htb!]
\centering
\epsscale{1.}
\plotone{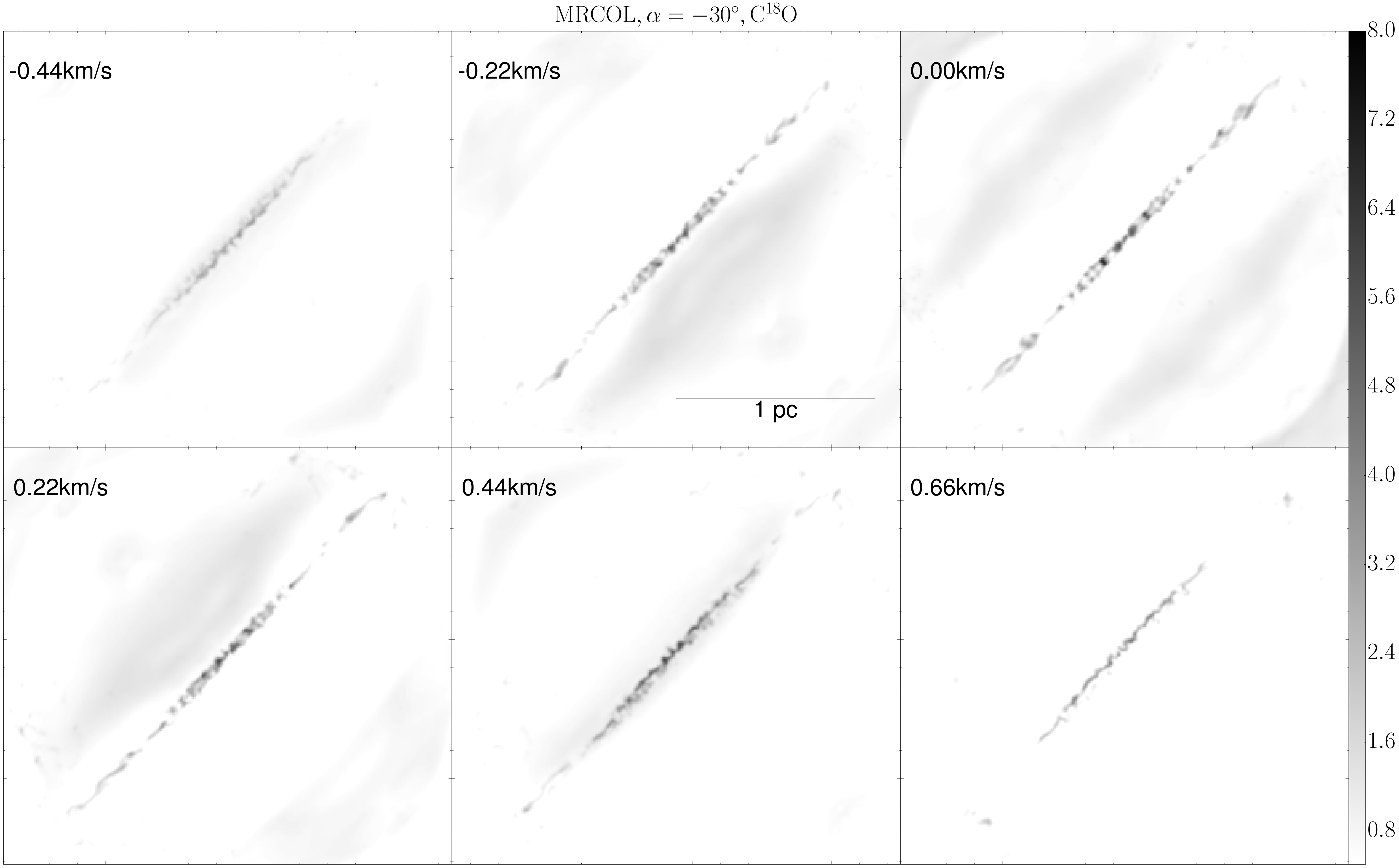}
\caption{
RT model of \mbox{MRCOL} at t=0.3 (0.6 Myr). 
The line of sight is shown in Figure \ref{fig:ic}
with a rotation of $\alpha$=-30\arcdeg. Then,
there is an additional rotation around the 
line of sight to tilt the filament by 43 degree
to match the position angle of the Stick.
The cube is smoothed to match the observation
beam of 9\arcsec\ (two simulation cells).
The color scale is the same as Figure \ref{fig:chan18}.
\label{fig:rtchan18}}
\end{figure*}

\begin{figure*}[htb!]
\centering
\epsscale{1.}
\plotone{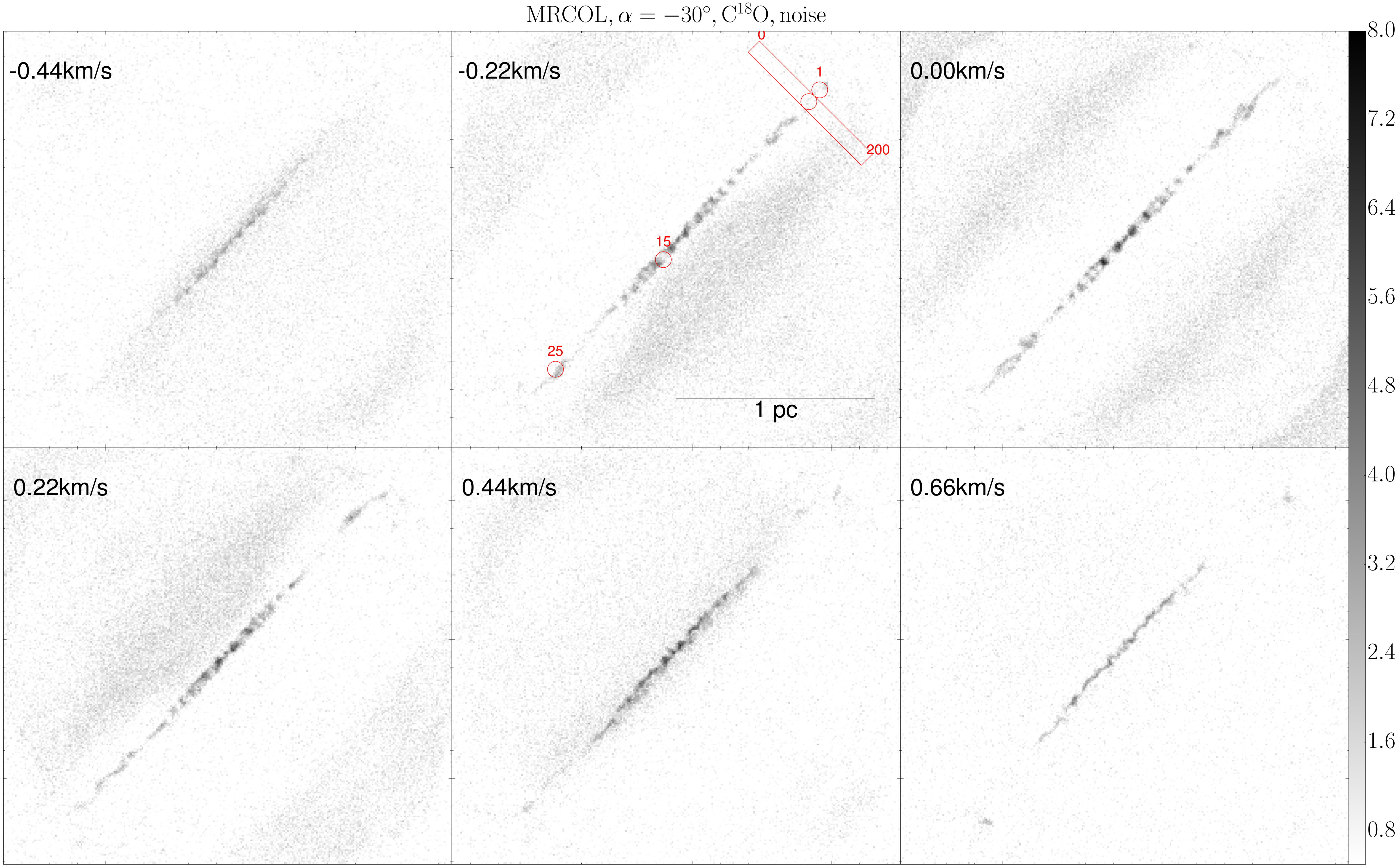}
\caption{
Same as Figure \ref{fig:rtchan18}, but with an 
rms noise of 0.47 K per 0.22 km s$^{-1}$
added to the cube. The noise is the same as the
CARMA-NRO Orion data for C$^{18}$O(1-0).
The red circles show the 25 positions for PV-cuts
(only showing four of them),
similar to those in Figure \ref{fig:c18omom0}.
The red rectangle shows the PV-cut with a size of
200\arcsec$\times$20\arcsec, the same as Figures
\ref{fig:c18omom0} and \ref{fig:stickpv}.
\label{fig:rtchan18noise}}
\end{figure*}

\begin{figure*}[htb!]
\centering
\epsscale{1.}
\plotone{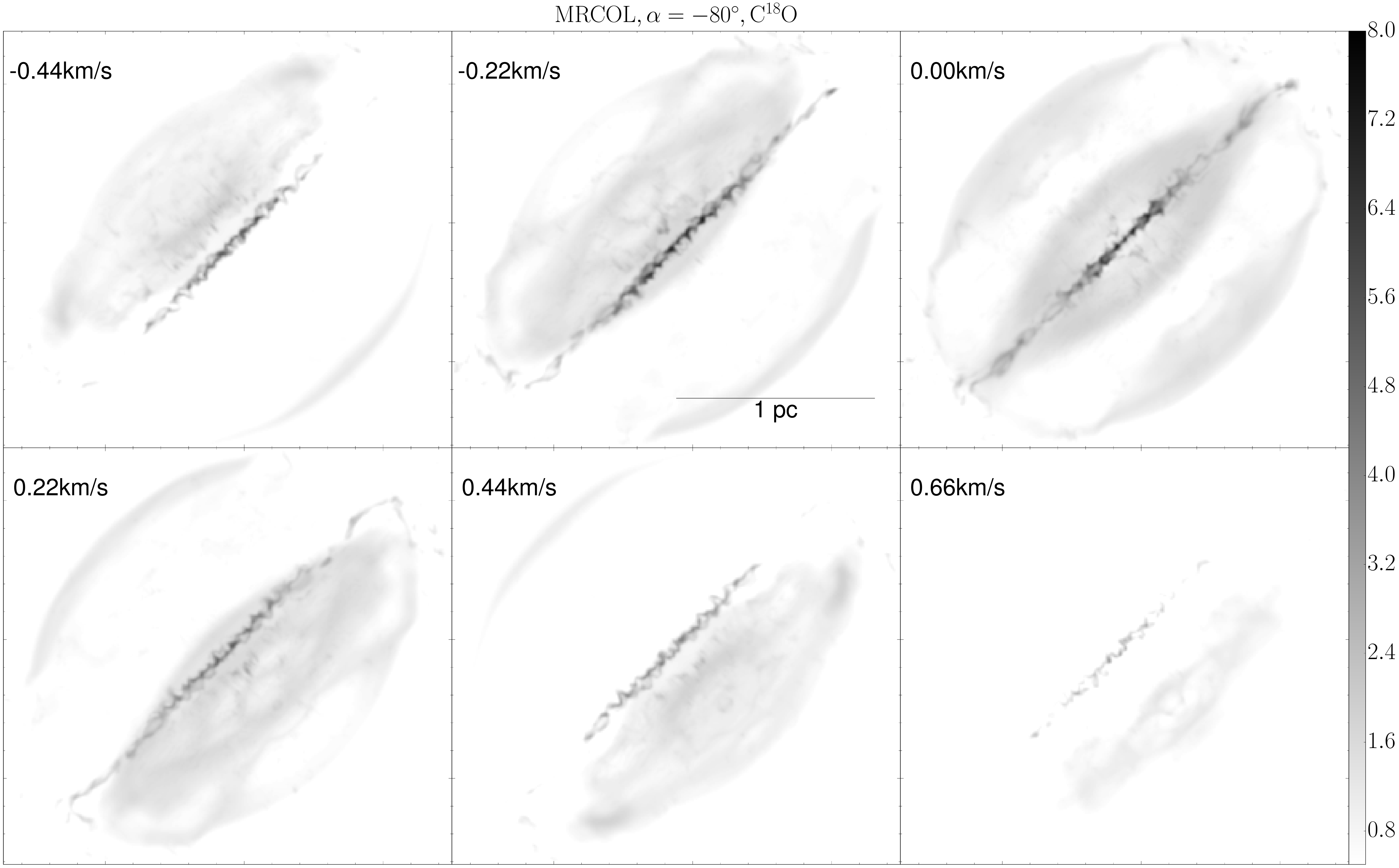}
\caption{
Same as Figure \ref{fig:rtchan18}, 
but with a rotation of
$\alpha$=-80\arcdeg.
\label{fig:rtchan18_-80}}
\end{figure*}

\begin{figure*}[htb!]
\centering
\epsscale{1.}
\plotone{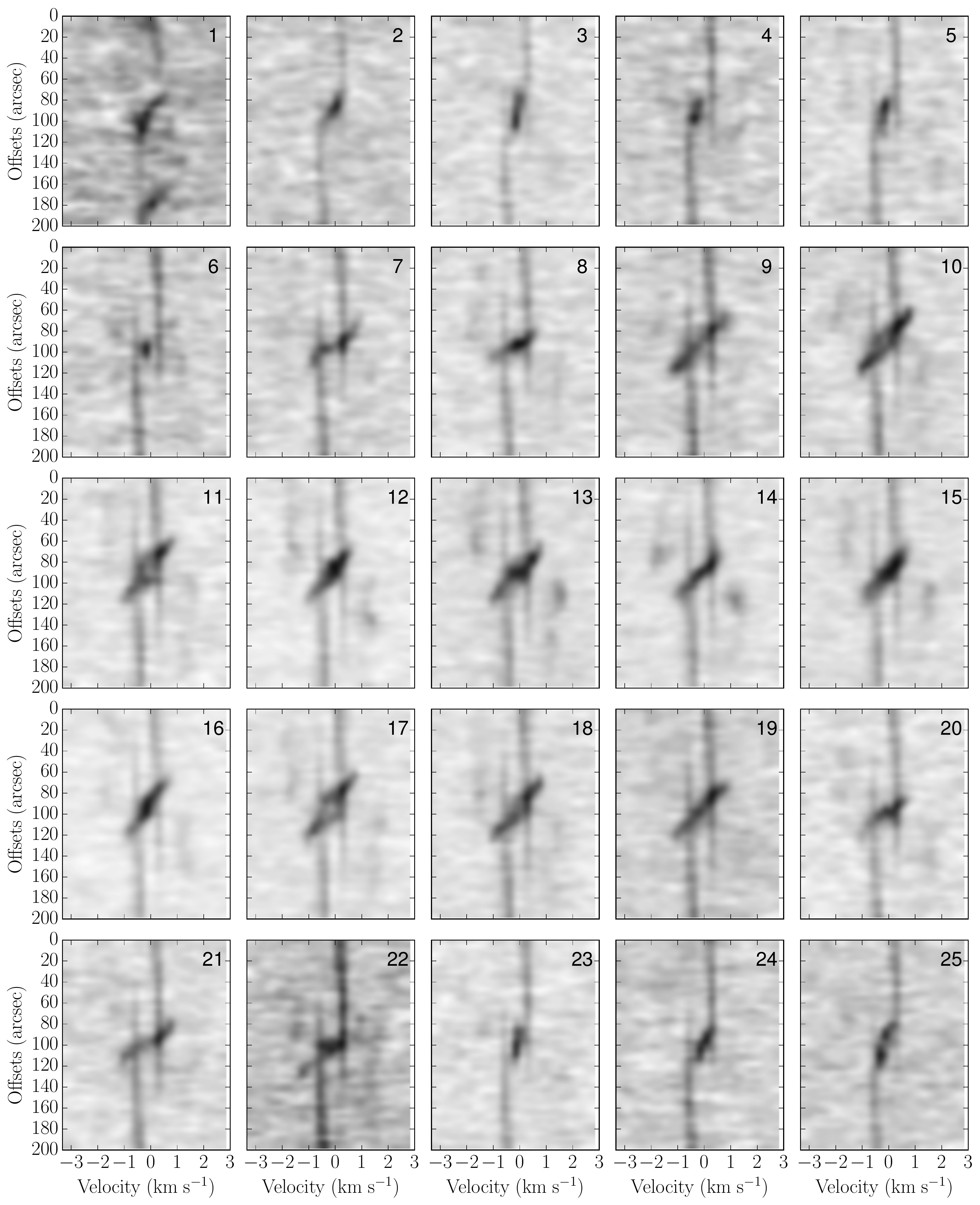}
\caption{
PV-diagrams for the RT cube with observation noise for \mbox{MRCOL} with $\alpha$=-30\arcdeg. The format is the same as Figure \ref{fig:stickpv}. The PV-cuts are shown in Figure \ref{fig:rtchan18noise}.
\label{fig:rtpv}}
\end{figure*}

\begin{figure*}[htb!]
\centering
\epsscale{1.1}
\plotone{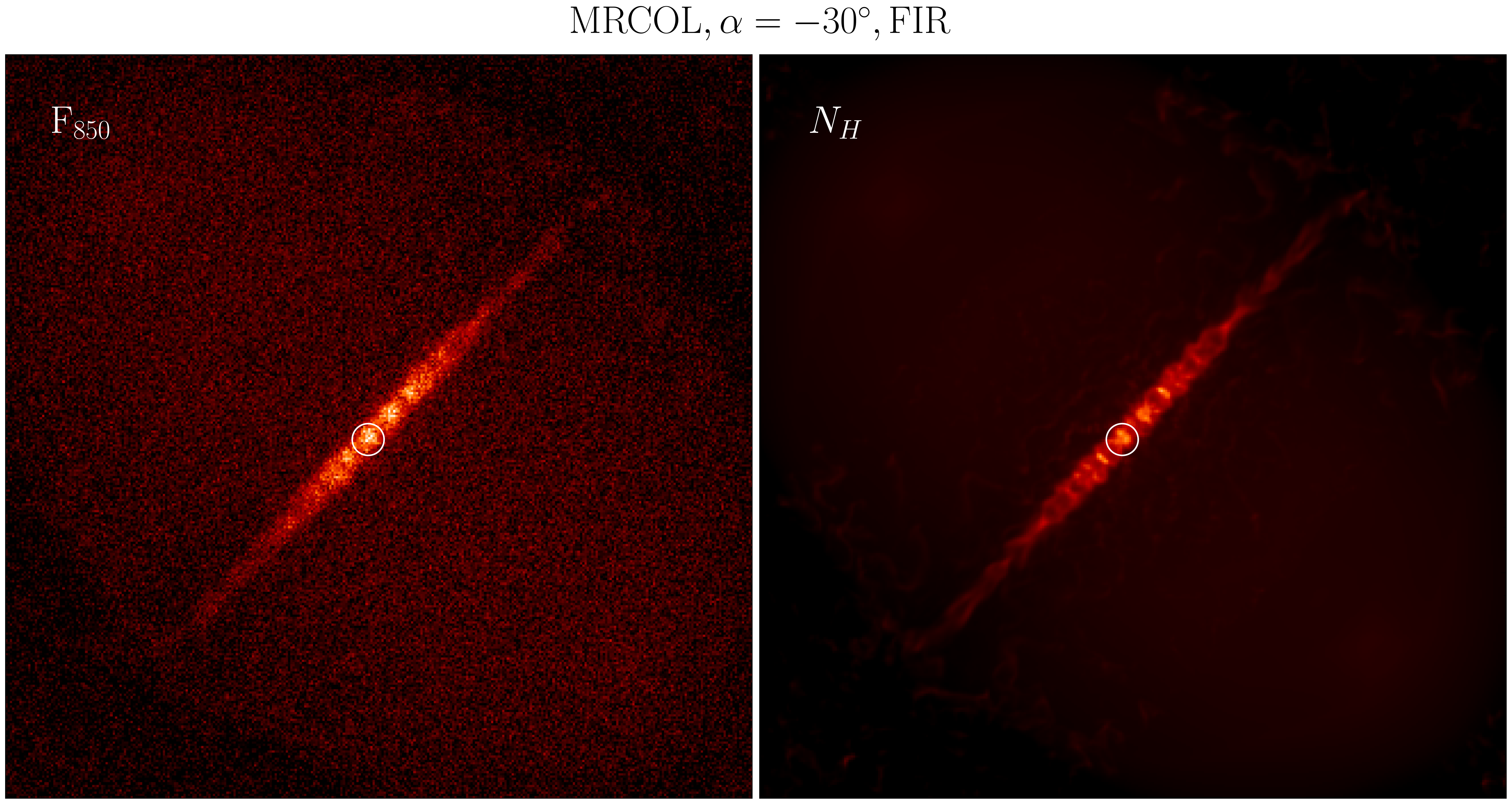}
\caption{
RT model for FIR and sub-mm wavelengths to be compared with {\it Herschel} and JCMT observations. The left panel shows emission at 850 $\mu$m (smoothed to 14.5\arcsec, adding a rms noise of 1.8 MJy sr$^{-1}$). The right panel shows the column density map. The white circle shows a core with 0.1 pc diameter. Its virial status is analyzed in \S\ref{subsec:kirkcores}.
\label{fig:fir}}
\end{figure*}

\begin{figure*}[htb!]
\centering
\epsscale{1.15}
\plotone{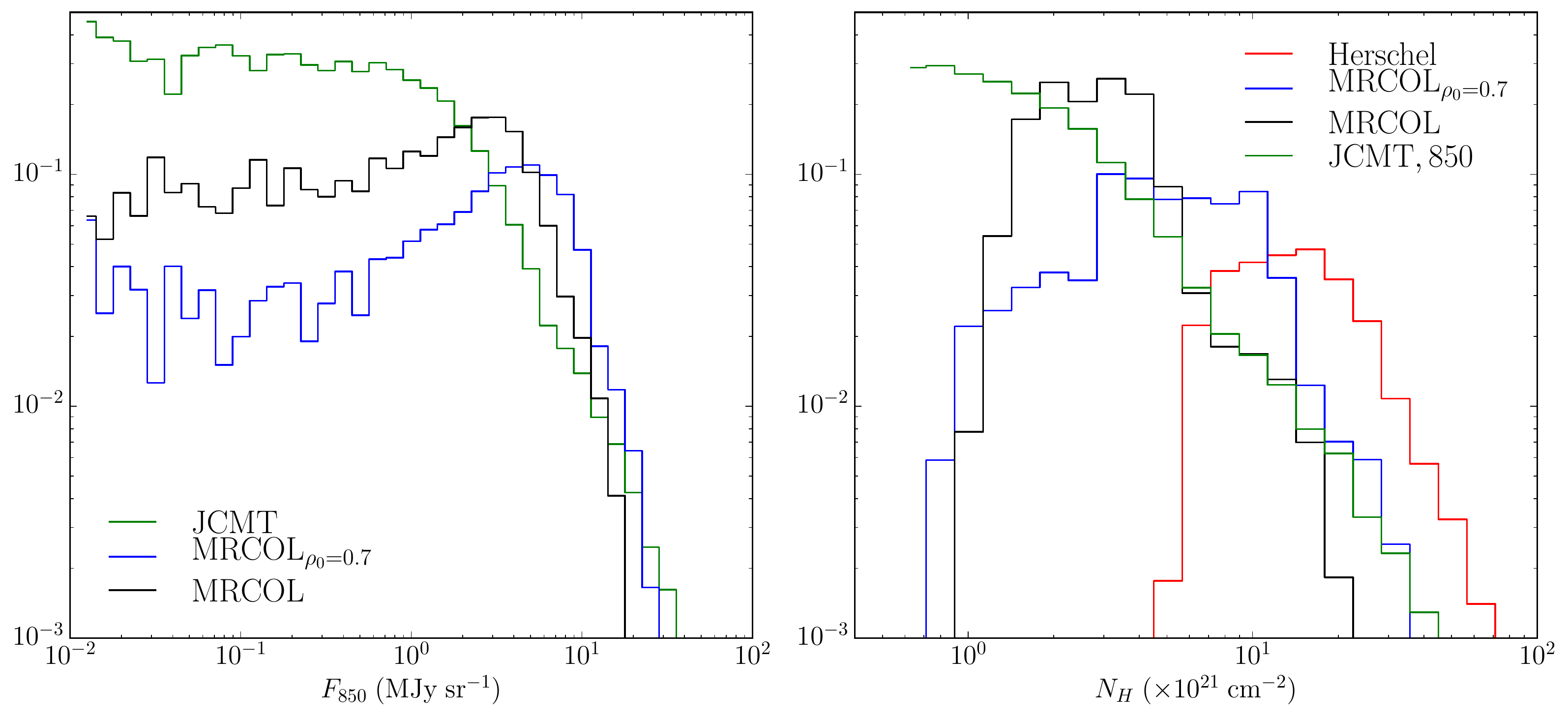}
\caption{
Histograms for the two images in Figure \ref{fig:fir}.
Each panel compares the observed data with the fiducial model
\mbox{MRCOL} ($\alpha$=-30\arcdeg) and the high density model \mbox{MRCOL$_{\rho_0=0.7}$}. The green lines represent the JCMT 850$\mu$m data (L16), the red column density histogram was derived from the {\it Herschel} observations (SK15). 
\label{fig:hist}}
\end{figure*}

Figure \ref{fig:rtchan18} shows the channel maps from the RT model for \mbox{MRCOL} at t=0.3 (0.6 Myr). The model ``observes'' the data at the Orion distance (390 pc).  We rotate the simulation data cube to match the Stick observation. First, we face the x-y plane along the negative z-axis. We rotate the cube about the y-axis by $\alpha=-30\arcdeg$ (see Figure \ref{fig:ic}). As a result, Clump2 is in front of Clump1 along the new line of sight. Then, we rotate the new cube about the new z-axis to match the Stick's diagonal appearance. Finally, we invert the z-axis velocity component to follow the line-of-sight convention. We apply the RT model to the final cube to get the PPV cube. The simulation cell size corresponds to a physical scale of 0.0078 pc. The CARMA-NRO Orion C$^{18}$O data beam size corresponds to a physical scale of 0.015 pc, i.e., twice the simulation cell size. So the PPV cube is smoothed to a resolution of 0.015 pc. Figure \ref{fig:rtchan18} shows the smoothed channel maps. In Figure \ref{fig:rtchan18noise}, we add a 0.47 K noise to the cube (the same as the CARMA-NRO Orion data, K18). 

In Figures \ref{fig:rtchan18} and \ref{fig:rtchan18noise}, we clearly see the ring/fork/spike structures along the straight filament. The filament is also very narrow with an aspect ratio $\ga~20$ and spans a velocity range of $\sim$1.3 km s$^{-1}$, similar to the Stick. In \S\ref{app:A} and Figure \ref{fig:simfork}, we show a zoom-in view of Figure \ref{fig:rtchan18} and highlight the structures. Of course, they are not exactly the same, which can be caused by irregular shapes of the incoming clumps, additional line-of-sight gas between the Stick and us, etc.

In Figure \ref{fig:rtchan18_-80}, we show the RT model for a new projection ($\alpha$=-80\arcdeg). This figure shows more pronounced ring structures along the filament. 
At this point, we use the $\alpha$=-30\arcdeg\ projection as the fiducial RT result but do not attempt to find the ``best'' projection. 

We make PV-diagrams for the fiducial model \mbox{MRCOL} and fiducial projection $\alpha$=-30\arcdeg, following the format in Figure \ref{fig:stickpv}. Note that we use the  cube with the added noise shown in Figure \ref{fig:rtchan18noise}. These PV-diagrams are shown in Figure \ref{fig:rtpv}. Again, we select 25 evenly distributed points on the filament (Figure \ref{fig:rtchan18noise}). Then, we make the PV-cut through the point perpendicular to the filament. The length of the PV-cut is 200\arcsec, and the width is 20\arcsec.
The result in Figure \ref{fig:rtpv} matches the observation (Figure \ref{fig:stickpv}) very well. Two velocity components appear, with a separation of about 1 km s$^{-1}$.
The filament is in the middle of the two components. 

In Figure \ref{fig:fir}, we show two more images from the simulations that can be compared with observations. The left panel shows the 850 $\mu$m flux density from the simulated filament, which can be compared with the observed dust map shown in  Figure \ref{fig:omc6}(i).  The right panel shows the column density which is comparable to  {\it Herschel} images in Figure \ref{fig:omc6}. 
Noise has been added to these images, with the same level as the observations.
Basically, both the model and the observations show a straight, narrow, long filament. Both show dense cores in the filament. The model images show two major circular cores in the central part of the filament which show sub-structures in the column density image. 

Figure \ref{fig:hist} shows a quantitative comparison of  the logarithmic probability distribution functions (PDFs) between the model and the observations.  The left panel compares the 850 $\mu$m RT images with the 850 $\mu$m JCMT  map. The right panel compares the column densities, where the red line corresponds to the {\it Herschel} observations (SK15) and the green line to the 850$\mu$m data (L16) after it has been converted into column density using the same assumptions ($T_\mathrm{dust}=15K$, OH5 opacities).
We note that the column density distributions from the two observed maps (JCMT and Herschel) are not consistent. The {\it Herschel} observations indicate higher column densities and a much narrower distribution than the 850 $\mu$m data. Both probably suffer from  observational limitations. Low column densities in the {\it Herschel} maps are truncated due to the sensitivity limits while part of the lower value tail of the 850 $\mu$m data is due to the rms noise. Large scale structures in the JCMT data also suffer some flux loss (L16). Still differences at high column densities remain unexplained, thus  we use the observational constraints only as rough guidelines.

The black histogram shows the PDF of the fiducial RT model \mbox{MRCOL} ($\alpha$=-30\arcdeg). In the right panel one can see that the model column density is lower than {\it Herschel} results and even than the 850 $\mu$m data for the Stick when considering the highest densities. 
Here we run another simulation with a factor of 1.4 higher initial density (hereafter \mbox{MRCOL$_{\rho_0=0.7}$}). In Figure \ref{fig:hist}, we show the new results with the blue histograms.
For the \mbox{MRCOL$_{\rho_0=0.7}$} model the column density is within a factor of 2 of the {\it Herschel}-based column density and well in agreement with the 850 $\mu$m data when assuming that part of the low-column density excess stems from noise. Also, the new model (\mbox{MRCOL$_{\rho_0=0.7}$}) preserves other observing features (rings/forks and PV-diagram appearance).

For a detailed comparison between the column-density PDFs, many factors can contribute to a difference, including but not limited to turbulence, collision velocity, magnetic field strength, initial clump density. In the following section \S\ref{subsec:ph}, we discuss the effect of a few of them.
Overall, the simulated filament matches the Stick very well. Adding noise reduces the clarity of the sub-structures in channel maps, although we can still see most of them.

\subsection{Physical Mechanism of the Filament Formation}\label{subsec:ph}

We first return  to our 2D test in Figure \ref{fig:mr2d} to understand the 3D result. As shown in the right panel of the figure, the magnetic islands accumulate a significant amount of dense gas. Expanding this to 3D, we see in Figure \ref{fig:mr3d} that the dense gas in the islands extends to cylinders perpendicular to the plane where the magnetic fields lie (i.e., the x-y plane). In the \mbox{MRCOL} run, where instead of applying velocity perturbations we have two colliding clumps,
a similar process occurs in the x=0 plane. The anti-parallel B-field is in the x-z plane, so the cylinder (i.e., the filament) forms  along the y-direction. The main difference is that \mbox{MRCOL} does not form many parallel filaments as in \mbox{MR\_3D}.

\begin{figure}[htb!]
\centering
\epsscale{1.1}
\plotone{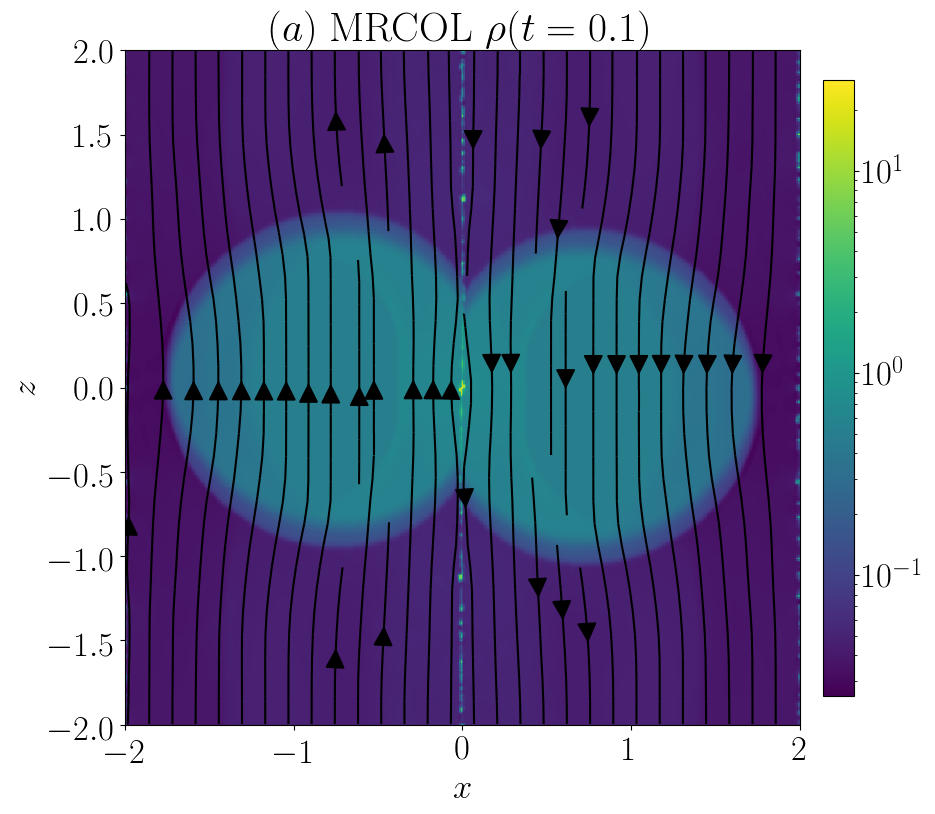}
\plotone{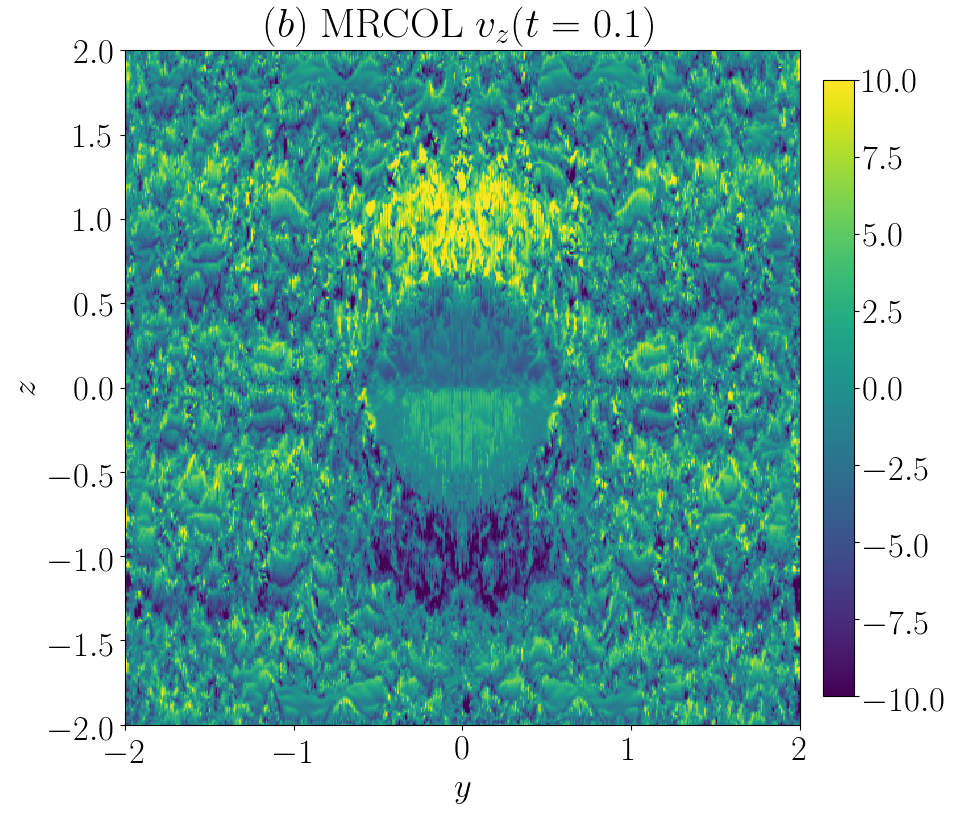}
\caption{
{\bf (a):} A snapshot of \mbox{MRCOL} at t=0.1 (0.2 Myr). The color scale is logarithmic and shows the density in the y=0 plane. The arrows show the B-field. See \S\ref{subsec:2dtest} for unit conversions.
{\bf (b):} A snapshot of \mbox{MRCOL} at t=0.1 (0.2 Myr). The color scale is linear and shows $v_z$ in the x=0 plane. See \S\ref{subsec:2dtest} for unit conversions.
\label{fig:mrcol_t0p1_xz_Bstream}}
\end{figure}

Note that in \mbox{MR\_2D} and \mbox{MR\_3D}, the MR happened because we applied a velocity perturbation. Besides, the two simulations did not have collisions at the x=0 plane. However, in \mbox{MRCOL}, we did not apply any perturbation in the collision plane. Figure \ref{fig:mrcol_t0p1_xz_Bstream}(a) shows again the x-z plane density from \mbox{MRCOL} at t=0.1 (same as Figure \ref{fig:mrcol}(c)). Here we include B-field streamlines. It is quite clear  that MR happened at z=$\pm$0.6 where field lines joined.
They joined where the clump surfaces were about to contact. The B-fields bent toward the x=0 plane and the magnetic loop formed between z=-0.6 and z=0.6. The loop enclosed the compression pancake and dragged gas inward toward the filament. Figure \ref{fig:mrcol_t0p1_xz_Bstream}(b) shows the z-axis velocity in the y-z (x=0) plane. In the pancake, we can see that $v_z>0$ for z$<$0 and $v_z<0$ for z$>$0, meaning that gas in the pancake is moving toward the filament. On the other hand, gas outside the pancake is moving away from the pancake. 

\begin{figure*}[htb!]
\centering
\epsscale{0.55}
\plotone{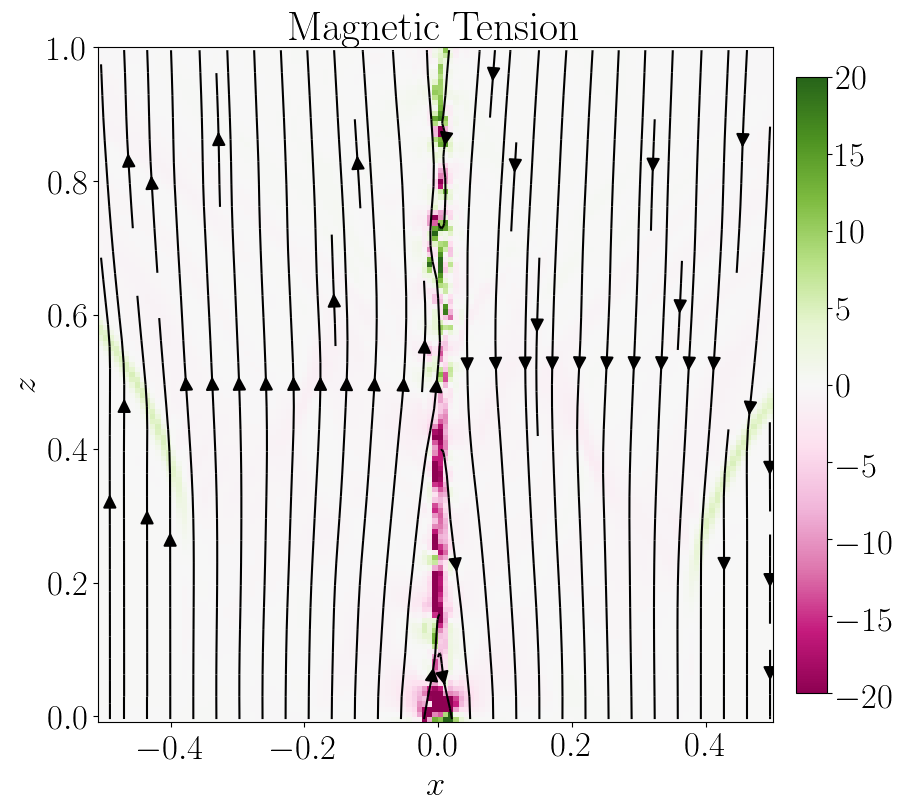}
\plotone{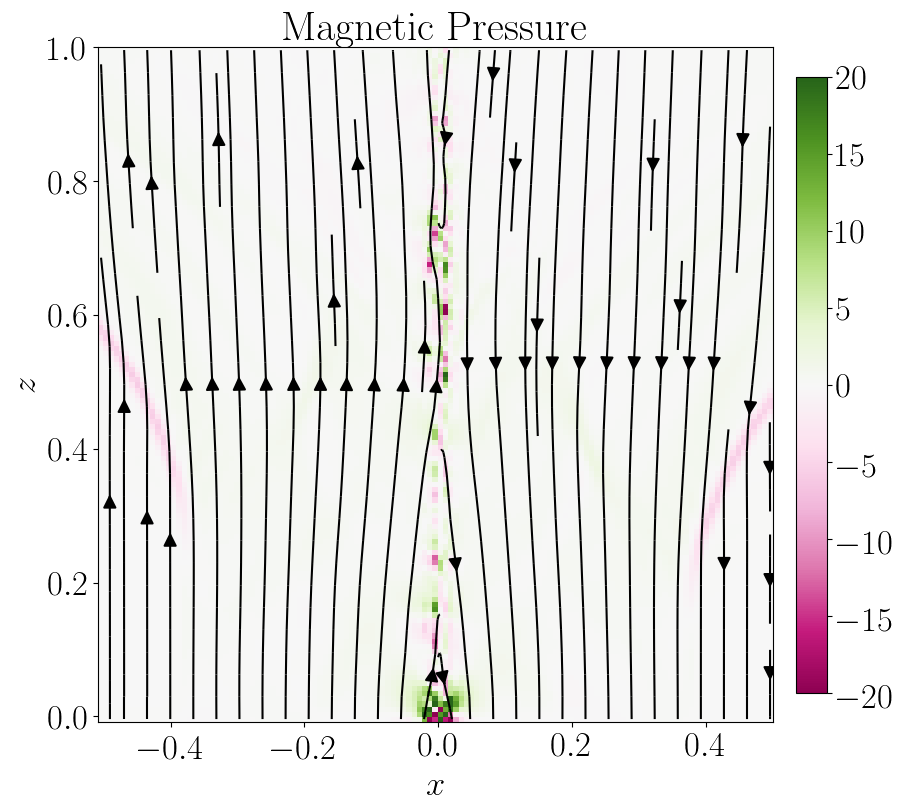}\\
\plotone{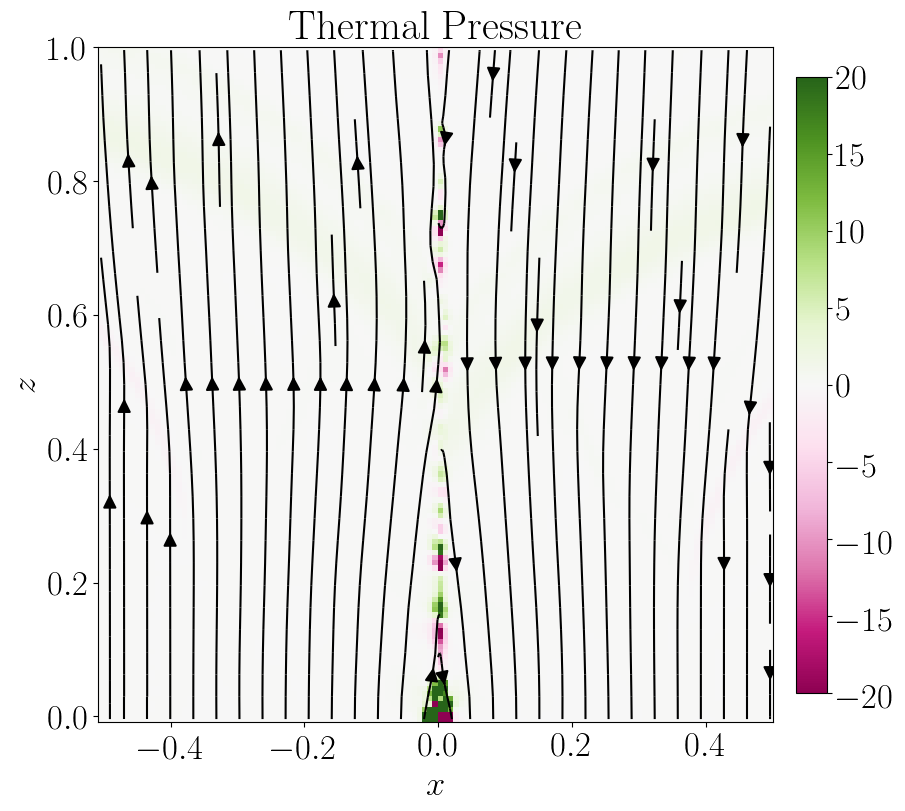}
\plotone{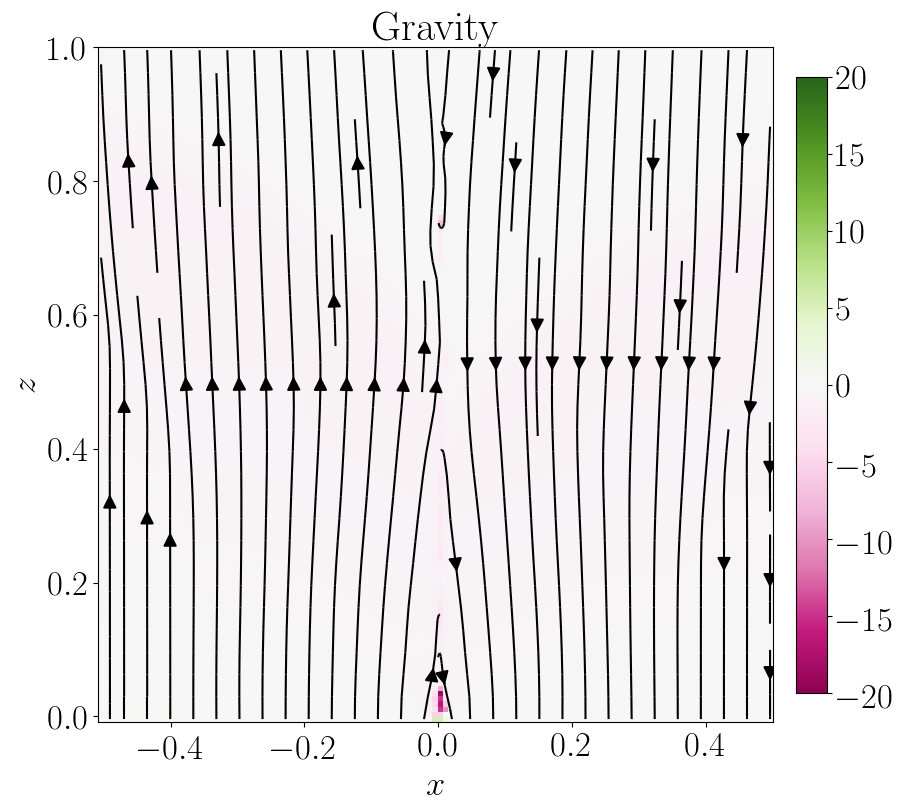}
\caption{
{\bf (a):} A snapshot of \mbox{MRCOL} y=0 plane at t=0.1 (0.2 Myr). The color scale is linear and shows the z-axis magnetic tension force $(\mathbf{B} \cdot \nabla)\mathbf{B}/4\pi$. The arrows show the B-field.
{\bf (b):} Same as (a), but for the magnetic pressure force $-\nabla(B^2/8\pi)$.
{\bf (c):} Same as (a), but for the thermal pressure term $-\nabla P$.
{\bf (d):} Same as (a), but for the gravity term $-\rho\nabla \phi$. All panels have the same color scale. See \S\ref{subsec:2dtest} for unit conversions.
\label{fig:force}}
\end{figure*}

Following \citet{2017stfo.book.....K}, we write the momentum equation below:
\begin{equation}\label{equ:force}
\begin{split}
&\frac{\partial}{\partial t}(\rho\mathbf{v}) + \nabla\cdot(\rho\mathbf{v}\mathbf{v}) = \\
& - \nabla P + \frac{1}{4\pi}(\nabla\times\mathbf{B})\times\mathbf{B} - \rho\nabla\phi \\
& = - \nabla P -\nabla\left(\frac{B^2}{8\pi}\right) + {(\mathbf{B} \cdot \nabla)\frac{\mathbf{B}}{8\pi}} - \rho\nabla\phi.
\end{split}
\end{equation}
Here, $\rho$ is the gas density, $\mathbf{v}$ is the velocity, $P$ is the thermal pressure, $\mathbf{B}$ is the magnetic field, and $\phi$ is the gravitational potential. There are four force terms on the right hand side, including the pressure term $-\nabla P$, the magnetic pressure force $-\nabla(B^2/8\pi)$, the magnetic tension force $(\mathbf{B} \cdot \nabla)\mathbf{B}/4\pi$, and the gravitational force $-\rho\nabla \phi$. The magnetic tension term plays a major role in the filament formation. It includes the tension force when the B-field tries to ``unbend'' itself. As shown in Figure \ref{fig:mrcol_t0p1_xz_Bstream}, MR creates B-field loops with sharp turns at the MR location. The highly curved loop tries to become ``well-shaped'' (a circle). Therefore, it pulls material into the central filament.

Figure \ref{fig:force} shows a detailed look of the contribution from different forces in the y=0 plane. We zoom in to  $-0.5<x<0.5$ and $0<z<1$. In this view, the filament (along the y-axis) cross-section is at x=0 and z=0. We show the z-axis component of the four force terms in equation \ref{equ:force}. The gradient of a quantity $\partial q/\partial x$ at cell i is approximated by $(q_{i+1}-q_{i-1})/(2\delta x)$ for each dimension ($\delta x$ is the cell size). We use the same color scale to show the relative importance of the forces. 

It can be seen that the magnetic tension force preferentially points toward the filament especially at $z\la0.5$. Whenever there is a magnetic loop produced by MR, the sharp turn creates a strong magnetic tension force. Their job is to pull everything to the center of the loop so that the loop becomes circular. The same thing happens in the other half of the pancake ($z<0$ not shown here). Meanwhile, the magnetic pressure, thermal pressure, and gravity forces do not do much compared to the magnetic tension, except for around the filament. At $z\sim0$, magnetic pressure and thermal pressure try to resist the incoming gas pulled by the tension force. Meanwhile, gravity tries to attract more gas to the filament. Outside the central area, the magnetic tension force dominates other forces and keeps sending material to the central region. This is the reason for the filament formation.

\begin{figure}[htb!]
\centering
\epsscale{1.1}
\plotone{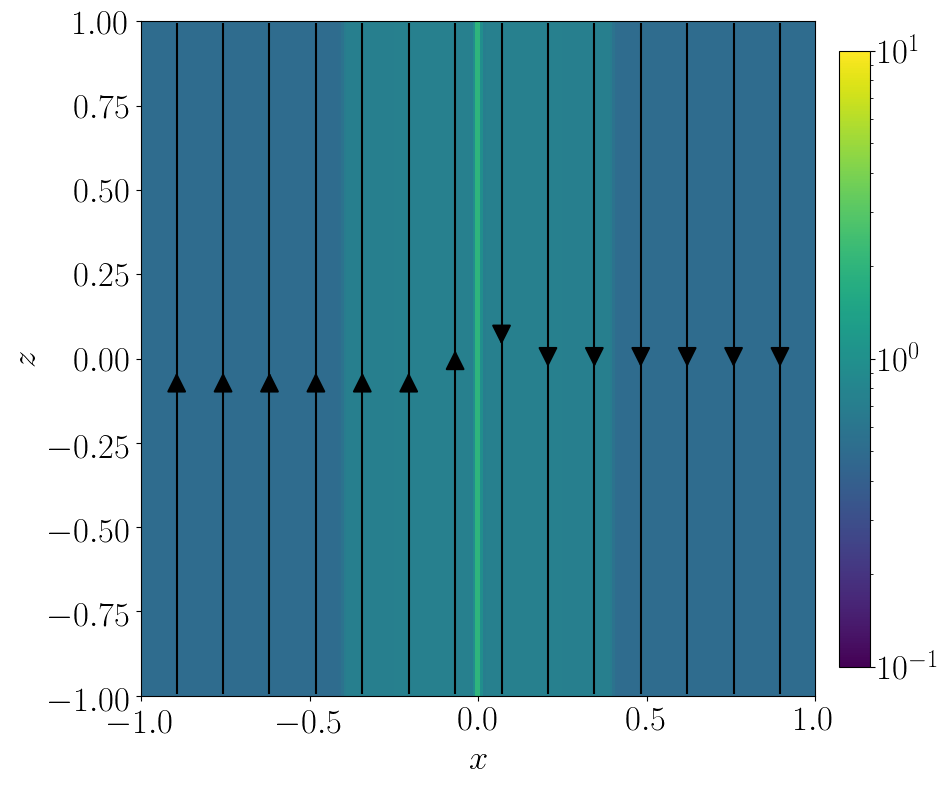}
\caption{
Same as Figure \ref{fig:mrcol_t0p1_xz_Bstream}(a) but for \mbox{MRCOL\_uniform}. See \S\ref{subsec:2dtest} for unit conversions.
\label{fig:mrcol_uniform}}
\end{figure}

On the contrary, if the two colliding sides are uniform gas with $\rho_1=\rho_2=\rho_{\rm amb}=0.5$ (hereafter \mbox{MRCOL\_uniform}), no filaments form. Figure \ref{fig:mrcol_uniform} shows a snapshot of \mbox{MRCOL\_uniform} density in the x-z (at y=0) plane at t=0.1. The difference between \mbox{MRCOL\_uniform} and \mbox{MRCOL} is that the former simulation has a uniform density of $\rho=0.5$ in the entire computation domain. The difference between \mbox{MRCOL\_uniform} and \mbox{MR\_3D} is that the former simulation has no x-axis velocity perturbation and the latter has no collision. Figure \ref{fig:mrcol_uniform} shows that no MR happens. Recall in \mbox{MR\_3D} multiple cylinders formed along the y-axis. Therefore, the density structure of the clump collision in \mbox{MRCOL} provides the initial perturbation to excite MR around the pancake. 

\begin{figure*}[htb!]
\centering
\epsscale{0.35}
\plotone{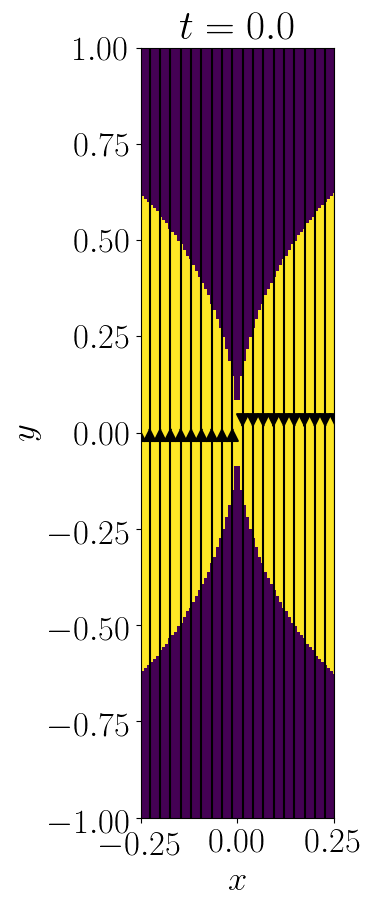}
\plotone{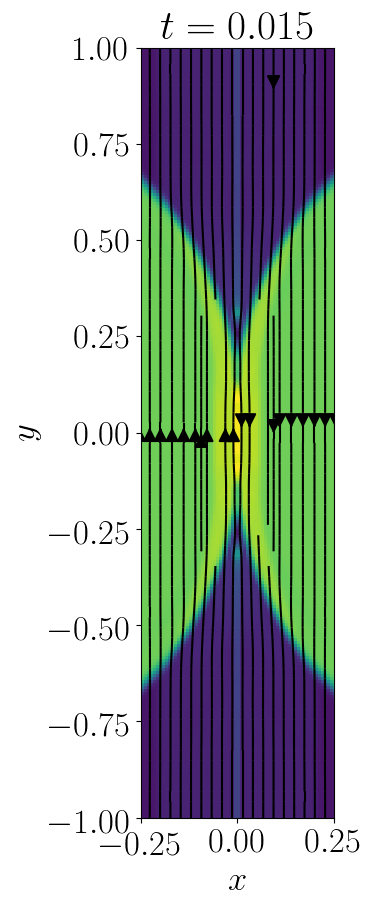}
\plotone{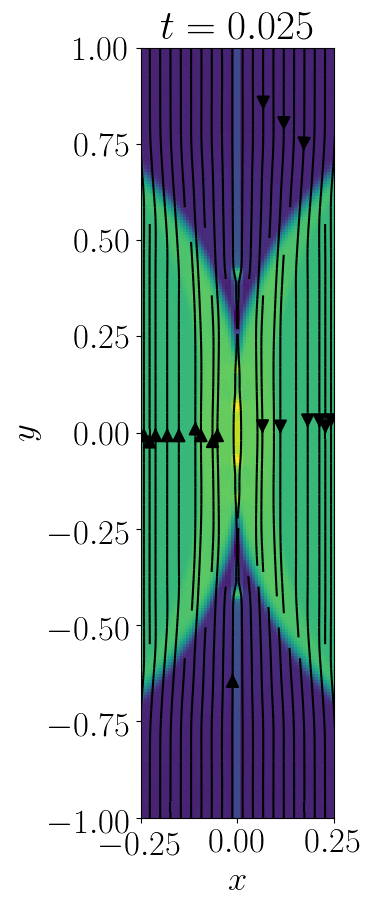}
\caption{
A snapshot of \mbox{MRCOL2D} density (color) and B-field (stream lines) at t=0 (left), 0.015 (middle), 0.025 (right). The x-y plane here should be compared with the x-z plane in \mbox{MRCOL}.
\label{fig:mrcol2d}}
\end{figure*}

To look into the details of how MR is triggered, we extract the y=0 plane from \mbox{MRCOL} and carry out a new 2D simulation (hereafter \mbox{MRCOL2D}) and monitor the MR triggering for t$<$0.1. For simplicity, this time we remove the shear velocity and only consider a head-on collision. Everything else follows the initial conditions of \mbox{MRCOL}. Figure \ref{fig:mrcol2d} shows three snapshots for \mbox{MRCOL2D} density at t=0, 0.015, 0.025. Again, the magnetic field is shown as the streamlines. At t=0, the initial condition is the same as the y=0 plane in \mbox{MRCOL} except there are no y-direction velocity components. At t=0.015, we see the B-field lines around x=0 start to come close at y $\approx\pm$0.25 (hereafter the merging point). The merging point is right behind a parcel of gas moving away from the compression pancake along the y-axis at x=0. The parcel at y $\approx$ 0.3 is moving toward positive y-direction, while the one at y $\approx$ -0.3 is moving toward negative y. The parcels are simply ejecta from the compression pancake. Such ejection was observed in previous studies \citep[e.g.,][]{1998ApJ...497..777K}. In our case, the parcel is simply accelerated by the high density in the pancake, since in isothermal case the pressure is simply computed as $P=\rho c_s^2$. 

If we think about how MR happens \citep{1958IAUS....6..123S,1957JGR....62..509P}, we essentially need differential orthogonal velocity in the reconnection interface to form ``X-shaped'' field lines. In our simulation \mbox{MRCOL\_uniform}, no velocity perturbation was introduced so no reconnection occurs, even though magnetic diffusion still takes place because we included $\eta_{\rm ohm}$. However, in the case of \mbox{MRCOL}, the ejecta carry dense gas parcels away. The region between the compression pancake and the dense parcel has a local minimum in pressure. Since the colliding clumps continue to bring in material, the local pressure minimum allows the development of a local maximum incoming velocity. As a result, B-field lines are bent inward and MR is triggered. Hereafter, we name this process ``collision-induced magnetic reconnection'' (CMR).

At t=0.025, MR happens at y $\approx\pm$0.25, and B-field lines break. One can see a loop forming and circling around the central compression pancake. It then tries to pull the pancake gas to y=0. The dense ejecta parcels are enclosed by other loops and are pulled away. Also, remember that the collision brings in more B-field lines to the collision interface. So the B-field strength at the interface is elevated. When MR happens, the central B-field loop is more capable of pulling gas inward, which is why \mbox{MRCOL} was able to enhance the density contrast by two orders of magnitude. 

In our setup of the clump-clump collision, the 2D symmetry in \mbox{MRCOL2D} is maintained in 3D. Because of magnetic coupling, ejected parcels still prefer moving along B-field lines. Therefore, the entire pancake is being pulled to the y-axis in \mbox{MRCOL}, resulting in the filament formation. The initial MR and the compression pancake trigger magnetic and compressive waves. These waves propagate radially in the x=0 plane, resulting in more MR and structure formation in the x=0 plane (see Figure \ref{fig:mrcol}(b)). Without MR, the clump collision produces a pancake that is axially symmetric. In the presence of MR, the symmetry is broken by the B-field and reduced to a filament. B-field cannot pull the gas into a spherical point. A filament is a natural result.

\begin{figure*}[htb!]
\centering
\epsscale{0.35}
\plotone{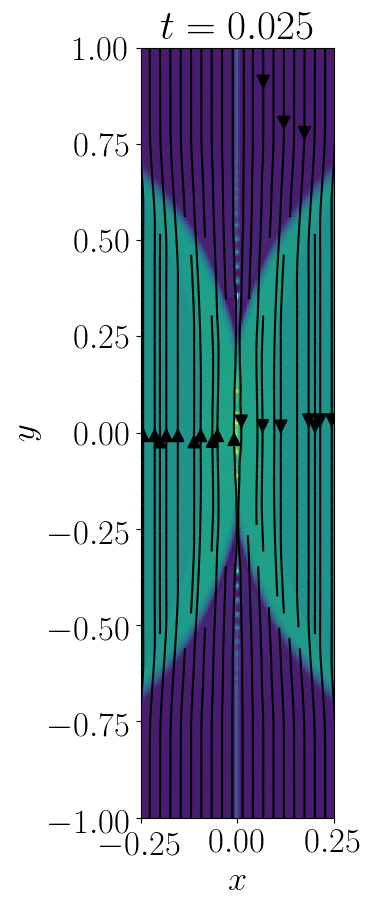}
\plotone{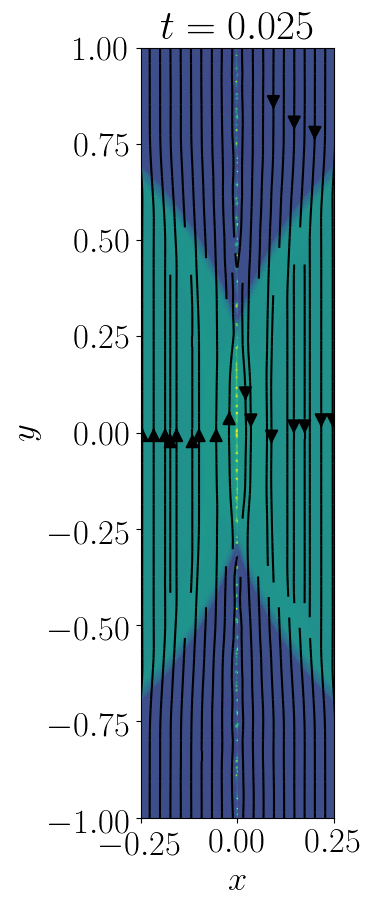}
\plotone{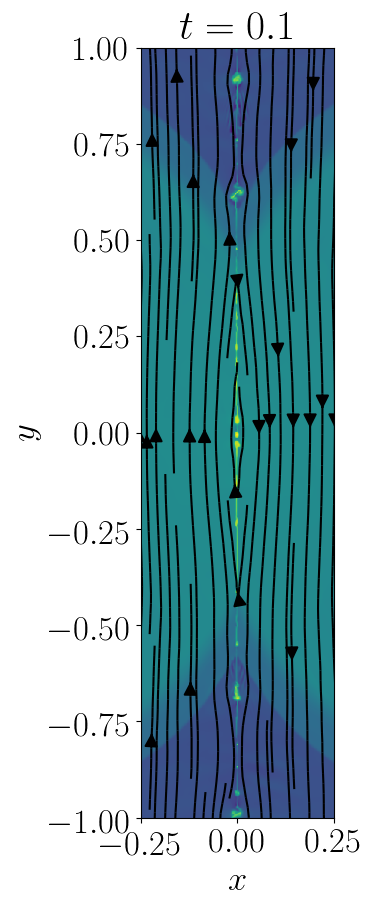}
\caption{
{\bf Left:} A snapshot of \mbox{MRCOL2D\_1024} density (color) and B-field (stream lines) at t=0.025. {\bf Middle and Right:} Snapshots of \mbox{MRCOL2D\_2048} density (color) and B-field (stream lines) at t=0.025 and t=0.1, respectively. 
\label{fig:mrcol2d_highres}}
\end{figure*}

To confirm the CMR, we run two more simulations with higher resolutions (1024$^2$ and 2048$^2$ grids). Hereafter we name them \mbox{MRCOL2D\_1024} and \mbox{MRCOL2D\_2048}. The former has a cell size of 0.004 pc, while the latter has 0.002 pc. The original \mbox{MRCOL2D} has 512$^2$ grids and a cell size of 0.008 pc. Figure \ref{fig:mrcol2d_highres} shows three snapshots from the two new simulations. The first two panels show t=0.025 density snapshots for \mbox{MRCOL2D\_1024} and \mbox{MRCOL2D\_2048}, respectively. Comparing the left and middle panels of  igure \ref{fig:mrcol2d_highres} with the right panel of Figure \ref{fig:mrcol2d}, we can see that all three simulations show the triggering of MR at x=0. The qualitative behavior between the three simulations (\mbox{MRCOL2D}, \mbox{MRCOL2D\_1024}, \mbox{MRCOL2D\_2048}) is the same. The difference in the high-resolution simulations is that more MR-induced local density peaks form. Interestingly, the merging of the slightly offset islands causes rotation. For instance, in the right panel of Figure \ref{fig:mrcol2d_highres}, we can see a spiral structure in the big island at y$\approx$0.62. This rotation could be the origin of the angular momentum of some filaments. 

\section{Discussion}

\subsection{Dense Core Status and Comparison with K17}\label{subsec:kirkcores}

K17 studied dense cores in Orion A and showed that most cores are not massive enough to stay bound. They suggested that the cores are confined by external pressure from the cloud weight. Their study included a few cores in the Stick. Since we argued that the Stick formed via CMR, it would be interesting to investigate how magnetic fields change the picture. We investigate this by identifying a core in the simulations to evaluate its dynamical status.

We pick out one core from the simulation to evaluate its dynamical status. The core is defined as a sphere enclosing the bright emission in Figure \ref{fig:fir} (the white circle). Its center of mass is at x=0, y=-0.05, z=0 and its radius is 0.05 (that is, a physical size of 0.1 pc). To quantify the core virial status, we compute the different virial terms for the core. Following \citet{2017stfo.book.....K}, we write the virial equation below:
\begin{equation}\label{equ:virial}
\begin{split}
\frac{1}{2}\ddot{I} = & \int_{V}\left(\rho v^2+3P\right)dV - \int_{S}\mathbf{r}\cdot\mathbf{\Pi}\cdot d\mathbf{S} \\
& + \frac{1}{8\pi}\int_{V}B^2dV + \int_{S}\mathbf{r}\cdot\mathbf{T_M}\cdot d\mathbf{S} \\
& - \int_{V}\rho\mathbf{r}\cdot\nabla\phi dV - \frac{1}{2}\frac{d}{dt}\int_{S}(\rho\mathbf{v}r^2)\cdot d\mathbf{S}
\end{split}
\end{equation}
Here, $I$ is the moment of inertia, $\mathbf{\Pi}$ is the fluid pressure tensor, and $\mathbf{T_M}$ is the Maxwell stress tensor. On the right hand side of equation \ref{equ:virial}, the first term (defined as $2\Omega_K$) includes the thermal and turbulent energy in the core. It tries to support the core from collapse. The second term (defined as -$\Omega_{KS}$) is the pressure surface term. It includes two terms:
\begin{equation}\label{equ:ks}
\begin{split}
-\int_{S}\mathbf{r}&\cdot\mathbf{\Pi}\cdot d\mathbf{S} = \\ & -\int_{S}\mathbf{r}\cdot(\rho\mathbf{vv})\cdot d\mathbf{S} -\int_{S}\mathbf{r}\cdot(P\mathbf{I})\cdot d\mathbf{S} \\ & = -\int_{S}\mathbf{r}\cdot(\rho\mathbf{vv})\cdot d\mathbf{S} - 4\pi R^3\overline{P_s}.
\end{split}
\end{equation}
Here $R$ is the radius of the core. The first term (-$\Omega_{KS1}$) relates to the momentum transfer through the surface. Its sign depends on whether gas is moving in or out of the core. The second term (-$\Omega_{KS2}$) is the surface pressure that tries to confine the core. This term was included in the K17 analysis. 

The third term (defined as $\Omega_B$) on the right hand side of equation \ref{equ:virial} is the magnetic pressure term that supports the core. The fourth term (defined as -$\Omega_{BS}$) is the magnetic surface pressure term. It also includes two terms:
\begin{equation}\label{equ:bs}
\begin{split}
\int_{S}&\mathbf{r}\cdot\mathbf{T_M}\cdot d\mathbf{S} = \\ & \int_{S}\mathbf{r}\cdot\left(\frac{1}{4\pi}\mathbf{BB}\right)\cdot d\mathbf{S} - \int_{S}\mathbf{r}\cdot\left(\frac{B^2}{8\pi}\mathbf{I}\right)\cdot d\mathbf{S} \\ & = \int_{S}\mathbf{r}\cdot\left(\frac{1}{4\pi}\mathbf{BB}\right)\cdot d\mathbf{S} - 4\pi R^3\left(\frac{\overline{B_s^2}}{8\pi}\right)
\end{split}
\end{equation}
The first term (defined as $\Omega_{BS1}$) relates to the magnetic tension discussed earlier. The second term (defined as -$\Omega_{BS2}$) is the magnetic pressure at the surface. The fifth term (defined as $\Omega_G$) on the right hand side of equation \ref{equ:virial} is the gravity term that tries to make the core collapse. The sixth term (defined as $\Omega_t$) is the moment of inertia that gets in or goes out of the core. A core in virial equilibrium has:
\begin{equation}\label{equ:veq}
\begin{split}
2\Omega_K & + \Omega_B + \Omega_t = -\Omega_G + \Omega_{KS} + \Omega_{BS} \\ & = -\Omega_G + \Omega_{KS1} + \Omega_{KS2} - \Omega_{BS1} + \Omega_{BS2}
\end{split}
\end{equation}

We compute each of these terms for the core. In particular, the momentum surface term $\Omega_{KS1}$ is computed as:
\begin{equation}\label{equ:ks1}
\begin{split}
\Omega_{KS1} = & \int_{S}\left(\frac{\rho}{r}\right)(x^2v_x^2+y^2v_y^2+z^2v_z^2 \\ & +2xy v_x v_y+2yz v_y v_z+2zx v_z v_x)dS    
\end{split}
\end{equation}
Here, (x, y, z) is the coordinate and ($v_x$, $v_y$, $v_z$) is the velocity. In practice, to estimate $\Omega_{KS1}$, we take the mean value of the integrand in a shell outside the core and multiply it by $4\pi R^2$. Similarly, the magnetic tension surface term $\Omega_{BS1}$ is computed as:
\begin{equation}\label{equ:bs1}
\begin{split}
\Omega_{BS1} = & \int_{S}\left(\frac{1}{4\pi r}\right)(x^2B_x^2+y^2B_y^2+z^2B_z^2 \\ & +2xy B_x B_y+2yz B_y B_z+2zx B_z B_x)dS    
\end{split}
\end{equation}
Here, ($B_x$, $B_y$, $B_z$) is the magnetic field. The $\Omega_t$ term is computed as:
\begin{equation}\label{equ:t}
\begin{split}
\Omega_{t} = - \frac{d}{dt}\int_{S}\left(\frac{\rho r}{2}\right)(xv_x+yv_y+zv_z)dS
\end{split}
\end{equation}

The result is, $2\Omega_K=0.027$, $\Omega_B=0.00058$, $\Omega_t=0.0014$, $-\Omega_G=0.0016$, $\Omega_{KS1}=0.014$, $\Omega_{KS2}=0.0064$, $-\Omega_{BS1}=-0.0016$, $\Omega_{BS2}=0.057$. The dominating supporting source is the kinetic energy term, $2\Omega_K$. The dominating confining source is the surface magnetic pressure term, $\Omega_{BS2}$. Circular field lines wrap the core. They pile up strong magnetic pressure that confines the core. This can be understood from equation \ref{equ:bs}. When circular fields wrap the core surface, $\mathbf{B}$ is perpendicular to $d\mathbf{S}$. Thus, the first term in equation \ref{equ:bs} diminishes. Then, all contributions to the magnetic surface term are from the magnetic surface pressure term $\Omega_{BS2}$ (=$B_s^2R^3/2$).

The surface pressure term $\Omega_{KS2}$, which was the main confining source in K17, is also confining the core but plays a minor role. The surface momentum transfer term $\Omega_{KS1}$ is actually twice as strong as the surface pressure confinement. Note that our simulation only includes local clumps. If we consider a larger cloud that encloses the clumps, as K17 considered, the surface pressure term might be stronger. Interestingly, gravity is not very important. It is more than an order of magnitude smaller than the kinetic energy term, similar to what was seen in K17.

What is the fate of the cores in the Stick filament? Can they eventually form stars? The key factor is probably cooling. In fact, we have shown that the largest dense cores in the Stick have a lower temperature than the filament and its environment. Also, cores (especially 99) show signs of CO depletion. Both suggest that the dense cores can cool efficiently enough to remove the excessive energy. They may eventually collapse and form stars. 

\subsection{CMR in Orion A}\label{subsec:oriona}

The flipped line-of-sight magnetic fields around Orion A have been interpreted as a helical field \citep[e.g.,][]{2016A&A...590A...2S}. Later, \citet{2019A&A...632A..68T} confirmed the flipped B-field and investigated several possibilities, and favored an interpretation of a bow-shaped B-field geometry. The flipped B-field motivated us to explain the Stick formation with CMR (Figure \ref{fig:ic}). 

Following this reasoning, it is possible that the entire Orion A formed through one or multiple collision events with CMR happening at different levels. The whole cloud, being swept by multiple supernovae \citep{2008hsf1.book..459B}, shows a large-scale velocity gradient \citep[in the ISF, e.g.,][]{2019ApJ...882...45K,2019MNRAS.489.4771G}. One possible scenario is that the expanding Barnard's Loop \citep[the most recent supernova event][]{Ochsendorf2015} collided with another cloud of gas. The colliding clouds could have carried anti-parallel B-fields. As a result, the cloud collision could have triggered MR and created a dense filamentary cloud, which is the Orion A we see today. Recently, \citet{2020PASJ..tmp..187L} also suggested a cloud-cloud collision scenario for Orion A formation but based on the gas-star kinematics. 

The question is, why we do not see other filaments like the Stick in Orion A? As we have shown in \S\ref{subsec:mrcolre} and Figure \ref{fig:mrcol_sameB}, no filament was able to form when the colliding clumps have parallel B-fields because MR requires anti-parallel B-fields. This variation in the relative B-field orientation reminds us of considering collision conditions, especially whether they favor MR or not. Following this idea, we can imagine two colliding giant diffuse clouds with clumpy structures, i.e., each cloud contains many clumps. The clumps have different shapes, sizes, densities, magnetic orientations, and magnetic field strengths. While some clumps in one cloud may collide with clumps in the other cloud, other clumps may not  pass through the other giant cloud without colliding with another clump. Therefore, conditions that favor \mbox{MRCOL} may be quite uncommon. The Stick is probably a rare case that happened via a process like that in \mbox{MRCOL}. A global simulation of a collision between two clumpy giant clouds is probably necessary to address these questions. Below we explore a few variations of clump-clump collision from \mbox{MRCOL} to cast some light on how (un)likely it is for a Stick-like filament to form.

\begin{figure}[htb!]
\centering
\epsscale{1.}
\plotone{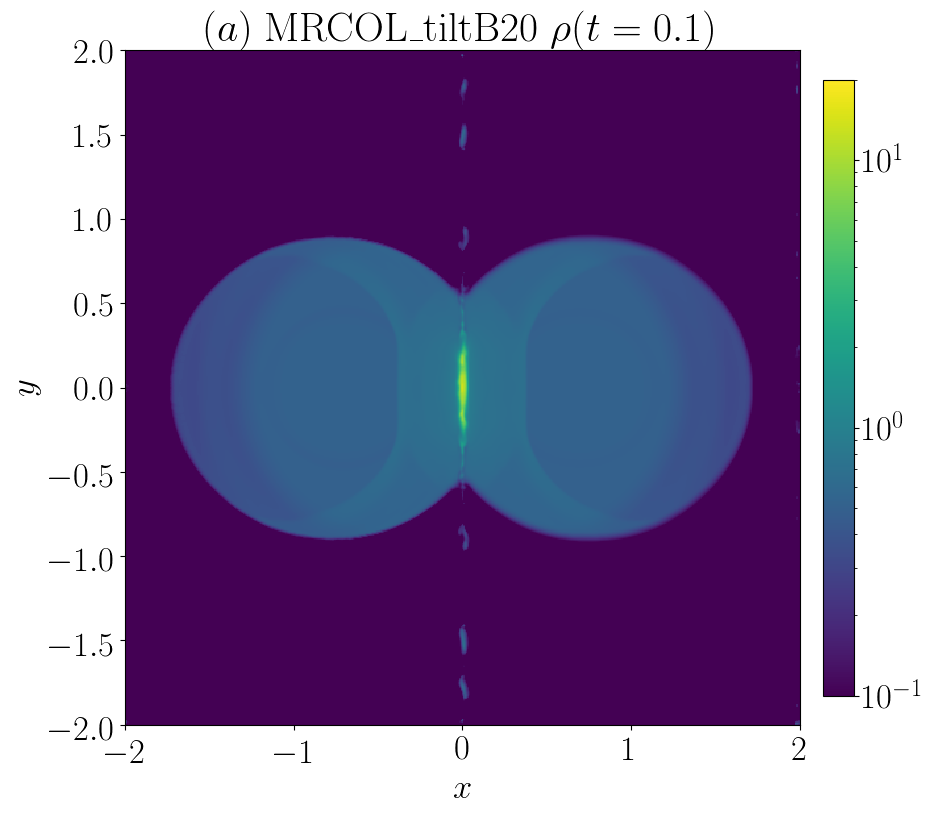}
\plotone{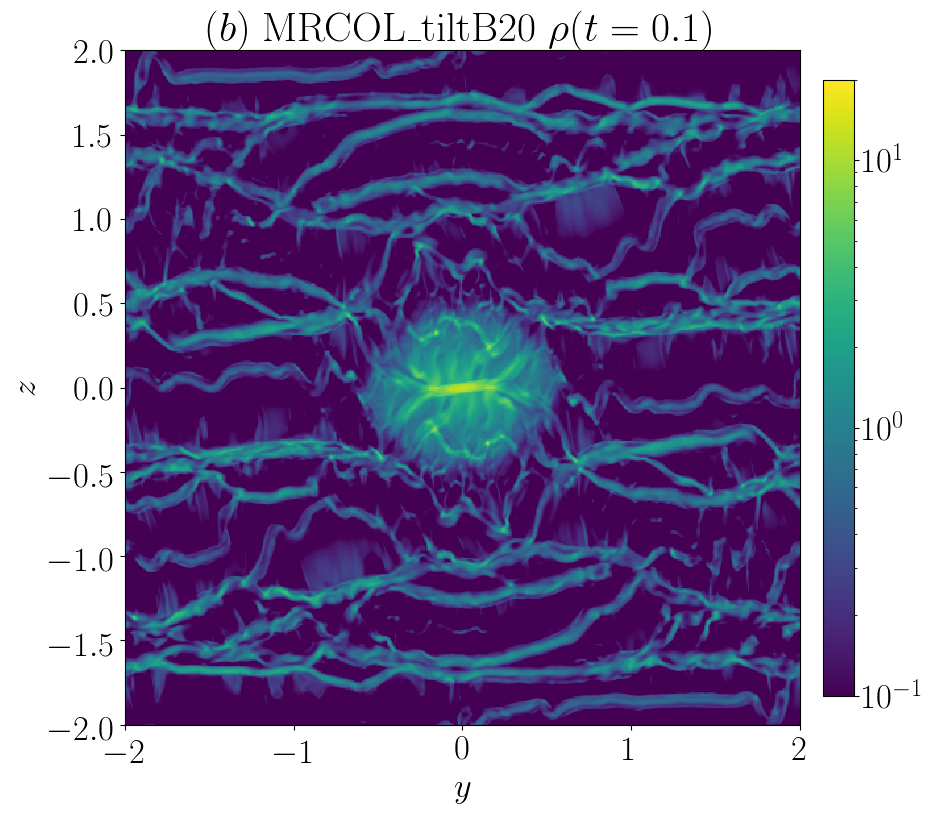}
\plotone{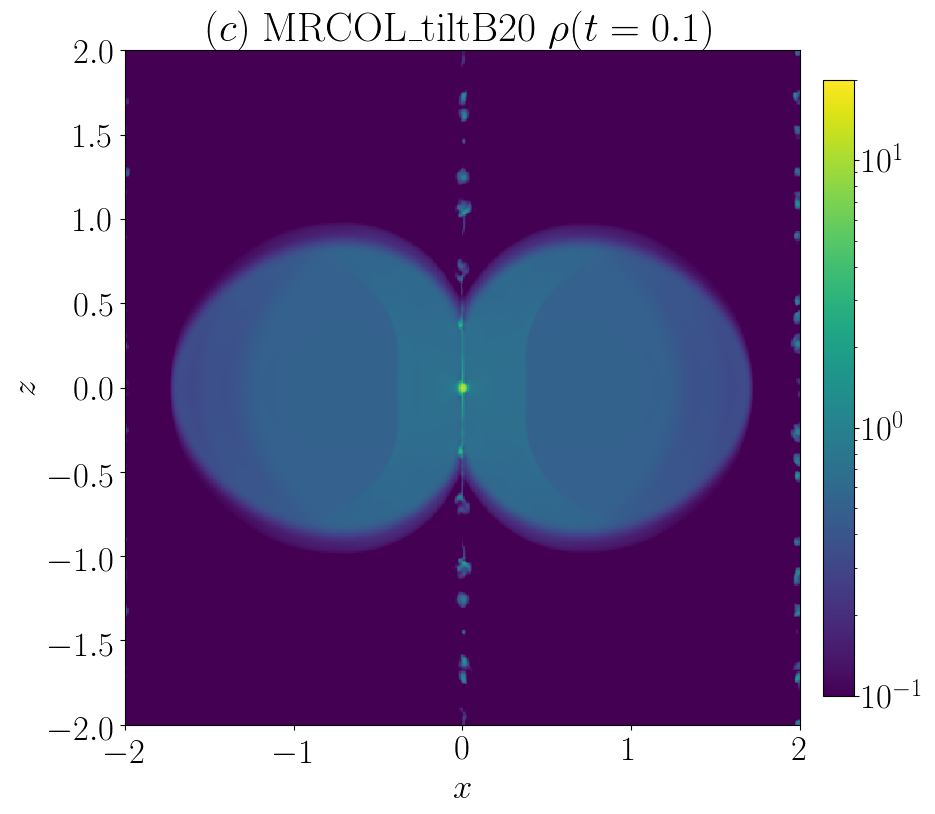}
\caption{
A snapshot of \mbox{MRCOL\_tiltB20} density at t=0.1 (0.2 Myr). Panels (a), (b), (c) show the z=0, x=0, y=0 planes, respectively. The color scale is logarithmic, ranging from 0.1-20. See \S\ref{subsec:2dtest} for unit conversions.
\label{fig:mrcol_tiltB20}}
\end{figure}

First, we run \mbox{MRCOL} again but with $B_2$ tilted by 20 degree (i.e.,  $B_1$ and $B_2$ are not exactly anti-parallel). We remove the shear velocity and only consider head-on collisions. Hereafter we name this simulation \mbox{MRCOL\_tiltB20}. Figure \ref{fig:mrcol_tiltB20} shows a snapshot of the result at t=0.1. There is still a dense filament formed in the pancake, as shown in panels (a) and (c). But in panel (b) the filament is not as straight as that in \mbox{MRCOL}. It is also shorter and more chaotic. The highest density in the cube only reaches 13.7, an order of magnitude smaller than \mbox{MRCOL} (at the same time). The results show that a slight deviation from the anti-parallel B-fields can ruin the filament formation. The curved filaments outside the pancake are interesting. They are long and roughly parallel to the central filament. They show rich sub-structures, including sub-filaments that are tangled together.

\begin{figure}[htb!]
\centering
\epsscale{1.}
\plotone{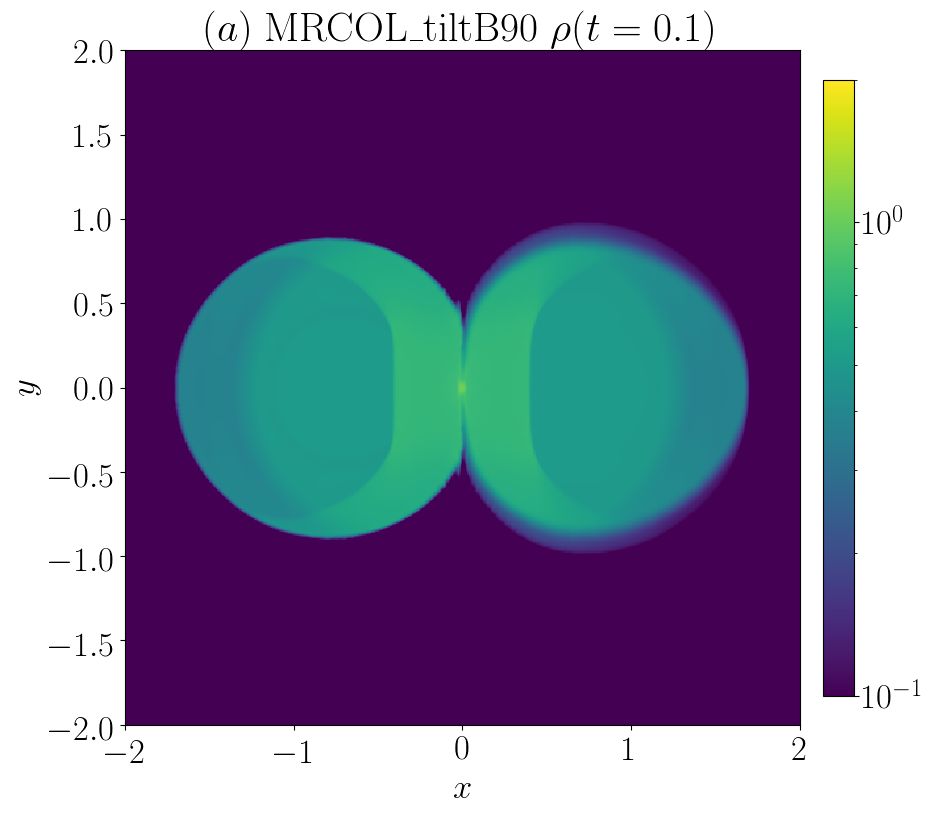}
\plotone{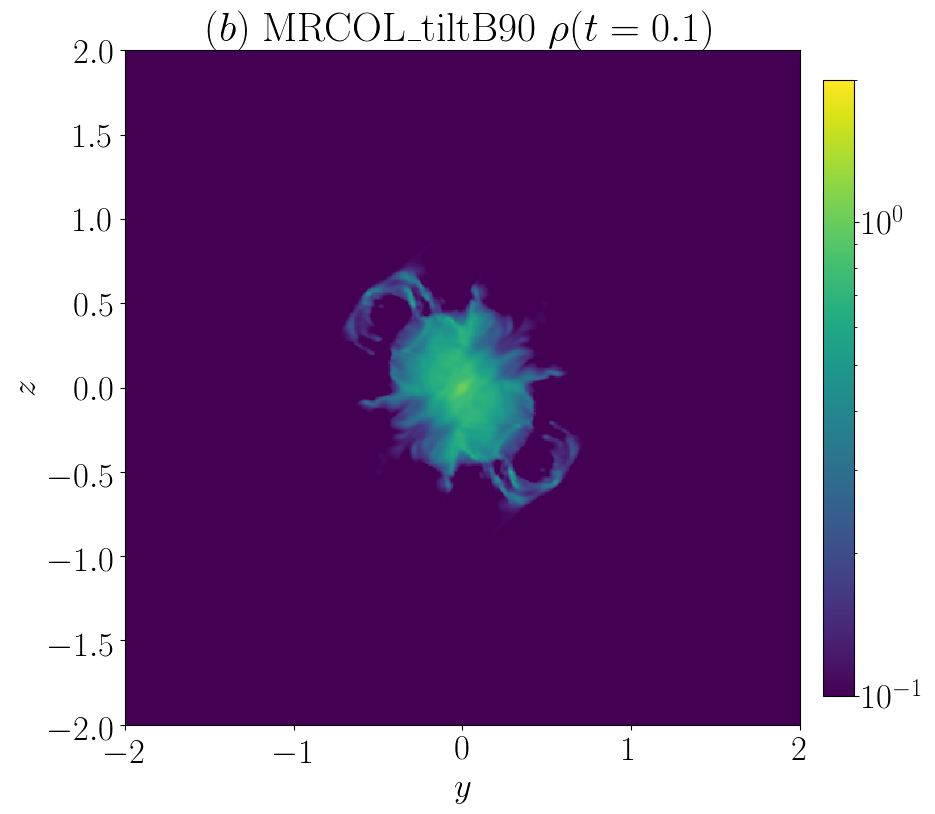}
\plotone{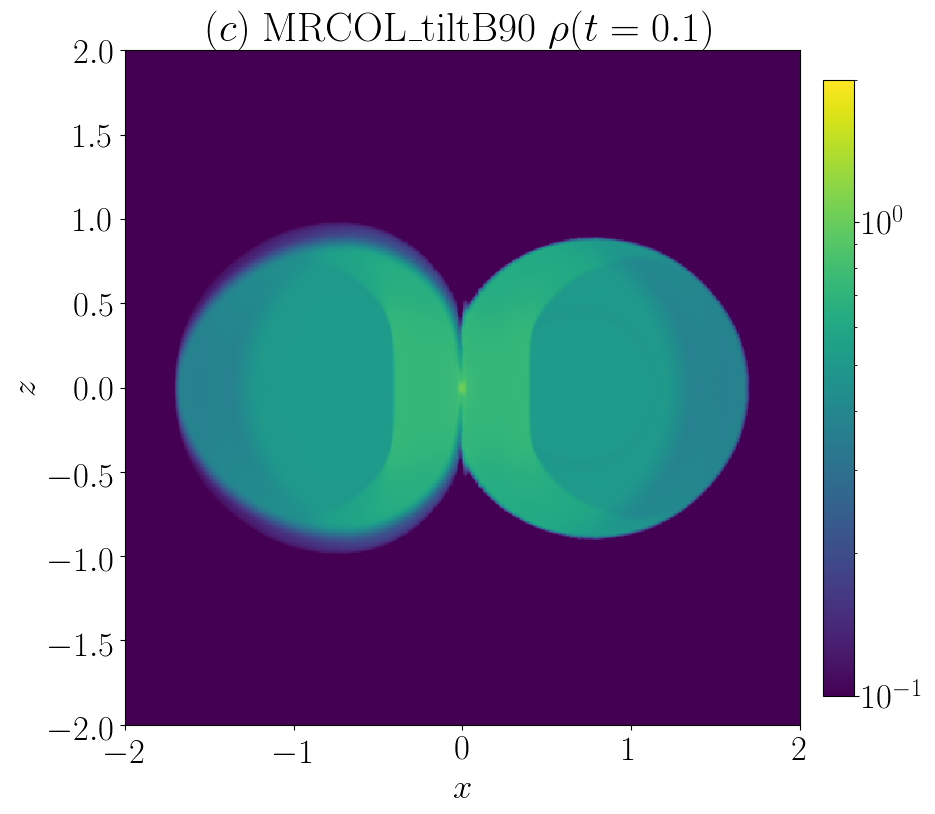}
\caption{
A snapshot of \mbox{MRCOL\_tiltB90} density at t=0.1 (0.2 Myr). Panels (a), (b), (c) show the z=0, x=0, y=0 planes, respectively. The color scale is logarithmic, ranging from 0.1-2. See \S\ref{subsec:2dtest} for unit conversions.
\label{fig:mrcol_tiltB90}}
\end{figure}

We next run another simulation similar to \mbox{MRCOL}, but with   $B_2$ tilted by 90 degree, and name this simulation \mbox{MRCOL\_tiltB90}. Figure \ref{fig:mrcol_tiltB90} shows the density snapshots at t=0.1. Compared to \mbox{MRCOL}, the major difference is that no dense filament forms. If we look at the x=0 plane shown in panel (b), we can see a peculiar structure forms in the midplane. It is roughly symmetric about the diagonal line. MR still happens in this case and causes the diagonal symmetry because one B-field line (in Clump1) points along the z-axis and another line (in Clump2) points along the y-axis (tilted 90 degree). MR creates hourglass-shaped field lines that are diagonally symmetric. In fact, in panel (b) we see an hourglass shape from lower left to top right. The maximum density reaches $\rho\sim1$, significantly lower than \mbox{MRCOL}. The perpendicular fields completely ruin the filament formation compared to \mbox{MRCOL}.

\begin{figure}[htb!]
\centering
\epsscale{1.}
\plotone{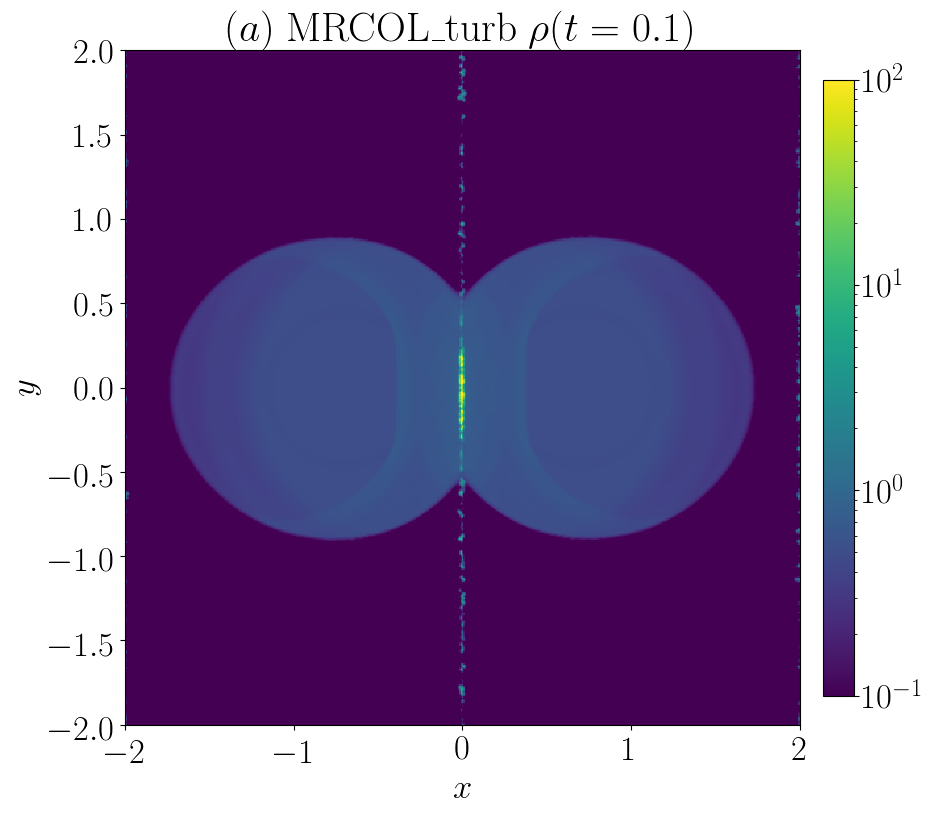}
\plotone{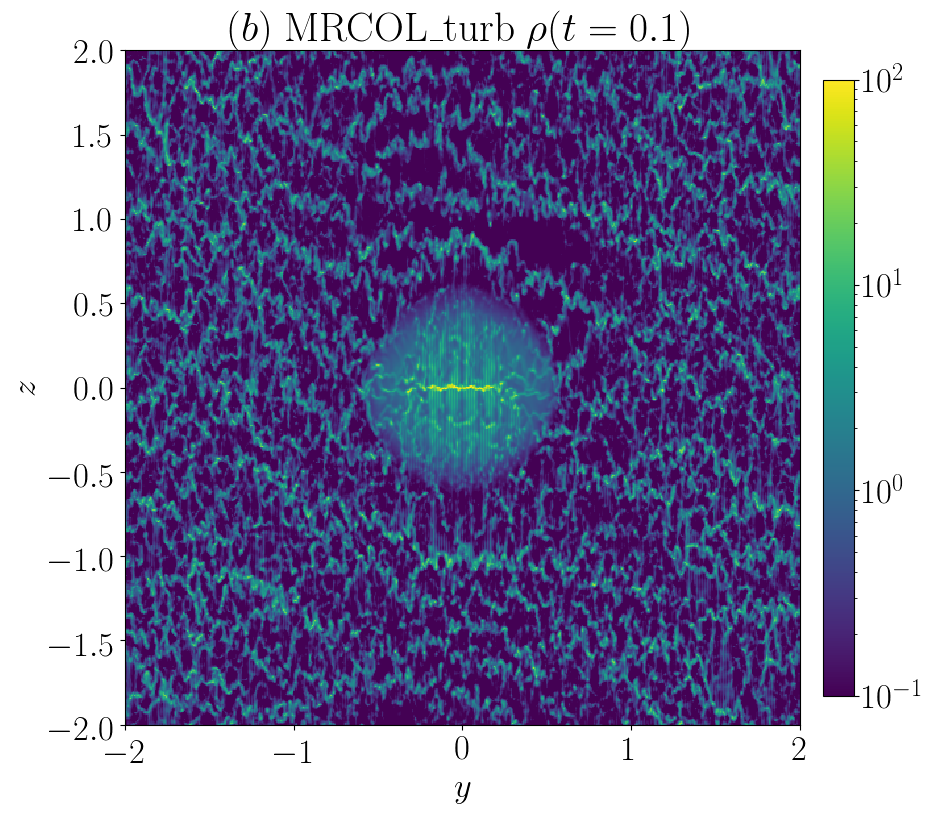}
\plotone{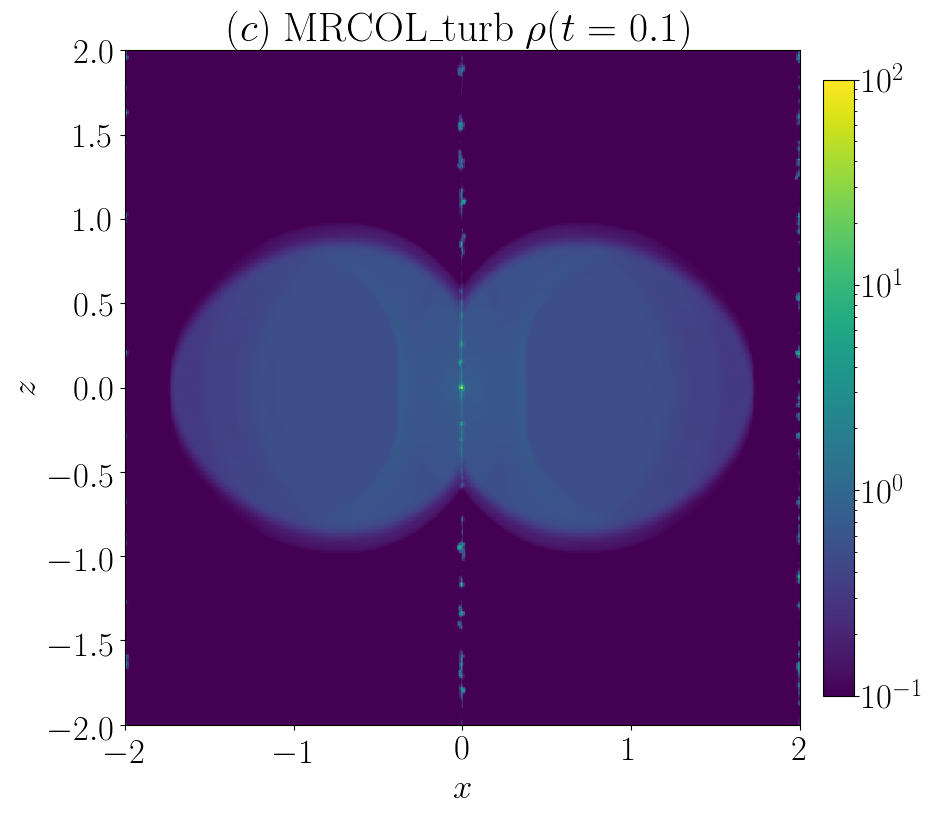}
\caption{
A snapshot of \mbox{MRCOL\_turb} density at t=0.1 (0.2 Myr). Panels (a), (b), (c) show the z=0, x=0, y=0 planes, respectively. The color scale is logarithmic, ranging from 0.1-100. See \S\ref{subsec:2dtest} for unit conversions.
\label{fig:mrcol_turb}}
\end{figure}

Next, we simulate another variation of \mbox{MRCOL} in which we include turbulence (hereafter \mbox{MRCOL\_turb}). Turbulence should be able to accelerate MR \citep{1999ApJ...517..700L,2009ApJ...700...63K} and interstellar clouds are turbulent \citep{Larson1981}. In \mbox{MRCOL\_turb}, we inject a small amount of turbulent energy (0.1 in code units) at the beginning of the simulation. The injection is at the largest scales with wavenumbers at $0<|{\bf k}|<2$ (in unit of $2\pi/L_{\rm box}$). No more turbulence is driven during the simulation. Athena++ allows us to distribute the turbulent energy between the compressive mode and the solenoidal mode. We simply allocate half of the energy injection to each of the modes. 

Figure \ref{fig:mrcol_turb} shows the result from \mbox{MRCOL\_turb}. Again, it is the snapshot at t=0.1. As shown in panel (b), the dense filament is able to form in the compression pancake. Interestingly, the highest density now reaches 223, a factor of 1.6 higher than \mbox{MRCOL}. However, this time the filament is confined at the inner region of the vague pancake, with a length shorter than 1. Recall in \mbox{MRCOL} the dense filament stretches out of the pancake and almost reaches the boundaries of the simulation box (Figure \ref{fig:mrcol}(b)). 

We have shown that the variation of one parameter, i.e., the B-field orientation, can result in very different outcomes. The inclusion of turbulence further enhance the difference. If we also consider variations in other physical parameters (density, size, temperature, to name a few), it would be of no surprise that only one  Stick formed in Orion A. Collisions between different clump pairs may also happen at different times because some clumps just come later. So there can also be an age spread in the dense gas and the stars. In our simulations, we did not include star formation because based on the observations (\S\ref{sec:obs}) the Stick is very young. However, in \mbox{MRCOL} and \mbox{MRCOL\_turb}, gas densities can reach $\sim$ 10$^5$ cm$^{-3}$, exceeding the star formation threshold \citep[e.g.,][]{2014MNRAS.444.2396C}.

Here we briefly discuss a few possible predictions for future observational tests. First, high-resolution Zeeman measurements of the line-of-sight magnetic fields around the Stick may be useful. Based on our model, the line-of-sight B-fields should be flipped. However, depending on the projection, the line-of-sight column can include both positive and negative fields, which is probably why \citet{1997ApJS..111..245H} had less Zeeman detections in the central regions of Orion A. Second, high-resolution polarization study should find relatively ordered field lines outside the Stick but more chaotic lines inside the Stick, as the filament is where MR happens. The MR-dominated compression pancake should have messy B-field orientations. 

\subsection{Dense Gas Production}\label{subsec:dg}

The formation of the filament in \mbox{MRCOL} is essentially a way of producing dense gas. It is different from the dense gas creation by supersonic turbulence \citep[e.g.,][]{2016MNRAS.457..375F,2016ApJ...822...11P}, where filaments primarily form from the shocked gas created by turbulence. Consequently, it would be interesting to see in future studies how CMR-produced dense gas changes the density PDF and star formation rate (SFR). It is also different from other  cloud-cloud collisions \citep[e.g.,][]{2020ApJ...891..168W}, where stronger B-field strengths suppress the SFR. In our model, in principle, the stronger the initial B-field strength the more capable the reconnected B-field loop is of pulling gas into the filament (\S\ref{subsec:ph}), potentially accelerating star formation. 

For instance, \mbox{MRCOL} is able to create dense gas with $n_{\rm H_2}=1.2\times10^5$ cm$^{-3}$ within $2.0\times10^5$ yr from colliding clumps with $n_{\rm H_2}=420$ cm$^{-3}$. The density increase is a factor of 276. Meanwhile, the time the clump pressurelessly collapses to gain the same factor is $\sim$0.94 times the free-fall time of the clump, which is $\sim1.5\times10^6$ yr. Therefore, it will take the clump seven times longer to reach the high density by free-fall than by CMR. The time factor can be larger if we explore different initial conditions. CMR is fast in producing dense gas.

However, as we briefly explored in \S\ref{subsec:oriona}, what happens in \mbox{MRCOL} may not be very common during a cloud-cloud collision event. A slight change in the relative B-field orientation can greatly reduce the dense gas production. It remains to be explored how much CMR may contribute to the overall SFR in giant molecular clouds and galaxies. There could be a favorable situation where CMR happens prevalently so as to create a (mini) starburst. Recall that in a disk galaxy, a molecular cloud is expected to collide with another cloud once every 1/5 of its orbital period \citep{2009ApJ...700..358T}. So the CMR mode of star formation can be crucial. 

An interesting example is the comparison between the California Molecular Cloud (CMC) and Orion A, which have similar physical properties in many aspects. However, the latter has an order of magnitude higher SFR than the former \citep{2009ApJ...703...52L}. After some exploration \citep{2015ApJ...805...58K,2016ApJ...825...91L,2017A&A...606A.100L}, it was suggested that the lack of dense gas in the CMC was the cause of the low SFR. If Orion A indeed had CMR, then it can probably explain the origin of the excessive dense gas in Orion A because CMR can produce dense gas much faster than free-fall collapse. Moreover, when considering other dense molecular clouds (e.g., infrared dark clouds), we may also want to consider the possibility of CMR, especially for those clouds created through collision events \citep[e.g.,][]{2016ApJ...820...26F,2020A&A...638A..44B}.

\section{Summary and Conclusion}

In this paper, we have identified a unique filament (the Stick) in the Orion A cloud, based on {\it Herschel} data and molecular line data. Being next to the famous L1641-N cluster, the Stick filament stands out due to its ruler-straight morphology. The Stick has densities $n_{\rm H_2}\ga10^5$ cm$^{-3}$ and is in an environment with a temperature of $\sim$ 15 K.  Using the CARMA-NRO Orion data, we have found the molecular line counterpart of the Stick. The densest part of the Stick shows a moderate CO depletion factor of $\sim$5. The majority of the Stick is starless. These observational results show that the Stick is at a very early evolutionary stage.

The Stick shows very interesting kinematic features in the optically thin C$^{18}$O(1-0) line data. In particular, it shows ring/fork-like features in channel maps. In both C$^{18}$O and NH$_3$ position-velocity diagrams, we see double velocity components on both sides of the Stick. Their velocity separation is about 1 km s$^{-1}$. These kinematic features, especially the fork-like structures, motivate us to consider that the Stick formed via magnetic reconnection. We carry out numerical simulations to model the Stick formation and compare with the observations. 

We use the latest public version of the Athena++ code to simulate the compressible, isothermal, inviscid MHD fluid, with self-gravity and Ohmic resistivity. To match previous observational results of the magnetic fields around the Stick, we propose an initial condition of clump-clump collision, each clump carrying an anti-parallel B-field (Figure \ref{fig:ic}). In the fiducial model \mbox{MRCOL}, a narrow, straight, long filament forms at the collision point. The filament reaches a density of $n_{\rm H_2}=1.2\times$10$^5$ cm$^{-3}$ within $2\times10^5$ yr. The density is a factor of more than 200 times higher than the initial clump density ($n_{\rm H_2}=420$ cm$^{-3}$). The time for the clump to free-fall collapse to reach the high density is seven times longer.

Most importantly, the filament formed in \mbox{MRCOL} shows many ring/fork-like structures that are reminiscent of the Stick. We carry out radiative transfer models for the line emission and compare it with the molecular line maps of the region. The results are consistent with the observations regarding the morphology and the kinematics. In particular, the filament is narrow and long and shows rings and forks in the model channel maps. A transverse PV-diagram across the filament shows a similar two-velocity-component feature as shown in the C$^{18}$O and NH$_3$ PV-diagrams. 

Magnetic reconnection is the cause of the filament formation in \mbox{MRCOL}. We find that it happens naturally as a result of the clump collision. The region between the compression pancake and the ejected parcel forms a local pressure minimum. It allows faster incoming gas from both sides of the collision midplane and triggers magnetic reconnection. The collision-induced magnetic reconnection (CMR) gives rise to B-field loops circling the pancake. The loop fields pull the pancake to the central axis. Material accumulates along the central axis and form the filament. The triggering mechanism for the reconnection is independent of the numerical resolution, as demonstrated by our 2D explorations. Therefore, given favorable conditions (namely, anti-parallel B-fields), magnetic reconnection and dense filament formation are inevitable in a clump-clump collision event. 

We have studied the virial status of one core in the filament from the \mbox{MRCOL} simulation and compared it with the cores in the Stick filament studied by K17. We  find that the core in the simulations is not gravitationally dominated but confined by surface pressure. Unlike the findings in K17 that cores are confined by surface pressure exerted by the cloud weight, we find the core is predominantly confined by surface magnetic pressure due to the wrapping fields. The surface thermal pressure plays a minor role in confining the core, although including an enclosing cloud may enhance the surface pressure due to the cloud weight.

Meanwhile, what happens in \mbox{MRCOL} is not always favorable in reality, simply because colliding clumps not always have conditions that are favorable for MR to take place and form filaments, in particular the anti-parallel B-fields. Through a few tests, we find that a small deviation from the anti-parallel B-field setup will greatly reduce the production of dense gas/filaments, although the inclusion of turbulence may counteract the reduction. This may explain why the Stick is so unique in Orion A. We speculate that the entire Orion A formed through a cloud-cloud collision, and magnetic reconnection happened at different levels, meaning that some reconnection events formed dense, quasi-straight filaments while other events formed dense structures but with irregular shapes.   

CMR provides another way of creating dense gas at a rate much faster than free-fall. It may provide a new way to explain the fast production of dense gas in star-forming clouds and galaxies, as cloud-cloud collision can be frequent in disk galaxies. Future global simulations of collisions between clumpy clouds will help clarify how magnetic reconnection impacts properties of molecular clouds and star formation, including the density probability distribution function, and the star formation rate and efficiency.

\acknowledgments 
We thank the anonymous referee for constructive comments that helped improving the manuscript.
We thank Amelia Stutz for providing the {\it Herschel} Orion maps.
SK acknowledges helpful discussions with Xueying Tang, Zhaohuan Zhu, Jesse Feddersen, John Bieging, Yancy Shirley, Jens Kauffmann, Thushara Pillai, and thanks the Yale Center for Research Computing and the staff of their High Performance Computing (HPC) facilities and for their support. All simulations were run on the Yale HPC Grace cluster.
SK and HGA were (partially) funded by NSF award AST-1140063, which 
also provided partial support for CARMA operations.
V.O. and A.S-M. were supported by the Collaborative Research Centre 956, sub-projects C1 and A6, funded by the Deutsche Forschungsgemeinschaft (DFG), project ID 184018867.
RSK acknowledge financial support from the German Research Foundation (DFG) via the Collaborative Research Center (SFB 881, Project-ID 138713538) 'The Milky Way System' (subprojects A1, B1, B2, and B8). He also thanks for funding from the Heidelberg Cluster of Excellence STRUCTURES in the framework of Germany's Excellence Strategy (grant EXC-2181/1 - 390900948) and for funding from the European Research Council via the ERC Synergy Grant ECOGAL (grant 855130).
S.~Suri acknowledges support from the European Research Council under the Horizon 2020 Framework Program via the ERC Consolidator Grant CSF-648405.
Part of this research was carried out at the Jet Propulsion Laboratory, California Institute of Technology, under a contract with the National Aeronautics and Space Administration.

CARMA operations were also supported by
the California Institute 
of Technology, the University of
California-Berkeley, the University of Illinois at
Urbana-Champaign, the University of Maryland College Park, and the
University of Chicago. The Nobeyama 45 m telescope 
is operated by the Nobeyama Radio 
Observatory, a branch of the National Astronomical Observatory 
of Japan.
The James Clerk Maxwell Telescope has historically been operated by the Joint Astronomy Centre on behalf of the Science and Technology Facilities Council of the United Kingdom, the National Research Council of Canada and the Netherlands Organisation for Scientific Research. Additional funds for the construction of SCUBA-2 were provided by the Canada Foundation for Innovation. All of the SCUBA-2 data used in this paper can be downloaded from the Canadian Astronomical Data Centre (CADC) at http://www.cadc-ccda. hia-iha.nrc-cnrc.gc.ca/en/jcmt/ using the Proposal ID code MJLSG31.

\software{Astropy \citep{Astropy-Collaboration13}, Numpy \citep{numpy}, APLpy \citep{Robitaille12}, Matplotlib \citep{matplotlib}, SAOImageDS9 \citep{2003ASPC..295..489J}}

\facility{CARMA,NRO45,Herschel,Yale HPC}; 

\appendix
\restartappendixnumbering

\section{Additional Figures}\label{app:A}

\begin{figure*}[htb!]
\centering
\epsscale{1.}
\plotone{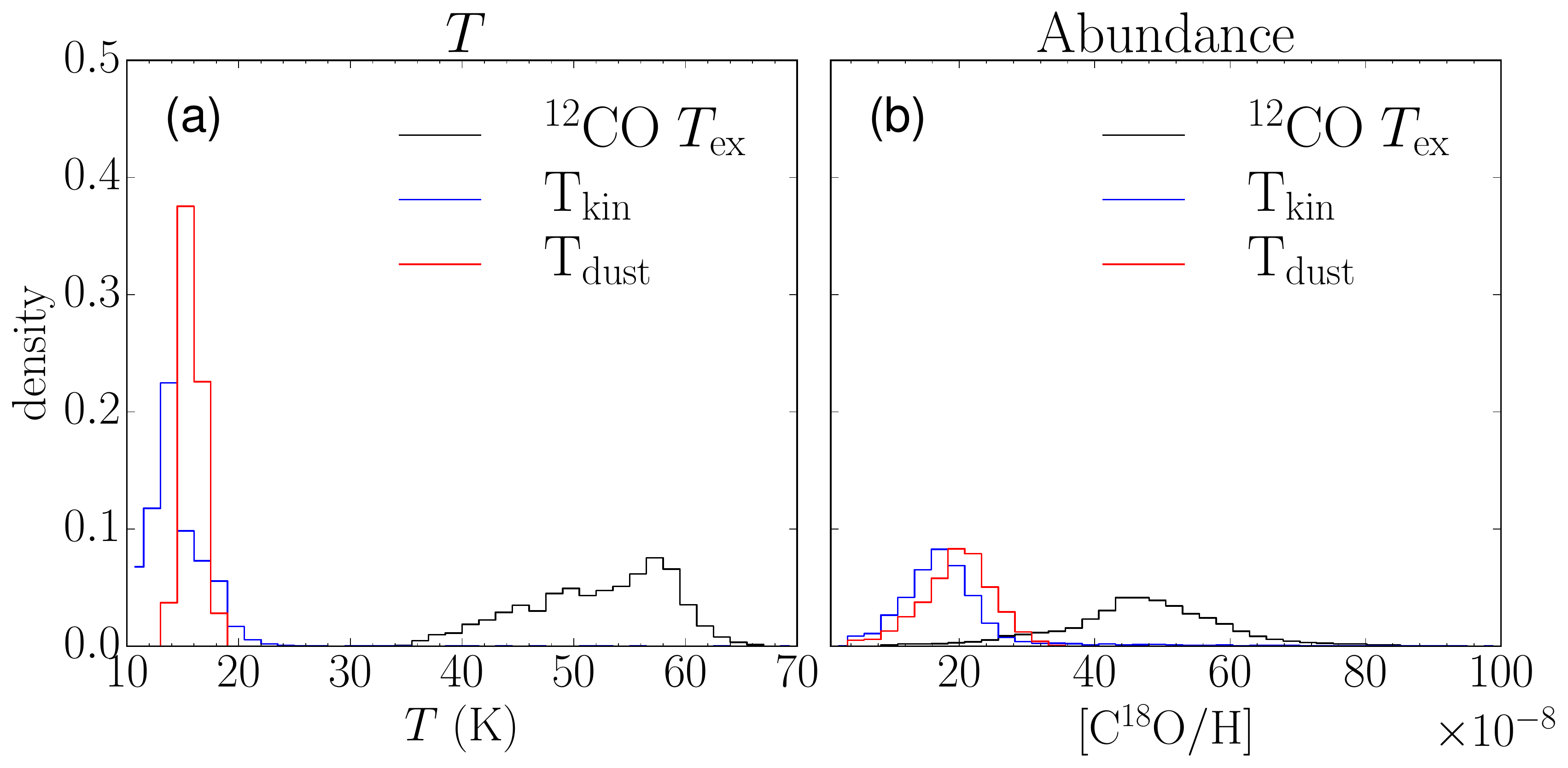}
\caption{
{\bf (a):}
Histograms of temperatures in the Stick body.
The $^{12}$CO excitation temperature is from 
K18. The dust temperature is from SK15
based on {\it Herschel} data. The kinetic temperature
is based on the NH$_3$ data from K17.
{\bf (b):} 
C$^{18}$O abundance map [C$^{18}$O/H]
computed using different temperatures.
\label{fig:hist_t_abu}}
\end{figure*}

In Figure \ref{fig:hist_t_abu}, we show histograms of different temperature estimations for the Stick body in panel (a) and different C$^{18}$O abundances based on the temperature assumptions in panel (b). Here we also include the NH$_3$ gas kinetic temperature from the GAS results (K17). Clearly, the gas kinetic temperature ($T_{\rm kin}$) matches better with the {\it Herschel} dust temperature ($T_{\rm dust}$), while the $^{12}$CO excitation temperature has higher values. The C$^{18}$O abundances based on $T_{\rm kin}$ and $T_{\rm dust}$ agree well, while that based on $T_{\rm ex}$ is higher by a factor of at least two (and is also wider). This comparison shows that in a region like Orion A that is heated by massive stars, the molecular column density can be overestimated due to the elevated excitation temperature from $^{12}$CO. It is not surprising since the C$^{18}$O morphology is quite different from the excitation temperature map (based on $^{12}$CO peak intensity map). The Stick must be embedded in molecular gas.

\begin{figure*}[htb!]
\centering
\epsscale{1.1}
\plotone{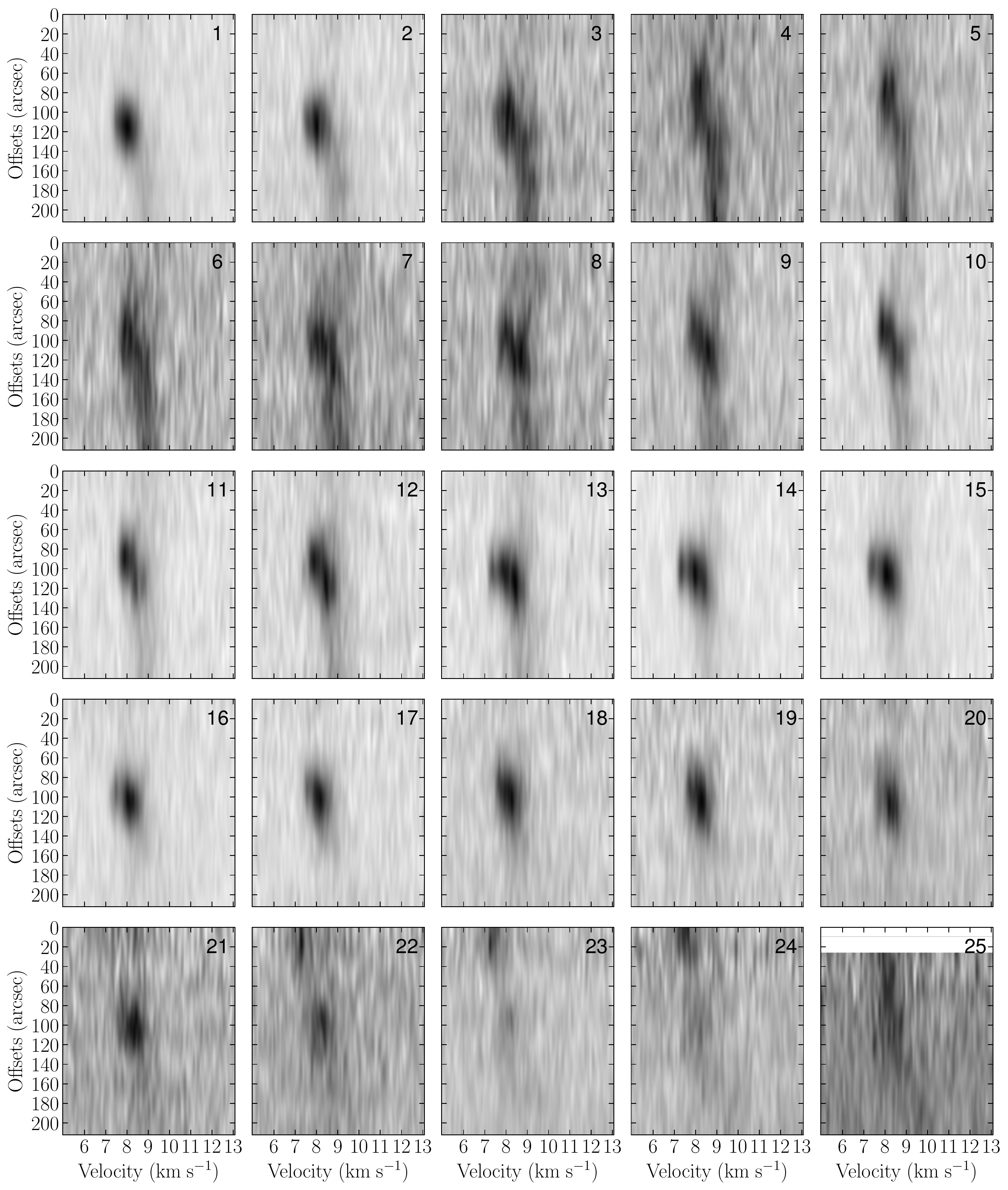}
\caption{
Same as Figure \ref{fig:stickpv} but for NH$_3$ data.
\label{fig:stickpvnh3}}
\end{figure*}

We obtained the publicly available NH$_3$(1,1) data cube from the GAS project \citet{2017ApJ...843...63F}.
In Figure \ref{fig:stickpvnh3} we show NH$_3$ PV-diagrams along the same PV-cuts as the ones shown in Figure \ref{fig:stickpv}. 
They show very similar velocity features as those in the C$^{18}$O PV-diagrams. In particular, we also see two velocity components. The difference is that the NH$_3$ data shows faint or no emission away from the filament. This is likely due to that NH$_3$ traces higher density gas than C$^{18}$O.

\begin{figure*}[htb!]
\centering
\epsscale{1.1}
\plotone{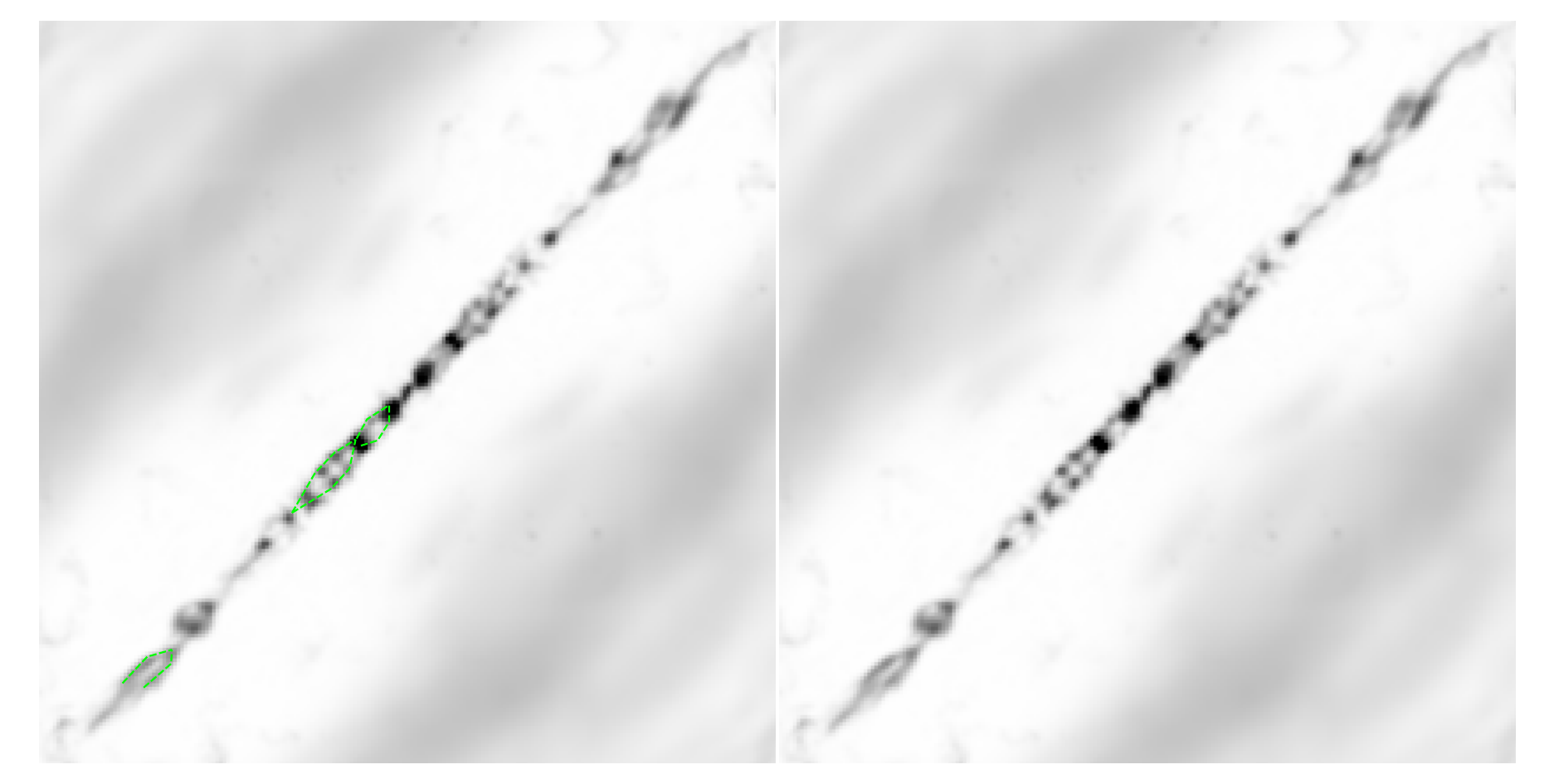}
\caption{
Zoom-in view of channel 0 km s$^{-1}$ from Figure \ref{fig:rtchan18}. The left panel has green dashed curves that mark the rings/forks mentioned in \S\ref{subsec:rt}.
\label{fig:simfork}}
\end{figure*}

Figure \ref{fig:simfork} shows the 0 km s$^{-1}$ channel extracted from Figure \ref{fig:rtchan18}. The two panels are the same except that the left one has green curves overlaid on the ring/fork structures. In the figure we highlight two rings (middle) and one fork (lower-left). A few more rings and forks are also visible to the upper-right. 

\newpage

\bibliography{ref}

\begin{thebibliography}{}
\expandafter\ifx\csname natexlab\endcsname\relax\def\natexlab#1{#1}\fi

\bibitem[{{Andr{\'e}} {et~al.}(2014){Andr{\'e}}, {Di Francesco},
  {Ward-Thompson}, {Inutsuka}, {Pudritz}, \& {Pineda}}]{2014prpl.conf...27A}
{Andr{\'e}}, P., {Di Francesco}, J., {Ward-Thompson}, D., {et~al.} 2014,
  Protostars and Planets VI, 27

\bibitem[{{Arzoumanian} {et~al.}(2011){Arzoumanian}, {Andr{\'e}}, {Didelon},
  {K{\"o}nyves}, {Schneider}, {Men'shchikov}, {Sousbie}, {Zavagno}, {Bontemps},
  {di Francesco}, {Griffin}, {Hennemann}, {Hill}, {Kirk}, {Martin}, {Minier},
  {Molinari}, {Motte}, {Peretto}, {Pezzuto}, {Spinoglio}, {Ward-Thompson},
  {White}, \& {Wilson}}]{2011A&A...529L...6A}
{Arzoumanian}, D., {Andr{\'e}}, P., {Didelon}, P., {et~al.} 2011, \aap, 529, L6

\bibitem[{{Astropy Collaboration} {et~al.}(2013){Astropy Collaboration},
  {Robitaille}, {Tollerud}, {Greenfield}, {Droettboom}, {Bray}, {Aldcroft},
  {Davis}, {Ginsburg}, {Price-Whelan}, {Kerzendorf}, {Conley}, {Crighton},
  {Barbary}, {Muna}, {Ferguson}, {Grollier}, {Parikh}, {Nair}, {Unther},
  {Deil}, {Woillez}, {Conseil}, {Kramer}, {Turner}, {Singer}, {Fox}, {Weaver},
  {Zabalza}, {Edwards}, {Azalee Bostroem}, {Burke}, {Casey}, {Crawford},
  {Dencheva}, {Ely}, {Jenness}, {Labrie}, {Lim}, {Pierfederici}, {Pontzen},
  {Ptak}, {Refsdal}, {Servillat}, \& {Streicher}}]{Astropy-Collaboration13}
{Astropy Collaboration}, {Robitaille}, T.~P., {Tollerud}, E.~J., {et~al.} 2013,
  \aap, 558, A33

\bibitem[{{Balfour} {et~al.}(2015){Balfour}, {Whitworth}, {Hubber}, \&
  {Jaffa}}]{2015MNRAS.453.2471B}
{Balfour}, S.~K., {Whitworth}, A.~P., {Hubber}, D.~A., \& {Jaffa}, S.~E. 2015,
  \mnras, 453, 2471

\bibitem[{{Bally}(2008)}]{2008hsf1.book..459B}
{Bally}, J. 2008, {Overview of the Orion Complex}, ed. B.~{Reipurth}, 459

\bibitem[{{Bally} {et~al.}(1987){Bally}, {Langer}, {Stark}, \&
  {Wilson}}]{1987ApJ...312L..45B}
{Bally}, J., {Langer}, W.~D., {Stark}, A.~A., \& {Wilson}, R.~W. 1987, \apjl,
  312, L45

\bibitem[{{Beuther} {et~al.}(2020){Beuther}, {Wang}, {Soler}, {Linz},
  {Henshaw}, {Vazquez-Semadeni}, {Gomez}, {Ragan}, {Henning}, {Glover}, {Lee},
  \& {G{\"u}sten}}]{2020A&A...638A..44B}
{Beuther}, H., {Wang}, Y., {Soler}, J., {et~al.} 2020, \aap, 638, A44

\bibitem[{{Clark} \& {Glover}(2014)}]{2014MNRAS.444.2396C}
{Clark}, P.~C., \& {Glover}, S. C.~O. 2014, \mnras, 444, 2396

\bibitem[{{Crapsi} {et~al.}(2005){Crapsi}, {Caselli}, {Walmsley}, {Myers},
  {Tafalla}, {Lee}, \& {Bourke}}]{2005ApJ...619..379C}
{Crapsi}, A., {Caselli}, P., {Walmsley}, C.~M., {et~al.} 2005, \apj, 619, 379

\bibitem[{{Crutcher}(2012)}]{2012ARA&A..50...29C}
{Crutcher}, R.~M. 2012, \araa, 50, 29

\bibitem[{{Endres} {et~al.}(2016){Endres}, {Schlemmer}, {Schilke}, {Stutzki},
  \& {M{\"u}ller}}]{Endres2016}
{Endres}, C.~P., {Schlemmer}, S., {Schilke}, P., {Stutzki}, J., \&
  {M{\"u}ller}, H. S.~P. 2016, Journal of Molecular Spectroscopy, 327, 95

\bibitem[{{Feddersen} {et~al.}(2020){Feddersen}, {Arce}, {Kong}, {Suri},
  {S{\'a}nchez-Monge}, {Ossenkopf-Okada}, {Dunham}, {Nakamura}, {Shimajiri}, \&
  {Bally}}]{2020ApJ...896...11F}
{Feddersen}, J.~R., {Arce}, H.~G., {Kong}, S., {et~al.} 2020, \apj, 896, 11

\bibitem[{{Federrath}(2016)}]{2016MNRAS.457..375F}
{Federrath}, C. 2016, \mnras, 457, 375

\bibitem[{{Friesen} {et~al.}(2017){Friesen}, {Pineda}, {co-PIs}, {Rosolowsky},
  {Alves}, {Chac{\'o}n-Tanarro}, {How-Huan Chen}, {Chun-Yuan Chen}, {Di
  Francesco}, {Keown}, {Kirk}, {Punanova}, {Seo}, {Shirley}, {Ginsburg},
  {Hall}, {Offner}, {Singh}, {Arce}, {Caselli}, {Goodman}, {Martin}, {Matzner},
  {Myers}, {Redaelli}, \& {The GAS Collaboration}}]{2017ApJ...843...63F}
{Friesen}, R.~K., {Pineda}, J.~E., {co-PIs}, {et~al.} 2017, \apj, 843, 63

\bibitem[{{Fukui} {et~al.}(2016){Fukui}, {Torii}, {Ohama}, {Hasegawa},
  {Hattori}, {Sano}, {Ohashi}, {Fujii}, {Kuwahara}, {Mizuno}, {Dawson},
  {Yamamoto}, {Tachihara}, {Okuda}, {Onishi}, \&
  {Mizuno}}]{2016ApJ...820...26F}
{Fukui}, Y., {Torii}, K., {Ohama}, A., {et~al.} 2016, \apj, 820, 26

\bibitem[{{Gardiner} \& {Stone}(2005)}]{2005JCoPh.205..509G}
{Gardiner}, T.~A., \& {Stone}, J.~M. 2005, Journal of Computational Physics,
  205, 509

\bibitem[{{Glover} \& {Clark}(2012)}]{GloverClark2012}
{Glover}, S. C.~O., \& {Clark}, P.~C. 2012, \mnras, 421, 116

\bibitem[{{Gonz{\'a}lez Lobos} \& {Stutz}(2019)}]{2019MNRAS.489.4771G}
{Gonz{\'a}lez Lobos}, V., \& {Stutz}, A.~M. 2019, \mnras, 489, 4771

\bibitem[{{Gro{\ss}schedl} {et~al.}(2018){Gro{\ss}schedl}, {Alves}, {Meingast},
  {Ackerl}, {Ascenso}, {Bouy}, {Burkert}, {Forbrich}, {F{\"u}rnkranz},
  {Goodman}, {Hacar}, {Herbst-Kiss}, {Lada}, {Larreina}, {Leschinski},
  {Lombardi}, {Moitinho}, {Mortimer}, \& {Zari}}]{2018A&A...619A.106G}
{Gro{\ss}schedl}, J.~E., {Alves}, J., {Meingast}, S., {et~al.} 2018, \aap, 619,
  A106

\bibitem[{{Hasegawa} \& {Herbst}(1993)}]{1993MNRAS.261...83H}
{Hasegawa}, T.~I., \& {Herbst}, E. 1993, \mnras, 261, 83

\bibitem[{{Heiles}(1997)}]{1997ApJS..111..245H}
{Heiles}, C. 1997, \apjs, 111, 245

\bibitem[{{Hennebelle}(2013)}]{2013A&A...556A.153H}
{Hennebelle}, P. 2013, \aap, 556, A153

\bibitem[{{Hunter}(2007)}]{matplotlib}
{Hunter}, J.~D. 2007, Computing in Science and Engineering, 9, 90

\bibitem[{{Joye} \& {Mandel}(2003)}]{2003ASPC..295..489J}
{Joye}, W.~A., \& {Mandel}, E. 2003, in Astronomical Society of the Pacific
  Conference Series, Vol. 295, Astronomical Data Analysis Software and Systems
  XII, ed. H.~E. {Payne}, R.~I. {Jedrzejewski}, \& R.~N. {Hook}, 489

\bibitem[{{Kirk} {et~al.}(2017){Kirk}, {Friesen}, {Pineda}, {Rosolowsky},
  {Offner}, {Matzner}, {Myers}, {Di Francesco}, {Caselli}, {Alves},
  {Chac{\'o}n-Tanarro}, {Chen}, {Chun-Yuan Chen}, {Keown}, {Punanova}, {Seo},
  {Shirley}, {Ginsburg}, {Hall}, {Singh}, {Arce}, {Goodman}, {Martin}, \&
  {Redaelli}}]{2017ApJ...846..144K}
{Kirk}, H., {Friesen}, R.~K., {Pineda}, J.~E., {et~al.} 2017, \apj, 846, 144

\bibitem[{{Klein} \& {Woods}(1998)}]{1998ApJ...497..777K}
{Klein}, R.~I., \& {Woods}, D.~T. 1998, \apj, 497, 777

\bibitem[{{Kong} {et~al.}(2015){Kong}, {Lada}, {Lada},
  {Rom{\'a}n-Z{\'u}{\~n}iga}, {Bieging}, {Lombardi}, {Forbrich}, \&
  {Alves}}]{2015ApJ...805...58K}
{Kong}, S., {Lada}, C.~J., {Lada}, E.~A., {et~al.} 2015, \apj, 805, 58

\bibitem[{{Kong} {et~al.}(2018){Kong}, {Arce}, {Feddersen}, {Carpenter},
  {Nakamura}, {Shimajiri}, {Isella}, {Ossenkopf-Okada}, {Sargent},
  {S{\'a}nchez-Monge}, {Suri}, {Kauffmann}, {Pillai}, {Pineda}, {Koda},
  {Bally}, {Lis}, {Padoan}, {Klessen}, {Mairs}, {Goodman}, {Goldsmith},
  {McGehee}, {Schilke}, {Teuben}, {Jos{\'e} Maureira}, {Hara}, {Ginsburg},
  {Burkhart}, {Smith}, {Schmiedeke}, {Pineda}, {Ishii}, {Sasaki}, {Kawabe},
  {Urasawa}, {Oyamada}, \& {Tanabe}}]{2018ApJS..236...25K}
{Kong}, S., {Arce}, H.~G., {Feddersen}, J.~R., {et~al.} 2018, The Astrophysical
  Journal Supplement Series, 236, 25

\bibitem[{{Kong} {et~al.}(2019){Kong}, {Arce}, {Sargent}, {Mairs}, {Klessen},
  {Bally}, {Padoan}, {Smith}, {Jos{\'e} Maureira}, {Carpenter}, {Ginsburg},
  {Stutz}, {Goldsmith}, {Meingast}, {McGehee}, {S{\'a}nchez-Monge}, {Suri},
  {Pineda}, {Alves}, {Feddersen}, {Kauffmann}, \&
  {Schilke}}]{2019ApJ...882...45K}
{Kong}, S., {Arce}, H.~G., {Sargent}, A.~I., {et~al.} 2019, \apj, 882, 45

\bibitem[{{Kounkel} {et~al.}(2018){Kounkel}, {Covey}, {Su{\'a}rez},
  {Rom{\'a}n-Z{\'u}{\~n}iga}, {Hernandez}, {Stassun}, {Jaehnig}, {Feigelson},
  {Pe{\~n}a Ram{\'\i}rez}, {Roman-Lopes}, {Da Rio}, {Stringfellow}, {Kim},
  {Borissova}, {Fern{\'a}ndez-Trincado}, {Burgasser},
  {Garc{\'\i}a-Hern{\'a}ndez}, {Zamora}, {Pan}, \&
  {Nitschelm}}]{2018AJ....156...84K}
{Kounkel}, M., {Covey}, K., {Su{\'a}rez}, G., {et~al.} 2018, \aj, 156, 84

\bibitem[{{Kowal} {et~al.}(2011){Kowal}, {de Gouveia Dal Pino}, \&
  {Lazarian}}]{2011ApJ...735..102K}
{Kowal}, G., {de Gouveia Dal Pino}, E.~M., \& {Lazarian}, A. 2011, \apj, 735,
  102

\bibitem[{{Kowal} {et~al.}(2009){Kowal}, {Lazarian}, {Vishniac}, \&
  {Otmianowska-Mazur}}]{2009ApJ...700...63K}
{Kowal}, G., {Lazarian}, A., {Vishniac}, E.~T., \& {Otmianowska-Mazur}, K.
  2009, \apj, 700, 63

\bibitem[{{Krumholz}(2017)}]{2017stfo.book.....K}
{Krumholz}, M.~R. 2017, {Star Formation}, doi:10.1142/10091

\bibitem[{{Lada} {et~al.}(2017){Lada}, {Lewis}, {Lombardi}, \&
  {Alves}}]{2017A&A...606A.100L}
{Lada}, C.~J., {Lewis}, J.~A., {Lombardi}, M., \& {Alves}, J. 2017, \aap, 606,
  A100

\bibitem[{{Lada} {et~al.}(2009){Lada}, {Lombardi}, \&
  {Alves}}]{2009ApJ...703...52L}
{Lada}, C.~J., {Lombardi}, M., \& {Alves}, J.~F. 2009, \apj, 703, 52

\bibitem[{{Lane} {et~al.}(2016){Lane}, {Kirk}, {Johnstone}, {Mairs}, {Di
  Francesco}, {Sadavoy}, {Hatchell}, {Berry}, {Jenness}, {Hogerheijde},
  {Ward-Thompson}, \& {The JCMT Gould Belt Survey Team}}]{Lane2016}
{Lane}, J., {Kirk}, H., {Johnstone}, D., {et~al.} 2016, \apj, 833, 44

\bibitem[{{Larson}(1981)}]{Larson1981}
{Larson}, R.~B. 1981, \mnras, 194, 809

\bibitem[{{Lazarian}(2014)}]{2014SSRv..181....1L}
{Lazarian}, A. 2014, \ssr, 181, 1

\bibitem[{{Lazarian} {et~al.}(2012){Lazarian}, {Esquivel}, \&
  {Crutcher}}]{2012ApJ...757..154L}
{Lazarian}, A., {Esquivel}, A., \& {Crutcher}, R. 2012, \apj, 757, 154

\bibitem[{{Lazarian} \& {Vishniac}(1999)}]{1999ApJ...517..700L}
{Lazarian}, A., \& {Vishniac}, E.~T. 1999, \apj, 517, 700

\bibitem[{{Le Bourlot}(1991)}]{LeBourlot1991}
{Le Bourlot}, J. 1991, \aap, 242, 235

\bibitem[{{Lewis} \& {Lada}(2016)}]{2016ApJ...825...91L}
{Lewis}, J.~A., \& {Lada}, C.~J. 2016, \apj, 825, 91

\bibitem[{{Li} \& {Goldsmith}(2003)}]{2003ApJ...585..823L}
{Li}, D., \& {Goldsmith}, P.~F. 2003, \apj, 585, 823

\bibitem[{{Lim} {et~al.}(2020){Lim}, {Nakamura}, {Wu}, {Bisbas}, {Tan},
  {Chambers}, {Bally}, {Kong}, {McGehee}, {Lis}, {Ossenkopf-Okada}, \&
  {S{\'a}nchez-Monge}}]{2020PASJ..tmp..187L}
{Lim}, W., {Nakamura}, F., {Wu}, B., {et~al.} 2020, \pasj, arXiv:2004.03668

\bibitem[{{Lombardi} {et~al.}(2014){Lombardi}, {Bouy}, {Alves}, \&
  {Lada}}]{2014A&A...566A..45L}
{Lombardi}, M., {Bouy}, H., {Alves}, J., \& {Lada}, C.~J. 2014, \aap, 566, A45

\bibitem[{{Lubow} \& {Pringle}(1996)}]{1996MNRAS.279.1251L}
{Lubow}, S.~H., \& {Pringle}, J.~E. 1996, \mnras, 279, 1251

\bibitem[{{Marinho} {et~al.}(2001){Marinho}, {Andreazza}, \&
  {L{\'e}pine}}]{2001A&A...379.1123M}
{Marinho}, E.~P., {Andreazza}, C.~M., \& {L{\'e}pine}, J.~R.~D. 2001, \aap,
  379, 1123

\bibitem[{{Miniati} {et~al.}(1997){Miniati}, {Jones}, {Ferrara}, \&
  {Ryu}}]{1997ApJ...491..216M}
{Miniati}, F., {Jones}, T.~W., {Ferrara}, A., \& {Ryu}, D. 1997, \apj, 491, 216

\bibitem[{{Miniati} {et~al.}(1999){Miniati}, {Ryu}, {Ferrara}, \&
  {Jones}}]{1999ApJ...510..726M}
{Miniati}, F., {Ryu}, D., {Ferrara}, A., \& {Jones}, T.~W. 1999, \apj, 510, 726

\bibitem[{{Miville-Desch{\^e}nes} {et~al.}(2010){Miville-Desch{\^e}nes},
  {Martin}, {Abergel}, {Bernard}, {Boulanger}, {Lagache}, {Anderson},
  {Andr{\'e}}, {Arab}, {Baluteau}, {Blagrave}, {Bontemps}, {Cohen},
  {Compiegne}, {Cox}, {Dartois}, {Davis}, {Emery}, {Fulton}, {Gry}, {Habart},
  {Huang}, {Joblin}, {Jones}, {Kirk}, {Lim}, {Madden}, {Makiwa}, {Menshchikov},
  {Molinari}, {Moseley}, {Motte}, {Naylor}, {Okumura}, {Pinheiro
  Gon{\c{c}}alves}, {Polehampton}, {Rod{\'o}n}, {Russeil}, {Saraceno},
  {Schneider}, {Sidher}, {Spencer}, {Swinyard}, {Ward-Thompson}, {White}, \&
  {Zavagno}}]{2010A&A...518L.104M}
{Miville-Desch{\^e}nes}, M.~A., {Martin}, P.~G., {Abergel}, A., {et~al.} 2010,
  \aap, 518, L104

\bibitem[{{Nagai} {et~al.}(1998){Nagai}, {Inutsuka}, \&
  {Miyama}}]{1998ApJ...506..306N}
{Nagai}, T., {Inutsuka}, S.-i., \& {Miyama}, S.~M. 1998, \apj, 506, 306

\bibitem[{{Nakamura} {et~al.}(2012){Nakamura}, {Miura}, {Kitamura},
  {Shimajiri}, {Kawabe}, {Akashi}, {Ikeda}, {Tsukagoshi}, {Momose}, {Nishi}, \&
  {Li}}]{2012ApJ...746...25N}
{Nakamura}, F., {Miura}, T., {Kitamura}, Y., {et~al.} 2012, \apj, 746, 25

\bibitem[{{Ochsendorf} {et~al.}(2015){Ochsendorf}, {Brown}, {Bally}, \&
  {Tielens}}]{Ochsendorf2015}
{Ochsendorf}, B.~B., {Brown}, A. G.~A., {Bally}, J., \& {Tielens}, A. G.~G.~M.
  2015, \apj, 808, 111

\bibitem[{{Ossenkopf}(2002)}]{Ossenkopf2002}
{Ossenkopf}, V. 2002, \aap, 391, 295

\bibitem[{{Ossenkopf} \& {Henning}(1994)}]{1994A&A...291..943O}
{Ossenkopf}, V., \& {Henning}, T. 1994, \aap, 291, 943

\bibitem[{{Padoan} {et~al.}(2016){Padoan}, {Pan}, {Haugb{\o}lle}, \&
  {Nordlund}}]{2016ApJ...822...11P}
{Padoan}, P., {Pan}, L., {Haugb{\o}lle}, T., \& {Nordlund}, {\r{A}}. 2016,
  \apj, 822, 11

\bibitem[{{Parker}(1957)}]{1957JGR....62..509P}
{Parker}, E.~N. 1957, \jgr, 62, 509

\bibitem[{{Planck Collaboration} {et~al.}(2016){Planck Collaboration}, {Ade},
  {Aghanim}, {Alves}, {Arnaud}, {Arzoumanian}, {Aumont}, {Baccigalupi},
  {Banday}, {Barreiro}, {Bartolo}, {Battaner}, {Benabed}, {Benoit-L{\'e}vy},
  {Bernard}, {Bern{\'e}}, {Bersanelli}, {Bielewicz}, {Bonaldi}, {Bonavera},
  {Bond}, {Borrill}, {Bouchet}, {Boulanger}, {Bracco}, {Burigana}, {Calabrese},
  {Cardoso}, {Catalano}, {Chamballu}, {Chiang}, {Christensen}, {Clements},
  {Colombi}, {Colombo}, {Combet}, {Couchot}, {Crill}, {Curto}, {Cuttaia},
  {Danese}, {Davies}, {Davis}, {de Bernardis}, {de Rosa}, {de Zotti},
  {Delabrouille}, {Dickinson}, {Diego}, {Donzelli}, {Dor{\'e}}, {Douspis},
  {Ducout}, {Dupac}, {Elsner}, {En{\ss}lin}, {Eriksen}, {Falgarone},
  {Ferri{\`e}re}, {Finelli}, {Forni}, {Frailis}, {Fraisse}, {Franceschi},
  {Frejsel}, {Galeotta}, {Galli}, {Ganga}, {Ghosh}, {Giard},
  {Giraud-H{\'e}raud}, {Gjerl{\o}w}, {Gonz{\'a}lez-Nuevo}, {G{\'o}rski},
  {Gregorio}, {Gruppuso}, {Guillet}, {Hansen}, {Hanson}, {Harrison},
  {Hern{\'a}ndez-Monteagudo}, {Herranz}, {Hildebrandt}, {Hivon}, {Hobson},
  {Holmes}, {Huffenberger}, {Hurier}, {Jaffe}, {Jaffe}, {Jones}, {Juvela},
  {Keskitalo}, {Kisner}, {Knoche}, {Kunz}, {Kurki-Suonio}, {Lagache},
  {Lamarre}, {Lasenby}, {Lawrence}, {Leonardi}, {Levrier}, {Liguori}, {Lilje},
  {Linden-V{\o}rnle}, {L{\'o}pez-Caniego}, {Lubin}, {Mac{\'\i}as-P{\'e}rez},
  {Maffei}, {Mandolesi}, {Mangilli}, {Maris}, {Martin},
  {Mart{\'\i}nez-Gonz{\'a}lez}, {Masi}, {Matarrese}, {Mazzotta}, {Melchiorri},
  {Mendes}, {Mennella}, {Migliaccio}, {Mitra}, {Miville-Desch{\^e}nes},
  {Moneti}, {Montier}, {Morgante}, {Mortlock}, {Munshi}, {Murphy}, {Naselsky},
  {Nati}, {Natoli}, {N{\o}rgaard-Nielsen}, {Noviello}, {Novikov}, {Novikov},
  {Oppermann}, {Pagano}, {Pajot}, {Paladini}, {Paoletti}, {Pasian}, {Perrotta},
  {Pettorino}, {Piacentini}, {Piat}, {Pierpaoli}, {Pietrobon}, {Plaszczynski},
  {Pointecouteau}, {Polenta}, {Pratt}, {Puget}, {Rachen}, {Rebolo}, {Reinecke},
  {Remazeilles}, {Renault}, {Renzi}, {Ricciardi}, {Ristorcelli}, {Rocha},
  {Rosset}, {Rossetti}, {Roudier}, {Rubi{\~n}o-Mart{\'\i}n}, {Rusholme},
  {Sandri}, {Savelainen}, {Savini}, {Scott}, {Soler}, {Stolyarov}, {Sutton},
  {Suur-Uski}, {Sygnet}, {Tauber}, {Terenzi}, {Toffolatti}, {Tomasi},
  {Tristram}, {Tucci}, {Valenziano}, {Valiviita}, {Van Tent}, {Vielva},
  {Villa}, {Wade}, {Wandelt}, {Yvon}, {Zacchei}, \&
  {Zonca}}]{2016A&A...586A.136P}
{Planck Collaboration}, {Ade}, P.~A.~R., {Aghanim}, N., {et~al.} 2016, \aap,
  586, A136

\bibitem[{{Robitaille} \& {Bressert}(2012)}]{Robitaille12}
{Robitaille}, T., \& {Bressert}, E. 2012, {APLpy: Astronomical Plotting Library
  in Python}, Astrophysics Source Code Library, , , ascl:1208.017

\bibitem[{{Smith} {et~al.}(2020){Smith}, {Tre{\ss}}, {Sormani}, {Glover},
  {Klessen}, {Clark}, {Izquierdo}, {Duarte-Cabral}, \&
  {Zucker}}]{2020MNRAS.492.1594S}
{Smith}, R.~J., {Tre{\ss}}, R.~G., {Sormani}, M.~C., {et~al.} 2020, \mnras,
  492, 1594

\bibitem[{{Soler}(2019)}]{2019A&A...629A..96S}
{Soler}, J.~D. 2019, \aap, 629, A96

\bibitem[{{Stone} {et~al.}(2020){Stone}, {Tomida}, {White}, \&
  {Felker}}]{2020ApJS..249....4S}
{Stone}, J.~M., {Tomida}, K., {White}, C.~J., \& {Felker}, K.~G. 2020, \apjs,
  249, 4

\bibitem[{{Stone}(1970{\natexlab{a}})}]{1970ApJ...159..277S}
{Stone}, M.~E. 1970{\natexlab{a}}, \apj, 159, 277

\bibitem[{{Stone}(1970{\natexlab{b}})}]{1970ApJ...159..293S}
---. 1970{\natexlab{b}}, \apj, 159, 293

\bibitem[{{Stutz} \& {Gould}(2016)}]{2016A&A...590A...2S}
{Stutz}, A.~M., \& {Gould}, A. 2016, \aap, 590, A2

\bibitem[{{Stutz} \& {Kainulainen}(2015)}]{2015A&A...577L...6S}
{Stutz}, A.~M., \& {Kainulainen}, J. 2015, \aap, 577, L6

\bibitem[{{Suri} {et~al.}(2019){Suri}, {S{\'a}nchez-Monge}, {Schilke},
  {Clarke}, {Smith}, {Ossenkopf-Okada}, {Klessen}, {Padoan}, {Goldsmith},
  {Arce}, {Bally}, {Carpenter}, {Ginsburg}, {Johnstone}, {Kauffmann}, {Kong},
  {Lis}, {Mairs}, {Pillai}, {Pineda}, \& {Duarte-Cabral}}]{2019A&A...623A.142S}
{Suri}, S., {S{\'a}nchez-Monge}, {\'A}., {Schilke}, P., {et~al.} 2019, \aap,
  623, A142

\bibitem[{{Sweet}(1958)}]{1958IAUS....6..123S}
{Sweet}, P.~A. 1958, in IAU Symposium, Vol.~6, Electromagnetic Phenomena in
  Cosmical Physics, ed. B.~{Lehnert}, 123

\bibitem[{{Tahani} {et~al.}(2019){Tahani}, {Plume}, {Brown}, {Soler}, \&
  {Kainulainen}}]{2019A&A...632A..68T}
{Tahani}, M., {Plume}, R., {Brown}, J.~C., {Soler}, J.~D., \& {Kainulainen}, J.
  2019, \aap, 632, A68

\bibitem[{{Tasker} \& {Tan}(2009)}]{2009ApJ...700..358T}
{Tasker}, E.~J., \& {Tan}, J.~C. 2009, \apj, 700, 358

\bibitem[{van~der Walt {et~al.}(2011)van~der Walt, Colbert, \&
  Varoquaux}]{numpy}
van~der Walt, S., Colbert, S.~C., \& Varoquaux, G. 2011, Computing in Science
  \& Engineering, 13, 22

\bibitem[{{van Zadelhoff} {et~al.}(2002){van Zadelhoff}, {Dullemond}, {van der
  Tak}, {Yates}, {Doty}, {Ossenkopf}, {Hogerheijde}, {Juvela}, {Wiesemeyer}, \&
  {Sch{\"o}ier}}]{vanZadelhoff2002}
{van Zadelhoff}, G.-J., {Dullemond}, C.~P., {van der Tak}, F.~F.~S., {et~al.}
  2002, \aap, 395, 373

\bibitem[{{Vishniac} \& {Lazarian}(1999)}]{1999ApJ...511..193V}
{Vishniac}, E.~T., \& {Lazarian}, A. 1999, \apj, 511, 193

\bibitem[{{Wakelam} \& {Herbst}(2008)}]{2008ApJ...680..371W}
{Wakelam}, V., \& {Herbst}, E. 2008, \apj, 680, 371

\bibitem[{{Wu} {et~al.}(2020){Wu}, {Tan}, {Christie}, \&
  {Nakamura}}]{2020ApJ...891..168W}
{Wu}, B., {Tan}, J.~C., {Christie}, D., \& {Nakamura}, F. 2020, \apj, 891, 168

\bibitem[{{Yang} {et~al.}(2010){Yang}, {Stancil}, {Balakrishnan}, \&
  {Forrey}}]{Yang2010}
{Yang}, B., {Stancil}, P.~C., {Balakrishnan}, N., \& {Forrey}, R.~C. 2010,
  \apj, 718, 1062

\end{thebibliography}
\bibliographystyle{aasjournal}

\end{document}